\documentclass[a4paper, 11pt, oneside]{report}  % Use the "Thesis" style, based on the ECS Thesis style by Steve Gunn
\usepackage{epsfig}
\usepackage{epstopdf}
\usepackage{amsmath,amssymb,amsthm}
\usepackage{dcolumn}
\usepackage{bm}
\usepackage{graphicx}
\usepackage{subfig}
\usepackage{caption}
\begin{document}

\title  {Theoretical tools for simulations of cluster dynamics in strong laser pulses}
\author  {Dragos Iustin Palade }
\maketitle
%% ----------------------------------------------------------------

%% ----------------------------------------------------------------
% The "Funny Quote Page"
\pagestyle{empty}  % No headers or footers for the following pages

\null\vfill
% Now comes the "Funny Quote", written in italics
\textit{``The fundamental laws necessary for the mathematical treatment of a large part of physics and the whole of chemistry are thus completely known, and the difficulty lies only in the fact that application of these laws leads to equations that are too complex to be solved.''}

\begin{flushright}
Paul Dirac
\end{flushright}

\vfill\vfill\vfill\vfill\vfill\vfill\null
\clearpage  % Funny Quote page ended, start a new page
%% ----------------------------------------------------------------

% The Abstract Page
\tableofcontents  % Write out the Table of Contents

%% ----------------------------------------------------------------

% Chapter 1

\chapter{Introduction} % Write in your own chapter title
\label{Introduction}

During the last decades, the topic of cluster physics became a genuine scientific and interdisciplinary field of research. This evolution was driven on one side by the fact that clusters provide a tool for tuning phenomena at nano or even atomic scale. On the other hand there was an academic boost in research due to evolution of experimental techniques which allowed for more precise experiments at nano-scale level. But perhaps the most important boost was given by the computer development. The fact is that since the invention of the transistor, the Moore's law \cite{mack2011fifty} has never stopped working and, in the $80's$, the computers were already capable of dealing with quite difficult and demanding numerical problems. That was the period when the exponential era of many-body quantum theories began with applications like Density Functional Theory (which was basically dormant since $1965$ when it was designed as practical tool) and mankind was finally able to simulate electron dynamics in the frame of different approximations in solid and in molecular physics. 

Having the numerical tool at hand, the theoreticians became interested in these wonderful objects, the atomic clusters, and in the $90's$, one could already find in literature a considerable number of studies and reviews both on experimental and theoretical methods in cluster physics.

The metal clusters took immediately the lead due to their simple structure and the interesting features exhibited in the optical spectra. New behaviour, not present in solid state physics, nor in molecular physics was seen and the interest in the interaction with electromagnetic fields ascended quickly in the hierarchy of subjects. 

Parallel with clusters another field of interest in physics gain considerable momentum in the same period: the lasers. This was driven by the Chirped Pulse Amplification (CPA) technique which was invented by G\'erard Mourou and Donna Strickland at the University of Rochester in the mid $80's$ \cite{strickland1985compression}. CPA putted the lasers on a track in which their intensity (until than almost stagnating) become exponentially growing in time and now, we are looking forward to a generation of Peta-Watt lasers \cite{eli}. 

Naturally, lasers entered almost instantaneously in cluster physics as an easily tunable experimental tool for optical studies from which a large variety of spectra could be obtained and interpreted in terms of cluster properties. The theoreticians took their job seriously and soon, large codes implementing various theories were putted at work to reproduce the experimental results. 

As cluster and laser physics developed together into this beautiful flower of science, we are now at an edge where there is so many to know about what it has been done, but there are also serious theoretical questions about what and how to investigate further. 

It is the author's opinion that a new view on the general many-body problem and in particular on the quantum many-body systems is needed. This could translate in an active research topic in conceptual and numerical aspects but also in reaching new frontiers for laser-cluster phenomena, perhaps with many applications. 

While this might be true, the purpose of the present work is not to give such a new insight, but rather to establish a good, structured picture of the existing work with its pure theoretical or numerical aspects and to connect them with the experimental data.

In literature, there is a series of reviews and books that discuss different aspects of the topic, but comprehensive studies on the theoretical approaches are rare. Usually, one has to come with a strong theoretical background or to span many different works to have a clear picture about the theory that is to be known. The truth is that the subject is so complex that not they, nor I can construct a complete presentation in such narrow spaces as reviews or academic theses. Perhaps, a series of books would cover it, but by the time one would write a volume, another volume should be written to cover what has been done in the mean time.

Therefore, in the present work, I will try to fix a logical hierarchy of the theoretical methods to be used, on how are they connected with various experimental quantities or with different laser regimes and to provide enough further references that could allow a logical follow up into the subject.

\section{Problem setting}
\label{problemsetting}

In a standard thesis, the author can usually write a few pages on the central problem to be discussed. In my case, it is a challenge to point out what part of the theoretical methods used in laser-cluster interaction phenomena are more important and should be the core of the present work. It is rather that the whole concept of theoretical work in such complex systems is important and this is what a virtual reader should focus on.

Without diminishing in any way the importance of the experimentalist, in my opinion, the theoretician has a special challenge to overcome in the sense that is mandatory to have a wide spectrum of "interdisciplinary" knowledge. For our topic, a moderate mathematical background is needed in terms of analysis, algebra, geometry, etc. This should be accompanied with a good understanding of the electro-magnetism, at least in the classical sense, since almost all interactions are of electro-magnetic nature. On the other hand, even if a lot of theoretical methods employ, for the sake of simplicity, classical or semi-classical methods, the presence of quantum effect is pregnant in cluster physics, and so, the need for a strong background in quantum statistics, is stringent. This quantum methods, superpose with the quantum chemistry and the theories from mesoscopic solid state physics in a strange way, but still, the topic remains self consistent. Not the least, a cluster can be viewed as a quantum plasma, especially at high temperatures, as it is the case in strong lasers fields, therefore, a good knowledge of plasma physics and in particular quantum plasmas is required, given the fact that many of the theoretical methods overlap, at least conceptually.

As it is hard to say which theoretical method is more important, it is also hard to extract from the field of laser-cluster interaction a particular matter, or a goal to be achieved above all others. Therefore, I stress again that the main purpose of this thesis is not to give insight (as usually is done) in some narrow problem presenting the achievements of the author. In contrast, it was designed to be a \emph{pastel} of how the zoo of theories grows on the physical problem and reflects \emph{how little} the author has achieved while studying this matter. 

Finally, if there would be to mention a certain point in which the existing theory slips, is the fact that in strong laser fields, the scales involved span a few orders of magnitude, fact which makes any attempt to study the system with quantum methods an impossible task, the last resort measure being the classical methods which miss essential quantum phenomena. The compromise has to be done between classical and quantum theories and a mid-path is, in my opinion, the main problem in the present state of the research.

Writing this thesis required a lot of reading, but beside all the references cited during the following pages (and many others) it is important to mention some fundamental books and articles that stood at my bedside during the last two years \cite{manfredi2005model,reinhard2008introduction,haas2011introduction,fennel2010laser,bonitz1998quantum,jeckelmann739h} and some related (well written) PhD thesis \cite{deiss2009simulation,kandadai2012interaction,di2014dynamics,nogueira2008relativistic,mikaberidze2011atomic,walters2009atoms}

\section{Applications}

For the author, the most important aspect of the field of laser-cluster coupling is that it represents on of the richest systems in which the problem of radiation-matter interaction can be studied from a quantum many body perspective.

But the truth is that the scientific interest in the subject was just partially driven by this academic interest. The applicative possibilities were the ones that drove strongly the field. Just for a visual impact we should note some references which discuss the applications of clusters in medicine \cite{Kollmer2004153,duguet2006magnetic,jain2008noble,zharov2005synergistic,huang2007gold,bakry2007medicinal}, chemistry \cite{daniel2004gold, belloni2006nucleation, ferrando2008nanoalloys}, optics \cite{murphy2005anisotropic,schaadt2005enhanced}, microelectronic \cite{schon1995fascinating}, etc.
The list can go on, but it is less focused on the laser part and more on how sole clusters could be used. To have a more detailed enumeration of what can be achieved (until now) from clusters irradiated with strong lasers, I should mention:

\begin{enumerate}
	\item Production of shaped ion energy spectra \cite{bychenkov2005coulomb,kovalev2007quasimonoenergetic}
	\item Production of highly energetic charged ions \cite{ditmire1998explosion}.
	\item Ejection of hot electrons \cite{springate2003electron}
	\item Emission of extreme UV and X-ray photons \cite{dobosz1997absolute,kumarappan2001asymmetric,rozet2001state}
	\item Ultrafast X-ray diagnostics \cite{fukuda2004generation,fukuda2008soft}
	\item application of cluster explosion in inertial fusion confinement  \cite{ditmire1999nuclear}, \cite{holmlid2009ultrahigh},\cite{kishimoto2002high}
	\item Avalanches of ionized electrons can be used as a mechanism of damage in solids with applications in material processing, microfluidic devices, nano-surgery \cite{gattass2008femtosecond}
	\item A novel application which until now has been used only in small molecules  \cite{vager1989coulomb} would be to use the Coulomb explosion imaging approach to nano scale systems.
	\item High harmonic generation \cite{donnelly1996high,tisch1997investigation}.
\end{enumerate}

More detailed discussion and general situations can be found in \cite{daido2012review}. 

\section{Thesis structure}
\label{thesisstructure}

Beside the present Introduction \ref{Introduction}, this thesis is structured in $3$ chapters.

The second chapter \ref{Chapter1} is designed to give a short definition with appropriate pictures to the concept of the cluster. The picture is complemented by a shorter description on how laser pulses are modelled in the domain of interest. A gross classification of the main dynamical regimes of interaction is given with its quantities to be connected with the experimental and theoretical methods later on. 

The third chapter \ref{Chapter2} is the heart of this work. It gives us the structured picture of the existing theories used in cluster physics starting from first principle and axiomatic Quantum Statistics (Liouville-Von-Neumann equation) and going through different levels of approximation until the semi-classical and even classical methods are obtained. For logical reasons, a short, formal derivation in cascade is tried to be exposed in there. Since this is just a master thesis, one cannot hope for refined details about any of the presented theories (about many of them books have been written) but rather for a schematic image which should allow one to choose easily the appropriate method to be used.

Being a theoretician is hard... Your experiment is the numerical simulation. For this reason the fourth chapter \ref{Chapter3} exists. It is a representation of the numerical work (which, as always, took almost $80\%$ of the time devoted to the thesis) that has been done in the process of understanding and practising the theory. It was the author's intention to apply for different phenomena or regimes of study in cluster-laser physics as many theories as possible and to give insight about the numerical methods involved. If someone would read this work, that is the place where fancy coloured pictures are!

The last chapter \ref{Conclusions} provides a set of conclusions, future plans and perspective of the problems discussed in the thesis.

% Chapter 1

\chapter{Clusters $\&$ lasers} % Write in your own chapter title
\label{Chapter1}

\section{Clusters: the world of not too few, nor many enough}
\label{clusters}

Before entering in the technical details of laser-cluster interaction and the associated theoretical methods, it is necessary to have some understanding about what is a cluster in general and what are the characteristic properties to be expected in such systems. 

\emph{Clusters are, by definition, aggregates of atoms or molecules with regular and arbitrarily scalable repetition of a basic building block. Their size is intermediate between atoms and bulk. One could thus loosely characterize them with a formula of the form $X_n$ where $(3\le n\le 10^{5-7})$.} \cite{reinhard2008introduction}

In a pedestrian view, one could imagine that a cluster is some collection of atoms (many times the same type of atom) that stick together in some random shape. First question would be: why not to call them molecules? In the end, a molecule is also a (small) collection of atoms. Well, the difference is done by the larger number of atoms in a cluster and the properties therein. Perhaps the most pregnant difference is that the usual molecules has a small number of isomers and a precise (easy to identify by numerical means) configuration for the ground state. In cluster, this is no longer the case since even small clusters can present large number of isomers. For example $Ar_{13}$ it is known to have hundreds of isomers. On the other hand, although the quantum effects are present (and essential) in cluster dynamics, many times they behave in a coherent manner giving rise to collective phenomena, less present in molecules.

At the other extreme, one could ask: why not to consider very large clusters as being solid bulk and study them with the associate methods? Well, again it is a matter of quantum effects. It can happen that, even large, the arising quantum phenomena to cannot be neglected in a classical manner. But the most important difference between usual solid state systems (as crystalline lattices) and clusters (which are finite systems), is that the latter lacks the periodicity. There might be symmetries as in spherical clusters or fullerenes, but the translational symmetry (the usual solid state periodicity) is generally not present, therefore, it is hard to speak about band structure, etc.

\begin{figure}
	\centering
	\includegraphics[width=0.5\linewidth]{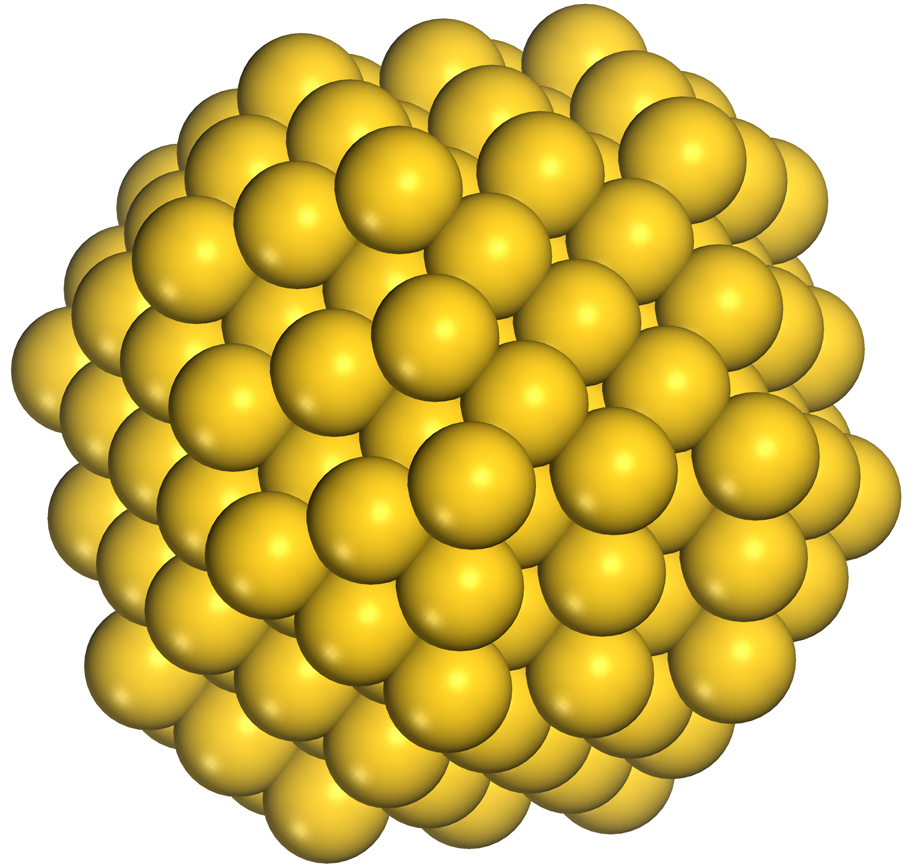}
	\caption{A pedestrian view on a gold cluster}
	\label{fig:gulp-nanocluster}
\end{figure}

Regarding the cluster material, as said before, it is standard to study clusters of a single type of atom. Historically, the first to emerge and by far the most studied are the metal clusters ($Na$, $Li$, etc.). This happened for many reasons: the quasi-free electrons, the easiness of producing them, the strong optical response, etc. Another class of clusters that attracted great interest during time is the class of fullerenes which are carbon compounds with high symmetries and astonishing physical and chemical properties. In the matter of intense laser-cluster interaction, has become customary in the past years to study the response of rare gas elements clusters ($Ar$, $Xe$, $Ne$, etc. ) especially due to the fact that they are a source of many exotic phenomena (the X-ray production, for example).

The theoretical approaches which will be discussed in Chapter \ref{Chapter2} will be used further in the numerical chapter \ref{Chapter3} of this thesis to investigate properties of some types of clusters, in particular the $Na_n$, $C_n$ and $Xe_n$. For this reason, I shall focus next on describing the main physical scales for this three cases.

\subsection{Characteristic scales}

In general is important to have a good intuitive sense about the magnitude of the observables in a physical system since it can give you insight about the applicability of various theories. In this matter, we should describe in this subsection some specific quantities characteristic to clusters. Similar discussions can be found in \cite{fennel2010laser,manfredi2005model,haas2011introduction}

\begin{enumerate}
	\item \textbf{The number of constituents}. The number of atoms in a cluster spans almost 6 orders of magnitudes. With this, problems or simplification arise. Generally, as in nuclear physics, there are specific clusters (the so called magic clusters) which are more abundant in universe and this can be seen also from mass spectrometry experiments. The existence of magic clusters is a consequence of the quantum nature of the electronic structure which allows for the presence of quantum shells.
	
	\item \textbf{Wigner-Seitz radius $r_s$}. The concept is used as the radius of a virtual sphere in which an atom from some material is contained in the bulk regime. Therefore, it is related with the type of material and has a weak physical meaning for (small) clusters. Still, many quantities can be expressed or normalized in terms of it.
	
	\item \textbf{The size}. As one can imagine, when you are dealing with a few atom cluster, the specific dimensions are $1\mathring{A}$. But as the number of atoms increases, one can find at the superior limit, clusters with a diameter of $\sim 10^2nm$. For spherical shaped clusters, the radius can be roughly expressed as $R_0=r_sN^{1/3}$, where $N$ is the number of atoms. From there a gross electronic density can be computed as $\rho_0\approx 3/(4\pi r_s^3)$.
	
	\item \textbf{The time scale}. The time scale is very important in the dynamical regime. In metal clusters for example, where the electrons are quasi free and they respond very easy (and fast) to external fields, the dynamics has a specific time of $\sim 1fs$. This approximate value is a general magnitude order for electrons in all kinds of clusters. This aspect reflects in experiments and numerical simulations, where the electron dynamics is usually investigated for at most a few hundreds of $fs$. On the other hand, the numerical time steps needed are around $10^{-3}fs$. The implications of this values are important for the laser-cluster interactions because it basically sets a specific range of frequencies at which the optical response of the cluster is resonating. As we will see in Sect. \ref{Laser-Cluster coupling regimes}, the dynamics of ions can be neglected for weak lasers (smaller then $I<10^{12}W/cm^2$). But when the power exceeds such values, the response of ions to the external fields is no longer negligible and we must deal with different ionic motions, from vibrations to escaping particles (as in a Coulomb explosion). Nonetheless, having in mind the ratio between the mass of a nucleon and an electron, it is clear that the ions have larger characteristic time scales with a generic value of tens of $fs$ or even $ps$.
	
	\item \textbf{Resonance frequency}. A significant part of the theoretical and experimental investigations of clusters is concerned with the optical or more generally, the dynamical response (linear or non-linear). The purpose is to find maxima in the response spectrum which mark the energetic resonances of the system. The general case requires complicated approaches, but a first glance on the magnitude of the resonance frequency (related to Mie plasmon in metal cluster) can be obtained modelling the cluster as a spherical metal drop and by Mie's theory \cite{mie1908beitrage}:
	
	$$\omega_{p}=(\frac{e}{4\pi\varepsilon_0m_er_s^3})^{1/2}$$ 
	
	\item \textbf{Landau fragmentation/collisions}. In clusters, this phenomenon, known from classical plasma physics is related with the coupling between plasmons (collective) and single-particle excitations with energies close to $\hbar\omega_{Mie}$. Roughly, the characteristic time of Landau damping can be expressed \cite{yannouleas1990microscopic, yannouleas1992landau} $\tau_L=(r_s^2N^{1/3})/v_F$, where $v_F=(\hbar/m)(9\pi/4)^{1/3}$ is the normalized Fermi velocity.
	
	\item \textbf{Electron-ion collisions}. Is a phenomenon which appears due to non zero temperatures in clusters and scales as $\tau_{ei}\propto T^{-1}$ at low temperature, due to electron-phonon scattering \cite{ashcroft1976solid}, while at high temperatures $\tau\propto T^{3/2}$ \cite{spitzer2013physics}. 
	
	\item \textbf{Electron-electron collisions}. Is a phenomenon related with the termalization of a cluster, or the evaporation of electrons. At law temperatures the characteristic time goes like $\tau_{ee}\propto T^{-2}$ accordingly with Fermi liquid theory \cite{pines1966theory}.
	
	\item \textbf{Ionization potentials}. This quantity is in general related with the energy of HOMO (the highest occupied molecular orbital) and it has lower values in metals. Still in general clusters, a wide range of values can be found, an roughly approximation being the order of $eV$. An interesting aspect is that the IP in a cluster can be dramatically different from the same IP in a single atom from the same element.
	
	\item \textbf{Critical laser intensity}. It represents the needed intensity of the laser field to induce ionization in a cluster. Even though tunnelling ionizations can appear, the quantity is closely related to the IP.
	
	\item \textbf{Thermal de Broigle wavelength $\lambda_B$}. Stays as a measure of the quantum effects in a thermal plasma. In an intuitive representation it can be seen as the spatial extension of a particle. When this dimension of the plasma increases and $\lambda_B$ becomes less than interparticle distance, then the quantum effects fade away. 
	
	\begin{eqnarray}
		\lambda_B=\frac{\hbar}{mv_T}\\
		v_T=\sqrt{\frac{k_BT}{m}}
	\end{eqnarray}

	\item \textbf{Fermi temperature}. The quantum statistics of free particles (Fermi Dirac) allows for a direct link between the density and the fermi energy, therefore, one can define a quasi-temperature similar with the classical case of ideal gas:
	
	\begin{eqnarray}
		T_F=\frac{(3\pi^2)^{2/3}\hbar^2\rho^{2/3}}{2mk_B}
	\end{eqnarray}
	
	\item \textbf{Regime parameters}. We can define the following quantum coupling parameters \cite{manfredi2005model}:
	
	\begin{eqnarray}
		\chi=\frac{T_F}{T}=\frac{(3\pi^2)^{2/3}}{2}(\rho\lambda_B^3)^{2/3}\\
		g_Q=\frac{E_{int}}{E_F}\sim (\frac{\hbar\omega_p}{E_F})^2
	\end{eqnarray}
	
	These two non-dimensional parameters define our regimes and consequently the theoretical methods to be used: $\chi\ll 1 \to $ classical, $\chi\gg 1 \to $ quantum, $g_Q\ll 1 \to $ collisional, $\chi\gg 1 \to $ collisionless.
\end{enumerate}

As a short picture about the values involved, in Tabel \ref{tab1} one can see some values of some quantities vs. specific elements.

\begin{center}
	\label{tab1}
	\begin{tabular}{| l | l | l | l | l |}
		\hline
		Element & IP[$eV$]  & $r_s[\mathring{A}]$ & $\hbar\omega_{Mie}[eV]$  & $I_{crit}[W/cm^2]$\\ \hline
		Na & 5.14 & $2.1$ & $2.8$ & $3\times 10^{12}$\\ \hline
		Ag & 7.58 & $1.59$ &  $0$ & $1\times 10^{13}$\\ \hline
		C & 11.26 & $1.21$ & $20$ & $6\times 10^{13}$\\ \hline 
	\end{tabular}
\end{center}

\section{Lasers}
\label{lasers}

From the perspective of laser-cluster interaction, we are less interested in how the laser field is produced, but rather in what "comes out" from it. Now, everybody knows that the laser field is an electromagnetic (EM) wave in its most pure definition but there are various parameters that can, or cannot be, neglected when speaking about its effects on molecular scale. 

First of all, the laser field is completely described by the knowledge of the two vectorial fields $\vec{E}$ electric and $\vec{B}$ magnetic which obey the Maxwell's equation. Second, a rough length scale of $1nm$ is much less than the spatial extent of the laser profile (which is usually gaussian in section), therefore, the dipolar approximation holds: the laser field is considered constant over the region of interest. On the other hand, the overall angular momentum and charge current in a cluster is small and consequently, its coupling with the magnetic component of the EM wave can be neglected. This considerations lead to a natural modelling of the laser in only one constant electric field:

\begin{equation}
	\label{laser1}
	\vec{\mathcal{E}}(\vec{r},t)=\mathcal{E}_0\hat{e}_zf(t)cos(\omega_{las}t+\phi(t))
\end{equation}

Now, it is clear that $\hat{e}_z$ is the versor of the $Oz$ direction in space which is chosen arbitrary just for a simplification of calculus (even though it maps to the linear polarization), $\mathcal{E}_0$ is the magnitude of the electric field (a constant) while the rest of the formula \ref{laser1} represents just time dependent terms. The $cos$ term models the oscillatory behaviour of the EM wave which has as a first parameter the laser frequency $\omega_{las}$ which relates to the energy of the photon carrier by the trivial $\hbar\omega_{las}$. The $\phi(t)$ term is a phase term which can be written roughly as $\phi(t)=\phi_{ce}+(\beta/2)t^2+(\gamma/3)t^3+\mathcal{O}(t^4)$ with $\phi_{ce}$ the carrier-envelope phase, $\beta, \gamma$ linear and quadratic chirps. 

The instantaneous intensity of the laser field is easily expressed $I(t)=c\varepsilon_0\mathcal{E}_0^2/2f(t)^2$. The function $f(t)$ models the temporal shape of the pulse. A common form for it is the Gaussian field profile $f(t)=exp(-(t/\tau)^2)$.

Now, as we shall see in Chapter \ref{Chapter2}, many quantum approaches on the electron dynamics do not work with the electric force acting on them, but with the effective potential in which the electron's wave function behaves. Therefore, it is useful to have an associated electric potential for the laser, and in the frame of dipolar approximation it can be expressed as:

$$V_{las}(\mathbf{r},t)=e\vec{\mathcal{E}}\mathbf{r}$$

As a key parameter for the classification of coupling regimes between lasers and electrons is the ponderomotive potential which reads:

$$U_p=\frac{e^2\mathcal{E}_0^2}{4m_e\omega_{las}^2}=9.33*10^{-14} I_0[W/cm^2](\lambda[\mu m])^2eV$$

This quantity is a measure of the relativistic effects to be expected. When $U_p$ becomes comparable with the rest energy of the electron, then the relativistic regime becomes active. In general, through this thesis, I do not refer to such extreme cases. 

It is worth mentioning that a more general description of the laser field is the modelling of a train of pulses \cite{gao2015analysis} and the relation \ref{laser1} generalize itself to :

\begin{eqnarray}
	\vec{\mathcal{E}}(\vec{r},t)=\mathcal{E}_0\hat{e}_z\sum_{\alpha=0}^{N-1}g(t)cos^2(\frac{t-t_{\alpha}-T/2}{T}\pi)\Theta((t-t_{\alpha})(T-t+t_{\alpha}))sin(\omega_{las}t)\\
	t_\alpha=\Delta\tau+\alpha T_{train}\\
	g(t)=exp[-4ln2\frac{(t-\Delta\tau-\mathcal{T})^2}{\mathcal{T}^2}]\\
	\mathcal{T}=NT_{train}/2
\end{eqnarray}

\section{Laser-Cluster coupling regimes}
\label{Laser-Cluster coupling regimes}

It is important to have some understanding about what different laser parameters bring on the table in terms of phenomena in clusters in order to have later a good sense about the theoretical method to be applied. The key feature that spans all the regimes to be discussed below is the (photo)ionization. This subsection was highly influenced by \cite{fennel2010laser}.

From an atomic perspective, the photo-ionization can be divided in two categories: \emph{vertical ionization} and \emph{optical field ionization} (OFI). The vertical ionization consists of a basic single or multiple absorption of photons by the electrons from the atom, and consequently, their transition on higher bound or free states. In general, a process of multiple photon ionization (MPI) which involves $\nu$ photons has a specific reaction rate that can be expressed as $\Gamma=\sigma_\nu I^\nu$ where $\sigma_\nu$ is the absorption cross-section and $I$ is the intensity of the laser field. On the other side, external fields with a slow time oscillation can be treated as static and so, the deformation of the atomic potential has a long enough life to allow ionization through tunnelling of the energetic barrier. This is known as OFI. An important (hystorical) parameter here is the Keldysh parameter \cite{keldysh1965ionization} $\gamma=\sqrt{E_{IP}/2U_p}$ which indicates the type of process that dominates: $\gamma\gg 1$ means MPI and $\gamma\le 1$ means OFI.

\begin{figure}[h]
	\centering
	\includegraphics[width=1.0\linewidth]{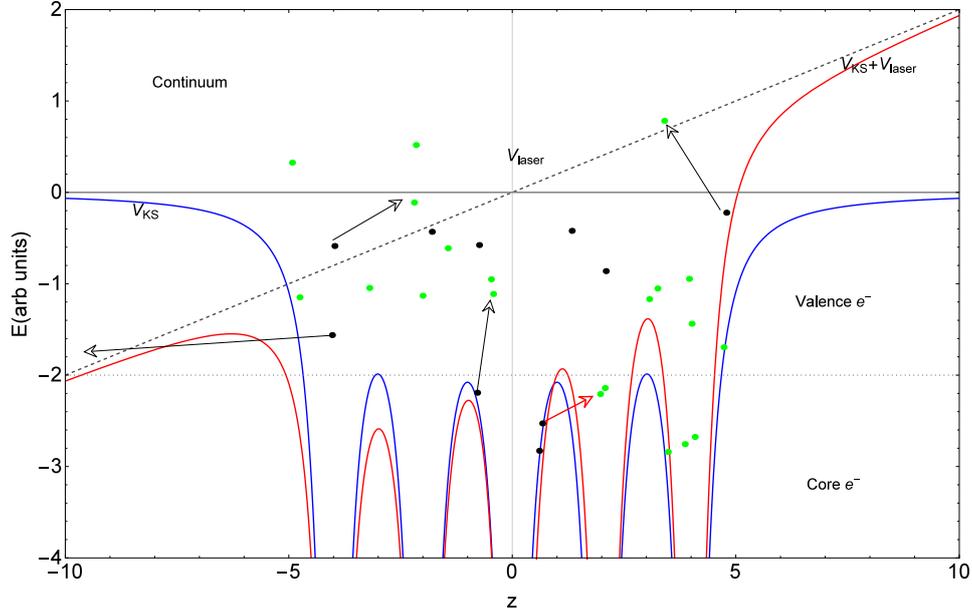}
	\caption{A schematic representation of the effective microscopic potential $V_{KS}$ (blue), the deformation induced by large external fields $V_{laser}$ (gray, dashed) and the main types of states: core electrons, valence electrons and continuum states; In green, occupied states/ in black, holes}
	\label{fig:ionizationmodel}
\end{figure}

When moving to clusters, the fact that in the system are more atoms makes the simple picture of atomic potential more involved. The interaction between atomic electrons and their overlapping deforms the effective potential in which they move and the details of ionic and electronic structure become crucial. The simple fact that two atoms are bind together, usually means that the potential is lowered between them and the effect known as charge-resonance-enhanced ionization (CREI) arises. This comes in package with an increase in the ionization rate.

Essentially, in clusters, there are two types of ionization: inner ionization and outer ionization. Inner ionization consists of vertical absorption of photons which excite electrons from bound states (core electrons) to valence states but still bounded to the cluster (quasi-free electrons). The outer ionization goes a step further and represents the transition of valence electrons into the continuum (free spectrum). The main channel of energy absorption is the direct transfer from laser field to the electrons. Still, during the dynamics of the electron cloud, the self-consistent Coulomb potential may play itself a role in the single particle excitation, creating metastable states in the (quasi) free spectrum. 

Following the laser parameters (in principle the intensity) we can draw three different regimes of interaction. 

\subsection{Linear regime}

Is characterized by laser fields with small intensities, usually bellow $10^9W/cm^2$. Correspondingly, the Keldysh parameter is very large since the ponderomotive potential is much lower than the ionization potential $\gamma\gg 1$. The processes are mainly frequency dependent. From them we distinct the photo-absorption related with the quantity called optical response (cross section) and the single photo-ionization related to the photo-electron spectroscopy (PES). From a theoretical point of view, this regime can be tackled with linearized methods. 

The optical spectra is constructed from the dielectric function $\epsilon(\omega)$ which in turn is proportional with the Fourier Transform of the total dipolar moment. A more suitable quantity to represent for non-linear features is the power spectrum $\mathcal{S}(\omega)$ \cite{calvayrac1997spectral}:

\begin{eqnarray}
	\sigma(\omega)\propto\omega\Im(\epsilon(\omega))\propto\omega\Im[\tilde{D}(\omega)]\\
	\mathcal{S}(\omega)\propto|\tilde{D}(\omega)|^2
\end{eqnarray}

As we shall see in Chapter \ref{Chapter3}, the PES can be computed from single particle methods recording the time evolution of the orbitals at a "far-away point". From a physical point of view, a PES spectrum is a picture of the density of states in the cluster. The main problem with the PES is that experimentally is hard to achieve for core electrons. But if one is interested in the single photon processes and linear regime, these types of excitations are not possible. 

More insight in the geometric geometrical structure and the active orbitals during the dynamics can be achieved in photo-emission spectra, from a spatial analysis. More precisely, it is investigated what is called \emph{Photo-electron Angular Distribution} (PAD). This can be quantified in the differential cross-section which is expanded in a series of Legendre polynomials:

\begin{equation}
	\frac{d\sigma}{d\Omega}=\frac{\sigma_{tot}}{4\pi}(1+\beta_2P_2(cos(\theta))+\beta_4P_4(cos(\theta)+...))
\end{equation}

At the first glance, it can be seen that the $\beta_2$ parameter is a reflection of the orientation of the emission over the polarization of the laser field. A $\beta_2=2$ means emission parallel with the laser electric field, $\beta_2=0$ an isotropic emission and $\beta_2=-1$ emission perpendicular on the polarization. An important issue with the PAD is that is photon dependent (in the sense of frequency) \cite{wopperer2012frequency}.

While PES and PAD give us information about the ground state structural properties of the cluster, a final tool in the study of single photon-linear regimes can be developed to investigate the dynamical features, namely the \emph{Time Resolved Photoelectron Spectra} (TRPES).

\subsection{Multiphoton regime}

Is characterized by intensities ranging in the $10^8<I<10^13 W/cm^2$. In terms of  Keldysh parameter $\gamma\le 1$. This time, all the characteristic parameters of the laser field ($\omega$, $I$, $E$, $f(t)$) become equally important in the dynamics. The MPI takes the leading role in the ionization and as a consequence, the phenomena of second harmonic generation appears.  The optical spectrum presents non-linear features. All the linearized theoretical approaches break at these intensities and full propagation schemes (Vlasov, TDDFT, TDHF, TDTF, etc.) must be employed.

When in this regime, the multiphoton processes are always accompanied by single-photon ones. In principle from PES experiments (or calculations) one can extract the single particle energy by the relation $E_{kin}=\nu\hbar\omega+\epsilon_0$. Reversely, if one knows $\epsilon_0$ and the laser frequency, the type of phenomena single/multi can be recovered computing $\nu$. By default, laser fields with frequencies bellow the ionization threshold can ionize the cluster only through  OFI which has a very small probability. 

A very robust phenomena captured by photo-ionization spectra is the plasmon. More details about it will be given in \ref{Chapter3}, but in principle, this is a collective phenomena characteristic to metal clusters. It remains a pregnant resonance even in non-linear regimes where the single particle features can be washed out by the laser intensity. Being a collective phenomena it represents a gate for the resonance and a good energy absorption of energy in the cluster. Various types of damping behaviours can fragment the plasmon and give a spectral picture with many peaks around $\omega_{mie}$.  

\subsection{Strong field domain}
\label{strongfield}

Is achieved in the $10^{13}<I<10^{19} W/cm^2$ range of intensity. Although there are studies which invoked the motion of ions during the dynamics, in general it is considered that the ions have a small amplitude and slow motion for $I<10^{12}W/cm^2$. When the laser exceeds this limit, the ionic background cannot be considered anymore frozen and its dynamics must be taken into account through theoretical methods related to the Molecular Dynamics.

\begin{figure}[h]
	\centering
	\includegraphics[width=1.0\linewidth]{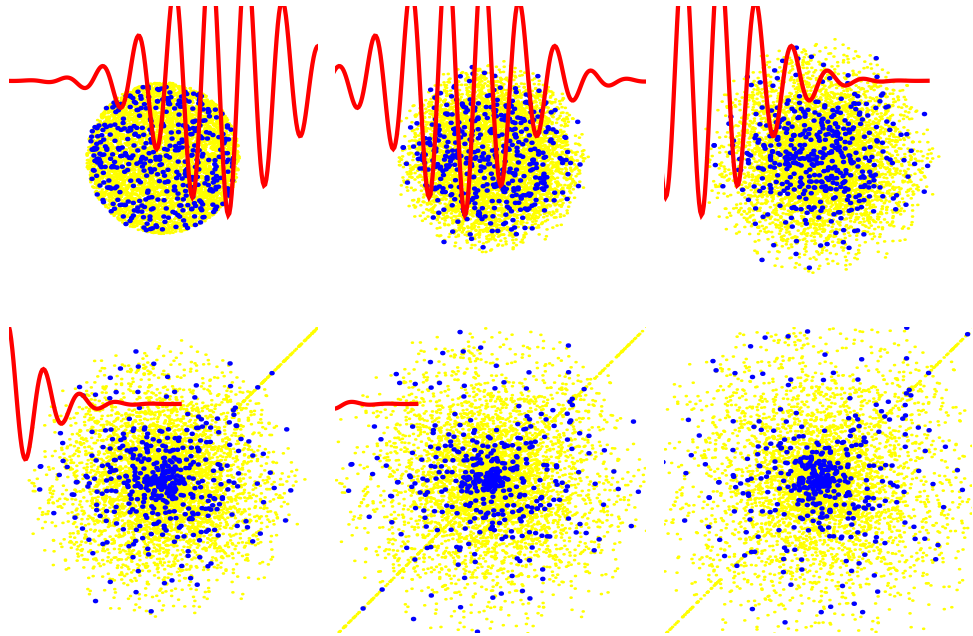}
	\caption{The cluster explosion is represented in a schematic view. The incoming laser pulse (in red) excites the ground state cluster (with yellow-electrons, blue-ions) which expels electrons and ions during the dynamics}
	\label{fig:dd}
\end{figure}

The \emph{guilt} for these violent dissociation of the cluster is a very efficient energy absorption. For intensities around $10^{15}W/cm^2$ the mean energy per atom can be of $\sim 100keV$. This energy absorption ionizes very rapidly the cluster stripping the valence electrons and giving some boost to the inner electrons and ions. Having a considerable positive charge, the cluster dezintegrates itself even if the laser pulse was short due to the Coulomb repulsion. This self-consistent interaction gives in turn the so called Coulomb explosion. In turn, the hot quantum plasma created ejects with high velocities electrons, ions and photons (from recombinations). A schematic picture of the cluster explosion is presented in \ref{fig:dd}.

From the ejected ions, a first feature is the presence of highly charged ions. Different studies have reported even ions with a $\sim +20$ charge. Beside high charge, the ions come in package with high energies at the level at even $1MeV$ \cite{ditmire1998explosion}. As amazing as it is, these energies have driven imediatelly the researchers to study the possibility of using them in nuclear fusion applications.

Regarding the electron dynamics, their inner ionization into a nano-plasma gives the possibility of recombination. In turn, the recombination of electrons with ions with empty core shells leads to the emission of photons, in particular, energetic photons. So, X-ray production becomes an active possibility together with high-order-harmonic generation.

% Chapter 1

\chapter{Theoretical methods} % Write in your own chapter title
\label{Chapter2}

The topic of laser-cluster interaction requires a wide set of knowledge from various fields of Physics. Nonetheless, the main requirement is the knowledge of Quantum Physics, coupled with some good understanding of the classical Electrodynamics. The present chapter, being a logical (hierarchical) picture of the existing theoretical methods used in cluster laser-physics, will follow a line which start with the fundamentals of Quantum Physics, more precisely, the basics of Quantum Statistics. In the end, the so-called classical methods will be discussed, but as the final point of a line of approximations.

Almost every many-body theory that will be discussed can be derived in more than a single way, following different representations of quantum mechanics, or different mathematical approaches. Still, the practical end is always the same, so we will stick with a self consistent presentation, in parallel, a sufficient number of further references being given.

Regarding the other realm of modern physics, namely the theory of relativity, we will not refer in general to this kind of effects since they become important only in the extreme ultra-intense lasers, with intensities above $10^{19}W/cm^2$. Still, almost any model which will be discussed can support relativistic versions or corrections without modifying the main idea of the theory. For example, methods as Density Functional Theory which are treated with Schroedinger-like equations will, naturally, be extended to Dirac-like equations. Furthermore, the QED effects will be also neglected during the presentation. This not just a choice of the author, but actively, this kind of studies in clusters are almost singular since the treatment is too complex, while the effects are negligible Also, while the main part of the dynamics in a cluster is taken upon the electrons, the spin will be taken into account only where is absolutely necessary. Otherwise, the extension of the theory to spin is almost trivial. 

For all these reasons, this chapter will start with a discussion about the fundamentals of Quantum Statistical Mechanics (QS).

\section{Quantum Liouville-von-Neumann equation}

Any science is defined by its set of postulates and its mathematical apparatus, therefore I shall start with the axiomatic logic of QS before treating any specific theory. Moreover, as we shall see later, even in the strongest external fields, a separation of the cluster will be at hand in nuclei (possibly ions) which are subject to (quasi) classical behaviour and electrons, subject to a more or less sophisticated quantal treatment. As the complex treatment is required by electrons, it is natural to discuss only theoretical methods that deal with systems of identical particles, in particular, fermions. 

From a historical perspective, the first rigorous axiomatic of quantum mechanics was established by P.M. Dirac \cite{dirac1930note,dirac1958principles} and it is a viable construction for the case of fully known, pure states. Unfortunately, during dynamics (especially strong dynamics) we deal with finite (possible large) temperatures, non-equilibrium phenomena in which the states of the system are no longer pure. Therefore, a more natural axiomatic, which should incorporate statistical aspects of the dynamics, is needed. This was brought on solid mathematical grounds by J. von \cite{john1955mathematical}.

While the introductory quantum mechanics associated with pure states of particles works with the concept of wave-function or the more abstract notion of vector in an Hilbert space, the physics of mixed states is described in terms of density operators, which will be denoted through the thesis with $\hat{\Gamma}$. There is no need or purpose to discuss the explicit axioms in here but some words should be said about the properties of the density operators and their dynamics.

First of all, $\hat{\Gamma}$ is defined on the Hilbert space $\mathcal{H}$ of the system (as are the vectors for the pure states) and there are some characteristics that any density operator must obey. First, the trace must be $1$, $Tr(\hat{\Gamma})=1$, or more generally, $N$ where $N$ is the number of particles from our system. Second, in order to have real observables associated with our system, $\hat{\Gamma}$ must be self-adjoint $\hat{\Gamma}=\hat{\Gamma}^{\dagger}$. Later, we will see how there are interpretations of $\hat{\Gamma}$  in terms of density of probability and for that reasons the operator $\hat{\Gamma}$ must be positively defined and bounded. In principle, if one knows $\hat{\Gamma}$ , then, any observable $A$ can be computed in terms of the associated operator $\hat{A}$ taking the trace in the Hilbert space of the system $\langle \hat{A}\rangle =Tr(\hat{A}\hat{\Gamma})$.

Beyond these properties of $\hat{\Gamma}$ and the first axioms of QS which are basically the same with those from the usual QM, it is important to start from the last of QS's axioms, the one that describes the quantum dynamics, namely, the Liouville-von-Neumann equation:

\begin{equation}\label{QLiouv}
	i\hbar\partial_t\hat{\Gamma}=[\hat{H},\hat{\Gamma}]
\end{equation}	

This is the quantum version of the classical Liouville equation. Both, are consequences of the Liouville theorem that states that in a Hamiltonian system the distribution function is constant along the trajectories, or, equivalently that in the phase space the volume element is conserved. While the parallel between classical and quantum Liouville eq. is stringent from a visual level, it remains a single (apparent) difference in the fact that the classical Poisson bracket is replaced with a commutator. This change is known as the correspondence principle and is a consequence of the quantization of the classical phase space in a Hilbert space.

Now, let us denote for future purposes our system as having $N$ identical (indistinguishable) particles at the coordinates $\mathbf{x}_N=\{x_1,x_2,...,x_N\}$ described by a field operator $\Psi(\mathbf{x}_N)$. Having this, the density operator can be formally written:

\begin{equation}
	\hat{\Gamma}(\mathbf{x}_N,\mathbf{x}_N')=\Psi(\mathbf{x}_N)\Psi^{\dagger}(\mathbf{x}_N')
\end{equation}

Another concept that soon will be of great use is the \emph{reduced density operator of order "s"} which is nothing else than a partial trace from the entire $\hat{\Gamma}$, on the sub Hilbert space of $N-s$ particles $\hat{\Gamma}_s=Tr_{s+1}^N\hat{\Gamma}$. If one prefers a representation of $\hat{\Gamma}$ in the coordinate space, than, it can write $\hat{\Gamma}_s(\mathbf{x}_s,\mathbf{x'}_s)$ (which in literature is found as \emph{distribution kernel of $\hat{\Gamma}$}) as:

\begin{equation}
	\hat{\Gamma}_s(\mathbf{x}_s,\mathbf{x}'_s)=\left( \begin{array}{c}N\\s\end{array} \right)\int \Gamma(\mathbf{x}_N,\mathbf{x}_N')dx'_{s+1}...dx'_Ndx_{s+1}...dx_N
\end{equation}

Now, let us assume that our hamiltonian operator describes a system with at most two body interactions $\hat{H}^N=\hat{p}_i^2/2m+\sum_iv(x_i)+1/2\sum_{i\neq j}\phi(|x_i-x_j|)$ (this is always the case for electrons in clusters). Taking this partial trace of the Liouville eq. \ref{QLiouv} we obtain an infinite hierarchy of equations, the so called BBGKY (Bogoliubov-Born-Green-Kirkwood-Yvon) hierarchy analogous to the one obtained in classical statistical mechanics :

\begin{eqnarray}\label{BBGKY}
	i\hbar\partial_t\hat{\Gamma}_s=[\hat{H}^s,\hat{\Gamma}_s]+\hat{Q}^s(\hat{\Gamma}_{s+1})\\
	\hat{H}^s=\sum_{i=k}^s(\frac{\hat{p}_k^2}{2m}+v(x_k))+\sum_{k\neq j}^s\phi(x_k-x_j)\\
	\hat{Q}^s(\hat{\Gamma}_{s+1})=(N-s)\sum_k^sTr_{s+1}\{[\phi(x_k-x_{s+1}),\hat{\Gamma}_{s+1}]\}
\end{eqnarray}

More or less cumbersome, the system is exact but infinite and there is no way to solve it. On the other hand, there is no use of solving it, since the knowledge of the entire density matrix is obsolete in practice. We are usually interested in macroscopic quantities as density, current, multipolar moments, etc. and this kind of information is accessible just from the first order reduced density operator $\hat{\Gamma}_1$. But one cannot solve the corresponding equation for $\hat{\Gamma}_1$ due to the $\hat{Q}^1$ term which is intrinsically dependent on the all other higher order density matrices.

The Liouville's eq. \ref{QLiouv} allows for a variable number of particles during the dynamics. This aspect is preserved in the BBGKY hierarchy and it becomes obvious that one can have creation or particles, usually in strong external fields. Other important properties of BBGKY are that the system can be formally solved, due to linearity and it contains an intrinsic mathematical time reversibility.  Regarding the conservation of energy, unlike other kinetic approaches (Boltzmann, Landau, or Balescu \cite{ropke1983linear} since they are derived under the assumption of zero three particle correlations), the system conserves the total energy.

There is some mapping between the density operator and the single particle representation of a statistical system through the relation \ref{oper-wave} in which $\hat{\Gamma}_1$ is expressed as a superposition of projectors in the single particle Hilbert space, weighted by some probability coefficients $\{p_j\}$.

\begin{equation}\label{oper-wave}
	\hat{\Gamma}_1=\sum_ip_i|\psi_i\rangle\langle\psi_i|
\end{equation}

\begin{figure}[!h]
	\centering
	\includegraphics[width=1.\linewidth]{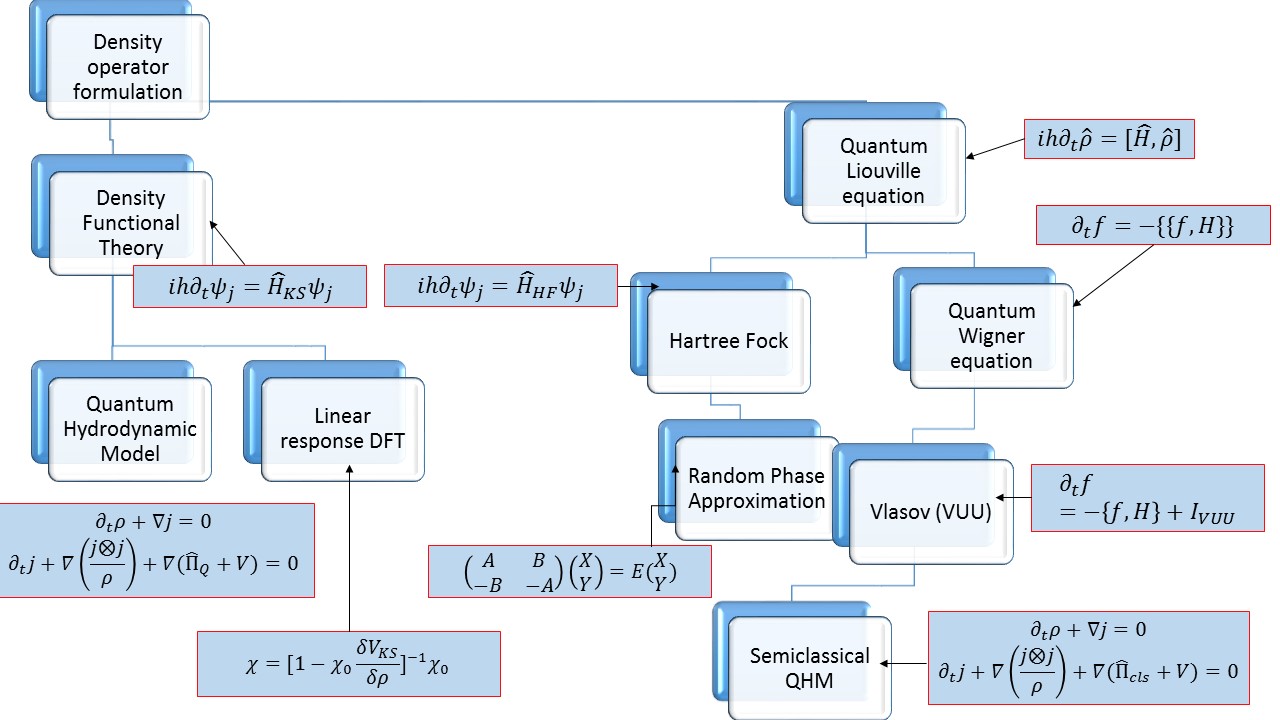}
	\caption{An hierarchical map of the main theoretical methods used in cluster physics}
	\label{fig:drawing-1}
\end{figure}

The BBGKY system of equation is the basis for part of the methods which will be discussed further. In particular we will start from the equation for $\hat{\Gamma}_1$ and use some approximations to decouple the hierarchical dependence with higher order densities. Just to have a visual picture about what is to emerge from these approximations, in Fig. \ref{fig:drawing-1} there is represented an organizational chart with the main theories and their relation with higher theories. Starting from the Von Neumann axiomatic and the density operator formulation of QS we see how we have on one hand the quantum Liouville's equation from which, passing trough the BBGKY hierarchy and some approximations two lower theories can be obtained: the Hartree-Fock (HF) theory and the Quantum Wigner (QW) equation. Both are at the same level, being equivalent conceptually but used under different representations. On a parallel level is the Density Functional Theory (DFT) which, apparently, is not related with QLiouville. This fact is not true, but the path on which DFT is derived has no direct connection with the latter mentioned equation. Both HF and DFT, being in practice, single particle methods, allow for a linearisation and a description of normal modes in terms of single particle wave functions and energies. From HF it is obtained the so called Random Phase Approximation (RPA) while from DFT the Linear Response DFT (LR-DFT), both having formally equivalent results.

A different theory can be extracted from QW Eq., taking the semi classical limit $\hbar\to 0$ and retaining the zero or the first order in $\hbar$, namely the Vlasov equation. This is equivalent with the classical Vlasov equation used in plasma physics, but quantum effects can be introduced through mean field potential (taken from DFT) or through collision operators that describe the semi classical quantization of the phase space (Uheling-Ulenbeck is one type of such operator used to mimic the Pauli principle).

Going even lower in the tree of approximations, one can derive a Quantum Hydrodynamic Model (QHM) from three different perspective: integrating the Vlasov equation on moments in the momentum space, using the abstract (functional) Euler-Lagrange equations of DFT or using a Madelung transform on the orbitals from the DFT/HF equations. The last level of approximation is designed for the intense laser-cluster interaction and large clusters, namely the nano-plasma model or classical molecular dynamics (CMD).

\begin{center}
	\label{tab2}
	\begin{tabular}{| l | l | l | l | l |}
		\hline
		Theory & E/N $(eV)$  & I ($W/cm^2$) & Regime & N\\ \hline
		(post)HF & 0-1 & $<10^{12}$ & $L/M\sim S$ & $<10^2$\\ \hline
		DFT & 0-1 & $<10^{15}$ & $L/M\sim S$ & $<10^3$\\ \hline
		RPA & 0-0.1 & $10^8$ & $L$ & $<10^3$\\ \hline
		TF (OF)  & $>0.1$ & $10^{16}$ & $L/M\sim S$ & $<10^6$\\ \hline
		Vlasov & $>0.1$ & $10^{16}$ & $L/M\sim S$ & $<10^4$\\ \hline
		MD & $>0.1$ & $10^{12}$ & $\sim M/S$ & $<10^6$\\ \hline
		RE  & $>1$ & $10^{15}$ & $S$ & $>10^4$\\ \hline
	\end{tabular}
	\captionof{table}{Schematic table with the gross applicability of the main theories to be discussed: Hartree-Fock \ref{Hartreefock}, Density Functional Theory \ref{Dftt}, Random Phase Approximation \ref{RPA}, Thomas Fermi (and Orbital Free extensions \ref{hydrod}), Vlasov \ref{Vlasovv}, Molecular Dynamics \ref{md} and Rate Equations (from nano-plasma model) \ref{nanoplasma}. $E/N$ stands for excitation energies, $N$ for the number of atoms in the cluster and $I$ for the intensity of the laser field. $L/M/S=$Linear/Moderate/Strong regimes.}
\end{center}

The tabel \ref{tab2} shows a schematic view of the applicability of each of these theories. As one can see, the large number of atoms coupled with large intensities has a poor representation. While there is no place or space to describe any of these theories in detail, all of them admit extensions and improvements which will be only mentioned with appropriate references.

\section{Quantum first principles in atomic systems}
\label{Quantum first principles in atomic systems}

Clusters, as atomic systems are quite complicated things to study. Think for a moment that you have an $X_n$ cluster which will contain by default in a neutral state $n$ nuclei and $nZ$ electrons moving around accordingly with the quantum mechanical rules.

As a first step in the investigation of dynamics (or ground state) we should write down a Hamiltonian $\hat{H}$ for our system. Accordingly with all we know from elementary classical and quantum mechanics the Hamiltonian should contain a kinetic and a potential energy term, the later given by the Coulomb interaction between electronic and nuclear charges. Let us denote with $\{\vec{R}_I\}=\mathfrak{R}$ the nuclear coordinates, with $\{\vec{r}_i\}=\mathfrak{r}$ the electron coordinates and with $\Psi(\mathfrak{R},\mathfrak{r})$ the total wave function associated with electron-nuclei system. With those, the Hamiltonian and the Schroedinger equation can be written as

\begin{eqnarray}
	\label{fullSCH}
	\hat{H}\Psi(\mathfrak{R},\mathfrak{r})=E\Psi(\mathfrak{R},\mathfrak{r})\\
	\label{fullhamiltonian}
	\hat{H}=\hat{T}_e+\hat{T}_N+\hat{V}_{ee}+\hat{V}_{nn}+\hat{V}_{en}\\
	\hat{T}_e=-\frac{\hbar^2}{2m}\sum_i\nabla_i^2\\
	\hat{T}_n=-\frac{\hbar^2}{2M_I}\sum_I\nabla_I^2\\
	\hat{V}_{ee}=\frac{e^2}{4\pi\varepsilon_0}\sum_{i,j}\frac{1}{|\vec{r}_i-\vec{r}_j|}\\
	\hat{V}_{nn}=\frac{e^2}{4\pi\varepsilon_0}\sum_{i,I}\frac{Z_IZ_J}{|\vec{R}_i-\vec{R}_J|}\\
	\hat{V}_{ne}=-\frac{e^2}{4\pi\varepsilon_0}\sum_{i,I}\frac{Z_I}{|\vec{R}_i-\vec{r}_i|}
\end{eqnarray}

Where $M_I$ is the mass of the $I-th$ nuclei, $m$ is the electron mass and $Z_I$ is the atomic number of the $I-th$ nuclei. Now, our simple $n$ atom cluster involves only for the ground state, to solve this Schroedinger \ref{fullSCH}  equation, which is a partial differential eigenvalue problem in $3n(1+Z)$ coordinates. Obviously, this is an impossible task both analytic or numeric. Therefore, stated as it is, the full atomic problem in a cluster is superfluous. The next section will present one of the basic approximation in any molecular, cluster or solid state physics problem, the Born-Oppenheimer approximation.

\subsection{Born-Oppenheimer approximation}

Now, having the \ref{fullhamiltonian} Hamiltonian written down and explained, we should see briefly what is the Born-Oppenheimer approximation \cite{born1927quantentheorie} and how it can be derived. 

\emph{BO: In an atomic system, the electrons-nuclei problem can be separated in two distinct problems. Consequently, the factorization $\Psi_{total}=\psi_{electron}\times\psi_{nuclei}$} holds.

Let us look first at the magnitudes of some constants that appear in the Hamiltonian. A nucleon is roughly $1800$ times heavier than an electron, so we can safely state that $M_I\gg m$. This means that in a classical view of our system, the nuclei move much slower than the electrons (being subject to forces of the same magnitude, Coulomb forces, but heavier particles).  For this reason we could, as a first step, neglect the kinetic energy of the nuclei and write the so called "clamped nuclei" Schroedinger equation:

\begin{eqnarray}\label{clamped}
	\hat{H}_{el}=\hat{T}_e+\hat{V}_{en}+\hat{V}_{ee}\\
	\hat{H}_{el}\psi_{el}(\mathfrak{r};\mathfrak{R})=E_{el}\psi_{el}(\mathfrak{r};\mathfrak{R})
\end{eqnarray}

Where, this time, the $\Psi_{el}(\mathfrak{r};\mathfrak{R})$ wave function stands for the electrons and the nuclei coordinates are considered only as parameters appearing in the electron-nuclei interaction (from there the $;$ sign). Now we go back to the total problem \ref{fullSCH} and use the expansion $\Psi(\mathfrak{r},\mathfrak{R})=\sum_k\psi^{el}_{k}(\mathfrak{r};\mathfrak{R})\chi_k(\mathfrak{R})$. We should not enter in the specific details of calculus, just mention that the electronic pseudo-solutions are orthogonal, therefore one can integrate over the electronic coordinates the resulting equation and obtain, formally:

\begin{eqnarray}\label{nuclear1}
	[T_n+E_k-E]\chi_k(\mathfrak{R})=\sum_l\mathbf{F}_{k,l}\chi_l(\mathfrak{R})\\
	\mathbf{F}_{k,l}=-\sum_I\frac{\hbar^2}{2M_I}[2\frac{\langle\psi_{k}|[\nabla_I,H_{el}]|\psi_{l}\rangle}{E_k-E_l}+\langle\psi_l\nabla_I^2\psi_k\rangle]
\end{eqnarray}

Now, reminding that $M_I\gg m$ and in general $|E_k-E_l|\gg 0$ we could safely assume that the right hand side terms are small compared to the left side ones, therefore, we can write an eigenvalue problem also for nuclei. But this was the whole purpose from the start: to separate the total Schrodinger problem in two eigenvalue problems for electrons and nuclei, coupled only parametrically trough the interaction potentials.

The BO approx. has common points with the adiabatic theorem \cite{born1928beweis} and some extended discussion can be done on the so called potential energy surfaces, given by the solution of the clamped Schroedinger eq.

In clusters, the Born-Oppenheimer approximation holds almost indefinitely, even though it might be questionable in the ultra intense laser regimes. But in general system the problem is not that fortunate and one could refer to specific articles that treats \cite{kouppel2007multimode}, \cite{pisana2007breakdown} the molecular problem beyond BO approximation.

\subsection{Dynamics of nuclei}
\label{md}

Now that we have seen how the full nuclei-electrons problem in a cluster can be separated in a problem of electrons and one for nuclei, let us go further and see what can be done more to simplify the treatment. In quantum chemistry, it is known by the name of \emph{Molecular dynamics} the study of ionic geometry (in ground state or dynamics) with methods which involve a more or less detailed level of quantum effects. These studies coupled intrinsically the motion of electrons in the problem. In contrast with that, the present subsection discusses just the motion of nuclei(ions) in a cluster.

To avoid any confusion, the BO approximation does not state that the nuclei and electrons are independent, in the sense that one could solve their problem separately, but rather that the problem is separable in a mathematical sense of nuclear and electronic variables. As we shall see soon, their dynamics is strongly correlated with the interaction potential which is dependent on both systems.

Starting from \ref{nuclear1} Schroedinger problem for nuclei in the frame of BO approximation, we extend it to time dependent processes:

\begin{equation}\label{nuclear2}
	-\sum_I\frac{\hbar^2}{2M_I}\nabla_I^2\chi+(V_{nn}+V_{ne})\chi=i\hbar\partial_t\chi
\end{equation}

We write some kind of Madelung transform \cite{madelung1927quantentheorie} for the total wave function of nuclei $\chi(\mathfrak{R},t)=A(\mathfrak{R},t)exp(iS(\mathfrak{R},t)/\hbar)$ and separate the \ref{nuclear2} equation in real and imaginary parts, obtaining a continuity and a phase equation:

\begin{eqnarray}
	\partial_tA(\mathfrak{R},t)^2+\sum_I\nabla_I(A(\mathfrak{R},t)^2\nabla_IS(\mathfrak{R},t)/2M_I)=0\\
	\partial_tS(\mathfrak{R},t)^2+\sum_I\frac{|\nabla_IS(\mathfrak{R},t)|^2}{2M_I}+V_{en}+V_{nn}+\sum_i\frac{\hbar^2}{2M_I}\frac{\nabla_I^2A(\mathfrak{R},t)}{A(\mathfrak{R},t)}=0\\
\end{eqnarray}

Now, since it is known that the nuclei have diameters of $fm$  order in comparison with our range of interest which is of $\mathring{A}$ order, we can approximate them in the classical limit of point like particles and go to the classical limit $\hbar\to 0$. Furthermore, let us denote the total potential interaction energy between nuclei and nuclei-electrons with $V(\mathfrak{R})$. Also, without entering in details, the remaining part of the equation for phase $S$ (apart from the time derivative) can be written as a Hamilton function in corespondence with the Hamilton-Jacobi formulation of classical mechanics, and denoting the momentum of the $I-th$ nuclei with $\vec{P}_I=\nabla_IS(\mathfrak{R},t)$ we obtain the classical equations of motion for nuclei:

\begin{eqnarray}\label{clasion}
	\frac{d\vec{R}_I}{dt}=\vec{P}_I\\
	\frac{d\vec{P}_I}{dt}=-\nabla_IV(\mathfrak{R},t)
\end{eqnarray}

As a conclusion of this particular subsection, we have obtained (under the assumption that we know the interaction energy of nuclei) the classical Newtonian like equation of motion.  Of course, the potential energy $V(\mathfrak{R},t)$ depends not only on the Coulomb energy between nuclei positions but also incorporates the electron-nuclei interaction which can be a quite disturbing term, depending on the method used to describe it. 

In the dynamical regime, supposing that we know the initial conditions for nuclei and electrons, we have only to solve the equations  \ref{clasion} with some associated numerical methods, again, presuming that we know at any time the $V_{en}$ part of $V$. The problem of initial condition is rather hard in the sense that an usual type of cluster simulation starts with the system in its ground state and acts at $t=0$ with an external laser field. Therefore, it is imperative to have full knowledge about the ground state configuration. 

The stationary cases of the equations of motion does not tell us anything new, just that the individual momentum nuclei must be null and remain that way, which means again that the $\mathfrak{R}$ configuration together with the electronic configuration must give the global minimum of $V$ which is a functional. 

Being an optimization problem which is by default highly non-linear, we should think from the start to tackle it with an iterative method. Beside the iterative aspect, there are many numerical methods to solve optimization problems in the world of mathematicians. Nonetheless, not many of them apply to a molecular problem, in particular through this thesis, Monte Carlo simulated annealing methods have been used. Generally this method is slower than others, but in the problem of molecular optimization is necessary since the potential energy has some special features: the convexity is not known a priori, the number of variables is large and the number of \emph{local} minima is also very large. Since we search for the global minimum, it  is necessary to use a method which is able to \emph{tunnel} through the local walls in the parameters space around a local minimum. More details about the numerics will be given in \ref{Chapter3}.

\section{Hartree Fock theory}
\label{Hartreefock}

Historically, Hartree-Fock (HF) theory, appeared in \cite{hartree1928wave} with the work of D. R. Hartree which assumed that the total wave function of a system of identical particles can be written as a product of single particle wave functions. Obviously, this assumption is wrong since it neglects the antisymmetric character of fermions therefore, the Pauli principle. His work has been refined later by Fock \cite{fock1930naherungsmethode} and Slater \cite{Slater1930note} which took into account the antisymmetric feature of the total wave function. The equations which were obtained under the energy minimization principle were concerning the ground state of a system, but soon \cite{dirac1930note} Dirac proposed a time dependent extension of the theory. 

Later, new representations or formulations (second quantization, many-body theory, quantum-electrodynamics, path integral, quantum field theory) have been invented to deal with the quantum mechanics of many-body systems and HF theory has been derived in many other ways, more different and rigorous than the original theory.

To continue the logical path started with the QLiouville's equations, we shall derive HF theory from density operator considerations. Recalling the BBGKY hierarchy \ref{BBGKY}, we take now the equation for the first order density operator 

\begin{eqnarray}\label{HFbbgky}
	i\hbar\partial_t\hat{\Gamma}_1=[\hat{H}^1,\hat{\Gamma}_1]+\hat{Q}^1(\hat{\Gamma}_{2})\\
	\hat{H}^1=\sum_{i=k}^s(\frac{\hat{p}_k^2}{2m}+v(x_k))\\
	\hat{Q}^1(\hat{\Gamma}_{2})=(N-1)Tr_{2}\{[\phi(x_k-x_{2}),\hat{\Gamma}_{2}]\}
\end{eqnarray}

Now, there is a standard abstract form (named cluster expansion \cite{kira2008cluster}) for the relation between two consecutive density matrices which for the first order, reads: $\hat{\Gamma}_2(r_1,r_1',r_2,r_2')=\hat{\Gamma}_1(r_1,r_1')\hat{\Gamma}_1(r_2,r_2')-\hat{\Gamma}_1(r_1,r_2')\hat{\Gamma}_1(r_1',r_2)+g_{12}$. We have used the position representation and the $g_{12}$ is called the correlation operator. Now, the first step in the HF approximation is to neglect correlations with higher order density matrices, e.g. $g_{12}=0$. This condition is well described in the quantum field theory by means of Feynmann diagrams where neglecting this type of correlations is equivalent with a \emph{mean field theory}, aspect which is essential in HF. 

If there were to be no spin, the first eq. from eqs. \ref{HFbbgky} could be  simplified to $i\hbar\partial_t\hat{\Gamma}_1=[\hat{H}^1+\hat{U}_H,\hat{\Gamma}_1]$, where $U_H$ is the mean field Hartree operator, $\hat{U}_H=Tr(\hat{\phi}_{12}\hat{\Gamma}_2)$. But, taking into account the Spin-Statistic theorem \cite{duck1998pauli} one can prove that the Hartree operator is defined by $\hat{U}_H=Tr(\hat{\phi}_{12},\hat{\Gamma}_2)\Lambda_{\pm}$ where $\Lambda_{\pm}$ is the anti-symmetrization operator. Further let us write down the equation obeyed by the $\Gamma_1$ and define the Fock operator:

\begin{eqnarray}\label{HFfull}
	i\hbar\partial_t\hat{\Gamma}_1=[\hat{F},\hat{\Gamma}_1]\\
	\hat{F}=\hat{H}^1+\hat{U}_H\end{eqnarray}

\begin{eqnarray}\label{Fockoperator}
	\hat{F}(x,x')=(\frac{\hat{p}^2}{2m^*}+v(x))\delta(x-x')+\delta(x-x')\int\frac{1}{x_{12}}\Gamma(x'',x'')dx''-\frac{1}{x_{11}}\Gamma(x,x')
\end{eqnarray}

Furthermore, the assumption that the second order density operator can be developed in a difference of product of first order density matrices is equivalent with the fact that $\hat{\Gamma}_1$ is idempotent. In turn, the idempotency implies that there is a natural set of eigenvalues $|\psi_i\rangle$ for $\hat{\Gamma}_1$ such that $\hat{\Gamma}_1=\sum_i|\psi_i\rangle\langle\psi_i|$. (Note that idempotent means $p_i\in\{0,1\}, \forall i\in\mathbb{N}$).

In the coordinate representation, this condition can be expressed through the fact that the total wave function is a Slater determinant and so, we have obtained through a factorization and an approximation, the original assumption used in the HF theory. With the above mentioned relations, the total energy in a HF system can be expressed only in terms of $\Gamma_1$ :

\begin{eqnarray}
	\label{HFenergy1}
	E_{HF}[\Gamma_1]=\int F(\mathbf{x}_1,\mathbf{x}_2)\Gamma_1(\mathbf{x}_1,\mathbf{x}_2)=\int (-\frac{\hbar^2}{2m^*}+v(\mathbf{x}_1))\gamma_1(\mathbf{x}_1,\mathbf{x}_1')|_{x_1=x_1'}d\mathbf{x}_1+\\\frac{1}{4\pi\varepsilon_0}\iint \frac{\gamma_1(\mathbf{x}_2,\mathbf{x}_2)\gamma_1(\mathbf{x}_1,\mathbf{x}_1)-\gamma_1(\mathbf{x}_2,\mathbf{x}_1)\gamma_1(\mathbf{x}_1,\mathbf{x}_2)}{|\mathbf{x}_{2}-\mathbf{x}_{1}|}d\mathbf{x}_1d\mathbf{x}_2
\end{eqnarray}

Now, if one uses the minimization principle $\delta E_{HF}[\Gamma]=0$ with the constrain $Tr[\Gamma_1]=N$, obtains after some algebra, $[\hat{F},\hat{\Gamma}_1]=0$, which is kind of a trivial information, since we knew that the ground state is a stationary state, thus, the lhs from eq. \ref{HFfull} is 0. But the meaning of this goes further. The fact that the Fock operator and the first order density operator commute, means that they have common eigenfunctions, i.e. the $\{|\psi_i\rangle\}$. Therefore, we could write down, using \ref{HFenergy1} the expression for energy in terms of orbitals

\begin{eqnarray}E_{HF}[\Gamma_1]=\sum_i\int \psi_i^*(\mathbf{x})(-\frac{\hbar^2}{2m^*}+v(\mathbf{x}))\psi_i(\mathbf{x})d\mathbf{x}+\\\frac{e^2}{4\pi\varepsilon_0}\sum_{i,j}\iint\psi_i(\mathbf{x}_1)\psi_i^*(\mathbf{x}_1) \frac{1}{|\mathbf{x}_{2}-\mathbf{x}_{1}|}\psi_j(\mathbf{x}_2)\psi_j^*(\mathbf{x}_2)d\mathbf{x}_1d\mathbf{x}_2-\\\frac{e^2}{4\pi\varepsilon_0}\sum_{i,j}\iint\psi_i(\mathbf{x}_1)\psi_j^*(\mathbf{x}_1) \frac{1}{|\mathbf{x}_{2}-\mathbf{x}_{1}|}\psi_j(\mathbf{x}_2)\psi_i^*(\mathbf{x}_2)d\mathbf{x}_1d\mathbf{x}_2\end{eqnarray}

Now, we use the constrains of $\langle \psi_i|\psi_j\rangle=\delta_{ij}$ with the energy minimization principle  $\delta(E_{HF}-\varepsilon_i\langle \psi_i|\psi_i\rangle)/\delta \langle\psi_i|=0$ to obtain the stationary HF equations. Similarly, we can go to the time dependent regime where $i\hbar\partial_t\hat{\Gamma_1}=[\hat{F}_{HF},\hat{\Gamma}_1]$ (or, using in stead of $\delta E=0$, the principle of action extremization $\delta \mathcal{A}=0$ where $\mathcal{A}=\int dt(\sum_j\langle\psi_j|\partial_t|\psi_j\rangle-E_{HF})$):

\begin{eqnarray}\label{HF}
	\hat{F}_{HF}|\psi_i\rangle=\varepsilon_i|\psi_i\rangle\\
	\label{TDHF}
	\hat{F}_{HF}|\psi_i\rangle=i\hbar\partial_t|\psi_i\rangle
\end{eqnarray}

There are further details, that should be discussed regarding the N-representability, the orthogonalization method, but this kind of discussion could drive the present subsection to an unnecessary length. The main points to be retained from this are that (TD)HF theory is a mean field theory that neglects statistical correlations between first order density matrix and higher order matrices and it is representable through a single particle set of Schroedinger like equation. This equations contain a natural term that gives the Coulomb self interaction of the electron density and a supplementary non-local potential known as exchange. The latter one is a pure quantum effect arising from the anti-symmetry condition for the total wave-function (density matrix) and, as we shall see, is the main reason for which HF involves a greater computational cost than DFT (for example).

It is worth mentioning that a true HF simulation should take into account the fermionic character of the orbitals beyond the anti-symmetric feature of the Slater determinant, including explicitly the spin. This is done working not with trivial wavefunctions $\psi_i$ but with spinors. 

As in any section of the present chapter, only the basic notions of the method are discussed. In practice, HF has a history of nearly $85$ years in which has been used preponderantly in nuclear physics. Since many systems have been found where the results were not accurate enough, extensions (the so called post HF theories) have been invented. The most straightforward one si the Configuration Interaction (CI) method which relaxes the condition of a single Slater $\Psi$ determinant for the total wavefunction to an expansion in a basis of Slater determinants $\Psi=\sum c_i\Psi_i$. Further, one can use a type of perturbation theory to extend the zeroth order Fock operator (eq. \ref{Fockoperator}) to a "perturbed correct" Hamiltonian and so arrive at the M\o ller-Plesset (MP) Perturbation Theory. Further, the Coupled Cluster (CC) theory uses an exponential cluster operator to improve the results. And the list can go on. The main purpose of all this extensions is to capture the electron correlation which was discarded from the start in the usual HF, since this quantity can be quite important in various systems. 

\section{Density Functional Theory}
\label{Dftt}

In the past decades, Density Functional Theory became the master method in many body simulations in a wide range of systems. Extensive studies can be found in nuclear physics \cite{stoitsov2007empirical}, atomic physics, molecular \cite{delley1990all}, cluster \cite{hobza1995density}, quantum plasma \cite{dharma1982density} , quantum dot \cite{hirose1999spin}, solid state, etc.

As we shall see, there are serious similarities with the HF theory, but there are also some fundamental differences which makes it faster in numerical simulations. This is the main reason for which is chosen over HF. 

Historically, the first genuine DFT theory appeared in the work of Thomas and Fermi \cite{thomas1927calculation} soon after Schroedinger equation was derived. They basically assumed that the density of kinetic energy of a fermionic system can be approximated with the \emph{only density dependent} expression derived analytically from the Homogeneous Electron Gas (HEG) model. This approximation gives, from a statistical perspective, an equation of state in the thermodynamic limit of large number of particles $N\to\infty$. We shall discuss in detail the TF approximation in a future section.

Some other extensions as the Bloch model which is just the time dependent versions of TF have been worked out in the next years, but essentially, it remained with the status of a simple, not reliable model, until 1964 when in a paper \cite{hohenberg1964inhomogeneous} was putted the idea of DFT on solid mathematical ground with the two theorems known nowadays as HK theorems. One year later, \cite{kohn1965self} designed a more practical way of implementing the DFT with a set of non-interacting particles obeying the KS equations. From that moment the road was free to extensions and applications. In the $90's$ \cite{runge1984density} have recreated the work from the $60's$ for time dependent phenomena and in coherence with the development of the computers, it became the tool for world wide scientists in a lot of  domains.

The central idea of DFT is that "everything" can be done just knowing the density of particles. In terms of $\hat{\Gamma}_1$, the density can be expressed as the diagonal part : $\rho(\mathbf{r})=\hat{\Gamma}_1(\mathbf{r,r})$. While the entire original construction was done on grounds of $\rho$, assuming that the external potential in which our system is placed is purely local $v(\mathbf{r})$ we shall take into consideration the fact that there are systems in which the the potential can be non-local $v(\mathbf{r},\mathbf{r}')$. The latter case has been treated by \cite{gilbert1975hohenberg}. For this reason we shall pass through the HK and G's theorems in parallel.

\subsection{Hohenberg Kohn theorems}

It is a logical (and mathematical) fact that having the $N$ number of particles fixed and the external potential $v(\mathbf{r})$ all the properties of the ground state (GS) are fully (uniquely) determined. In DFT this idea is somehow reversed, in the sense that the external potential is uniquely determined by the density.

Now, to state the first HK theorem and the first Gilbert theorem :

	\emph{HK: Between the external potential $v(\mathbf{r})$ and the density of particles $\rho(\mathbf{r})$ there is a bijective mapping in the sense that the density determines the potential up to a trivial constant.}	

	\emph{G: The external potential determines uniquely the density matrix $\hat{\Gamma}_1(\mathbf{r},\mathbf{r}')$}

To sketch the proof, let us presume by \emph{reduction at absurdum} that there are two total wave functions $|\Psi\rangle$ and $|\Psi'\rangle$ and their associated external potentials $v(\mathbf{r})$, $v'(\mathbf{r})$ give us the same $\Gamma_1$ and consequently the same $\rho$. If this is true, the ground state energy can be written for the two cases as:

\begin{eqnarray}
	E_{GS}=\langle\Psi|\hat{H}|\Psi\rangle\\
	E'_{GS}=\langle\Psi'|\hat{H}'|\Psi'\rangle
\end{eqnarray}

Now, since all the properties of the GS are defined by the densities, one can separate the energy functional in an universal unknown functional of density and the potential energy given by the external potential $E_{GS}=F[\rho]+\int\rho(\mathbf{r}) v(\mathbf{r})d\mathbf{r}$ or $E_{GS}=F[\Gamma_1]+\int\Gamma(\mathbf{r},\mathbf{r}') v(\mathbf{r},\mathbf{r}')d\mathbf{r}d\mathbf{r}'$ for non-local potentials. The Ritz variational principle says that the ground state energy is minimum, therefore:

$$E_{GS}-E'_{GS}=\int d\mathbf{r}d\mathbf{r}'\Gamma_1(\mathbf{r},\mathbf{r}')(v(\mathbf{r},\mathbf{r}')-v'(\mathbf{r},\mathbf{r}'))<0$$
$$E'_{GS}-E_{GS}=\int d\mathbf{r}d\mathbf{r}'\Gamma_1(\mathbf{r},\mathbf{r}')(v(\mathbf{r},\mathbf{r}')-v'(\mathbf{r},\mathbf{r}'))<0$$

But, adding this two relations we obtain $E_{GS}-E'_{GS}<E'_{GS}-E_{GS}$ which is absurd therefore, the hypothesis of the theorems hold true, both for local and non-local external potentials.

	\emph{HK: The true ground state density $\rho(\mathbf{r})$ minimizes the energy functional.}	

	\emph{HK: The true ground state density $\gamma_1(\mathbf{r},\mathbf{r}')$ minimizes the energy functional.}

Now, having the same external potential we have the $\rho(\mathbf{r})$ and $\gamma_1(\mathbf{r},\mathbf{r}')$ which minimize the energy functional $E[\rho/\gamma_1]$ and gives us the ground state configuration. If there would be another $\rho'$ and an associated $\gamma_1'$ that would minimize the energy, then by the variational principle, those minima would not be the ground state, therefore, the density matrix or density of particles that gives us the minima in energy are unique. 

The Gilbert theorems have been used, to generalize the original DFT to non-local potentials, which, as we shall see later, are very common in cluster physics.

Now, having this theorems that tells us that the Ground state density operator is uniquely determined by the external potential and is unique for the ground state, one can go back to the energy functional $E$ and apply the energy minimization principle with the constrain of constant number of particles (in the microcanonical ensemble) to obtain an Euler Lagrange equation. Indeed:

\begin{eqnarray}
	\delta\{E[\gamma_1]-\mu N\}=0\\
	\delta \{E[\rho]-\mu N\}=0
\end{eqnarray} 

Using the explicit expression for the energy and the number of particles $N=\int \rho(\mathbf{x})d\mathbf{x}=\iint \gamma_1(\mathbf{x},\mathbf{x'})d\mathbf{x'}d\mathbf{x}$ we obtain the following Euler Lagrange equations:

\begin{eqnarray}\label{abstractdft}
	\frac{\delta F[\rho]}{\delta \rho}+v(\mathbf{r})=\mu\\
	\frac{\delta F[\gamma_1]}{\delta \gamma_1}+v(\mathbf{r},\mathbf{r}')=\mu
\end{eqnarray}

The value of the DFT is that, in principle, if one would know the $F[\rho]$ or $F[\gamma_1]$ functionals exactly, then it would be straightforward to solve the above equations and to find the densities. Unfortunately, for this functionals only approximative expressions are known and will be discussed later. One might, wrongfully, think that could express the total energy as in eq. \ref{HFenergy1} from the HF. But let us remind that in there some specific factorization of the $\Gamma_2$ in terms of $\Gamma_1$ has been worked out plus an approximation of zero correlations. Here is not the case since one of the purpose of DFT is to capture as much as possible from the correlation effects. 

\subsection{Kohn-Sham method}
\label{Kohn-Sham}

In 1965 Kohn \& Sham have partially cured the problem of unknown $F$ functional in DFT (for local external potentials) introducing a system of non-interacting particles. In essence, one could separate a kinetic energy term and an interacting energy term in $F$ in such a way that $F[\rho]=T[\rho]+V_{ee}[\rho]$. Even trying to represent the density and the kinetic energy in terms of natural orbitals of $\gamma_1$ (which are not to be confunded with the ones from HF), one could at best write $\rho=\sum_ip_i\psi_i(r)$ and $T[\rho]=\sum_ip_i\langle\psi_i|\hat{p}^2|\psi_i \rangle$. But in here, the summation goes over an infinite number of orbitals. 

The idea of H\&K was to take a particular case of this representation, in such a way that $p_i=1, \forall i\le N$ and basically reframe the problem to a set of non-interacting particles. Still, the kinetic energy provided by this set of orbitals is not the true kinetic energy, nor the interacting energy $V_{ee}$ can be exactly reproduced. Therefore we introduce  $T_s[\rho]=\sum_i\langle\psi_i|\hat{p}^2|\psi_i \rangle$ the kinetic energy of the non-interacting KS orbitals and $\rho=\sum_ip_i\psi_i(\mathbf{r})$. With this, one can formally rewrite the total energy as $F_{HK}[\rho]=T_s[\rho]+T[\rho]-T_s[\rho]+V_{ee}[\rho]+J[\rho]-J[\rho]=T_s[\rho]+J[\rho]+E_{xc}[\rho]$. Where the $J[\rho]$ is the Coulomb interacting term like in the HF theory.

So what?, one might ask. Now the energy is even more complicated since you have the unknown $E_{xc}[\rho]$ plus a term which is at best representable by a system of fictitious particles. Well, now we will take into account the minimization principle for the total energy, with the associated normalization constrains for KS orbitals and performing the minimization with respect to orbitals. By some simple functional derivatives, one obtains the KS equations for a ground state system:

\begin{eqnarray}
	[-\frac{\hbar^2}{2m^*}\nabla^2+v_{KS}]\psi_j=E_j\psi_j\\
	v_{KS}=v_{ext}+v_{H}+v_{xc}[\rho]
\end{eqnarray}

To decript the terms from the effective KS potential $v_{KS}$ we see that the first is obviously the external one, which has to be specified for every system in part (in particular, for clusters on the ground state, is the potential created by the nuclei or ions). The second, is called Hartree potential in connection with the effective potential which can be found in HF theory and is subject to a Poisson equation (having Coulomb nature) $\nabla^2v_{H}=-4\pi\rho(\mathbf{r})$. Of course, talking about non-interacting particles, the total density can be easily reconstructed as $\rho(r)=\sum_j|\psi_j|^2$. The last term, which stands for exchange-correlation potential, can be 
written  formally as derivative of the exchange-correlation energy $v_{xc}=\delta E_{xc}/\delta\rho$, but apart from this we do not know its specific form. 

Now it should be clear that as HF, DFT allows us to solve $N$ single particle equations in the \emph{mean field} approximation from which any observable can be computed. The point in which the DFT becomes easier to be used than HF is exactly the term which is unknown. There has been a lot of search for good approximations on $v_{xc}$ and usually, it has a pure density dependent form, more or less simple or local. Still, being density dependent the calculation of this potential is much more easy to perform than the effect of the exchange operator from HF, therefore, the whole method is easier to implement numerically. 

Regarding the approximation used for $v_{xc}$ we do not enter in this subject, since it is only of practical use for cluster physics, but much of the physicists or chemists that work in the DFT branch are drawn in the search for better functional. I will just remind with appropriate references the well known approximations. First is the ideal approximation, LDA \cite{perdew1992accurate} in which xc potential is local in the total density $\rho$ therefore, very easy to compute numerically (generically the exchange part is taken as for the ideal gas $\propto\rho^{1/3}$). An extension of LDA is to consider also local functional for xc but which include further gradient corrections and this approximation goes by the name of GGA \cite{perdew1996generalized}. Going even further, one can use functionals which include laplacian of the density, MetaGGA or go the the non-local potentials \cite{langreth1981easily}. 

Through the simulations done for this thesis, I have used the Gunnarson-Lundqvuist \cite{gunnarsson1976exchange} due to its LDA form. Still, there are problems with such approximation that must be corrected by the so called Self-Interaction Correction \cite{perdew1981self}.

As a final remark, DFT should incorporate (like HF) the spin character of the orbitals. This can be including working with spinors instead of wave-functions, or simply assigning a spinorial $\sigma$ label to each $\psi_j\to\psi_{j,\sigma}$. The effects is that in practice there are xc functionals that take into account separated densities for spin $(\rho_{\downarrow},\rho_{\uparrow})$ and this can give certain differences in the numerical results, in particular on the total energy of the system.

\subsection{Time Dependent DFT}

Now we have seen how the DFT for local and non-local external potential has been constructed historically, we should go further to the time dependent version of this theory since, by default, the interaction with strong laser pulses is dynamic phenomenon. The basic work in this aspect has been carried out by Runge \& Ross in 1984 \cite{runge1984density} and essentially follows the same logic as the HK original DFT derivation. For this reasons we will not enter in too much detail, just perform a short parallel.

	\emph{RR:The bijective mapping between $v_{ext}(\mathbf{r},t)$ and $\rho(\mathbf{r},t)$ (or $\vec{j}(\mathbf{r},t)$) holds for time dependent systems in the sense that the eternal potential is determined uniquely up to a function of time.}	

	\emph{RR: The true density $\rho(\mathbf{r},t)$ minimizes the quantum action functional.}  	

The proof of the first theorem is more involved than in the stationary case, but the second one is applied in the same manner: a functional called action $\mathcal{A}[\rho]$ is defined and separated in a single particle manner which in turn gives us a set of Schrodinger like equations in the time dependent regime:

\begin{equation}
	\label{tdks}
	i\hbar\partial_t\psi_j=[-\frac{\hbar^2}{2m}\nabla^2+v_{KS}]\psi_j
\end{equation}

While formally, we have a Schrodinger equation and the effective KS potential is constructed in the same way as in the stationary case, the time dependent DFT problem is a bit more involved. The first heavy impediment is that in the dynamics, many of the functional approximations for $E_{xc}$ are not valid anymore, since they are derived under ground state considerentes. Therefore, the field of time dependent $v_{xc}$ approximations is a bit more dry, or at least gives worse results. A true functional should have memory properties in the sense that should be a time integral over a non-local time kernel, which would make any computational analysis far too complicated. For this reason, the LDA/GGA functionals are usually transferred in the dynamic case, at least for gross properties.

\section{Phase space and Vlasov limit}

Until now we have a picture of how from the density matrix formulation of quantum mechanics using Liouville von-Neumann equation and the reduced density matrix approach one can derive the BBGKY hierarchy. Further this hierarchy can be truncated at the zero level neglecting correlation and obtaining the  HF. 

Now, we should go further on another approach of parallel power with HF but contained in a different representation of quantum mechanics, namely the phase space representation. The pylons of this direction have been putted by Wigner in its seminal paper \cite{wigner1932quantum} on the quantum effects at thermodynamic systems. From there, a lot of work has been performed and now, there is a solid literature and mathematical apparatus that can be invoked in this direction. We shall start with a basic introduction in the formalism of the quantum phase space and than the Vlasov equation will be derived as a semiclassical limit.

\subsection{Phase space representation and Wigner-Weyl transform}

The search for alternative representations of quantum (statistical) mechanics has been always present in physics due to some intrinsic hope that a different representation would give access to the same physical reality from a different mathematical perspective. This is the case of the phase space representation which holds many similarities with the classical phase space statistical mechanics. 

Soon after Schrodinger's equation, Hermann Weyl found a way to map the classical functions from the PS to operators through a quantisation procedure \cite{weyl1927quantenmechanik} called \emph{Weyl quantization}. Conversely, in 1933 \cite{wigner1933quantum} found a way to map the quantum operators into classical-like functions in the PS. This bijection from quantum operators to classical functions is contained mathematically in the so called \emph{Wigner-Weyl transform}. Let us consider a function $f[\Phi]$ and its associated operator $\Phi[f]$. The transform reads :

\begin{eqnarray}
	\hat{\Phi}[f]=\frac{1}{(2\pi)^{-d}}\int f(q,p)e^{i[a(\hat{Q}-q)+b(\hat{P}-p)]} dqdpdadb\\
	f[\hat{\Phi}]=2\int dye^{-2ipy/\hbar}\langle q+y|\hat{\Phi}[f]|q-y \rangle
\end{eqnarray}

The position-momentum operators from the Weyl transform obey the Lie Algebra with the associated commutation relation: $[\hat{P},\hat{Q}]=\hat{P}\hat{Q}-\hat{Q}\hat{P}=-i\hbar\hat{\mathbf{1}}$. One could easily slip on the path of mathematical questions regarding the Weyl algebra \cite{block1981irreducible}, Moyal bracket \cite{bayen1978deformation}, etc. While being a theoretical thesis, still, it is not my purpose here to treat such subjects which have no practical value. 

Now, we have a quantum theory, from the BBGKY hierarchy and want to obtain a PS one, therefore, for us the Weyl transform is useless. We shall focus further on the Wigner representation which transform the \ref{BBGKY} equations in a PS ones. First of all, the $s$ reduced density matrix $\hat{\Gamma}_s$, after being dragged through a coordinate representation and than subject to a Wigner transform, gives us the evolution equation in PS:

\begin{eqnarray}
	f_s(\mathbf{r},\mathbf{p},t)=\int \frac{\mathbf{r'}}{(2\pi\hbar)^{3s}}e^{-i\mathbf{p}\mathbf{r'}/\hbar}\Gamma_s(\mathbf{r},\mathbf{r'},t)
	\\
	\{\partial_t+\frac{\mathbf{p}}{m^*}\nabla_{\mathbf{r}} \}f_s(\mathbf{p},\mathbf{r},t)=\frac{1}{i\hbar}Q[V,\Gamma_{s+1}]
\end{eqnarray}

The $Q$ term is quite tedious to show, but we will particularize the form for $f_1$:

\begin{eqnarray}
	\{\partial_t+\frac{\vec{p}}{m^*}\nabla_{\vec{r}} \}f_1(\vec{p},\vec{r},t)=\frac{1}{i\hbar}\int\frac{dr_1}{(2\pi\hbar)^3}dp_1e^{-i(p-p_1)r_1/\hbar}\mathcal{Q}\\
	\mathcal{Q}=V(x)|_{r-\frac{r_1}{2}}^{r+\frac{r_1}{2}}f_1(r,p_1,t)+\int dr_2dp_2\phi(x)|_{r-r_2+r_1/2}^{r+r_2-r_1/2}f_{12}(r,p_1,p_2,r_2,t)
\end{eqnarray}

Now, for brevity and elegance in writing we define the Moyal bracket as a mathematical operation which is basically a composition law which for two functions $f,g$ defined on a $2d$ Euclidean phase space like $\{\vec{r},\vec{p}\}$:

$$\{\{f,g\}\}=\frac{2}{\hbar}f(\mathbf{r},\mathbf{p})sin(\frac{\hbar}{2}(\overleftarrow{\nabla}_\mathbf{r}\overrightarrow{\nabla}_\mathbf{p}-\overleftarrow{\nabla}_\mathbf{p}\overrightarrow{\nabla}_\mathbf{r}))g(\mathbf{r},\mathbf{r})$$

Where the sense of the $\overleftarrow{}$ means that it acts on the left side of the expression, namely on $f$. If one, as in HF neglects the correlation and generally the coupling with higher order matrices, the so called Quantum Wigner equation is obtained in the compact form:

\begin{eqnarray}
	\label{qwigner}
	\partial_tf(\mathbf{r,p},t)=-\{\{f(\mathbf{r,p},t),H(\mathbf{r,p},t)\}\}
\end{eqnarray}

From now on, $f(\mathbf{r,p},t)\equiv f_1(\mathbf{r,p},t)$. This elegant form is interesting from several perspectives. First of all, we have been able to recast the whole Hamiltonian operator inside the Moyal bracket. Second, the similarities between eq. \ref{qwigner} and the classical kinetic equation (which also can be written for Hamiltonian systems in terms of Poisson brackets) is astonishing. If one looks closely to the Moyal bracket and takes the $\hbar\to 0$ limit, the classical Poisson bracket is obtained and the motion of particles becomes a classical one. The reason for the differences between classical and quantum can be understood in terms of phase space in the sense that the commutator $[\hat{p},\hat{x}]=i\hbar\hat{\mathbf{1}}$ is preserved in PS as volume. 

Further differences can be understood from the properties of the Wigner distribution function $f(\mathbf{r,p},t)$. First of all, it is a real function which has its norm (the integral over the PS) equal with trace of the density operator, so equal with $N$ the number of particles. The evolution equation is space-time symmetric and also Galilei covariant. From the definition of the Wigner transform, any physical quantity $A$ can be obtain integrating the product between $f(\mathbf{r,p},t)$ and the associated function for that quantity $A(\mathbf{r,p})$. In particular, the density is the integral of $f$ over the moment space $\int d\mathbf{p}f(\mathbf{r,p},t)=\langle \mathbf{r}|\hat{\rho}|\mathbf{r}\rangle=\rho(\mathbf{r})$.

Perhaps the most important property is related with the limits of $f$. It can be proven that due to the structure of the PS and the evolution equation, $-2/\hbar\le f(\mathbf{r,p},t) \le 2/\hbar$. As one can see, $f$ is not bounded bellow to zero which means that in can have local negative values. In this respect doesn't met the classical criteria of probability distribution function and the interpretations of it can be quite doubtful.

Even though,  knowing $f$ (in the frame of HF approximation) is equivalent with knowing basically any quantity of interest for our system, the path towards such knowledge is not just nontrivial, but usually impossible. The challenge of solving \ref{qwigner} even numerically is not feasible, therefore, further simplifications must be used.

\subsection{Vlasov equation}
\label{Vlasovv}

In the classical kinetic theory there is widely known the Boltzmann equation for the distribution function. This equation has its quantum correspondent but both have a fundamental flaw: the long range forces (as Coulomb force between electrons) are not described. The extension known from plasma physics is the Vlasov equation. The quantum correspondent can be derived from the Quantum Wigner equation. 

If one takes the Moyal bracket and expand it in powers of $\hbar$ and then retains only the first two terms, the following semi-classical equation it is obtained:

\begin{eqnarray}\label{qvlasov}
	(\partial_t+\frac{\mathbf{p}}{m}\nabla_\mathbf{r}-\nabla_\mathbf{r}V\nabla_\mathbf{p}-\frac{\hbar^2}{24m^3}\nabla_\mathbf{r}^3V\nabla^3_\mathbf{p})f(\mathbf{r,p},t)=0
\end{eqnarray}

Now, we have neglected in the derivation of Q Wigner Eq. \ref{qwigner} the rhs. of the zero order BBGKY equation. From a physical view, that part contained correlations of particles and terms associated with the so called collisions. If one would want to include such effect this can be done in an \emph{ad-hoc} manner with a somehow empirical term in the Vlasov Eq. which should take care of the Pauli blocking. The first approach on this part was done by \cite{uehling1933transport}. Using this, they wrote the so called VUU equation:

\begin{eqnarray}\label{VUU}
	(\partial_t+\frac{\mathbf{p}}{m}\nabla_\mathbf{r}-\nabla_\mathbf{r}V\nabla_\mathbf{p})f(\mathbf{r,p},t)=I[f(\mathbf{r,p},t)]\\
	I[\mathbf{r,p}]=\int\int d\mathbf{p}_2d\mathbf{p}_3d\mathbf{p}_4W_{1234}[f_1f_2f_3]\\
	W_{1234}=\frac{d\sigma}{d\Omega}\delta(\mathbf{p}_1+\mathbf{p}_2-\mathbf{p}_3-\mathbf{p}_4)\delta(\mathbf{p}_1^2+\mathbf{p}_2^2-\mathbf{p}_3^2-\mathbf{p}_4^2)
\end{eqnarray}

As one can see, the logical construction of the collision term is to block the presence of two particles in the same Heisenberg volume element of the PS: $(2\pi\hbar)^3$ consistent with the uncertainty principle. The Vlasov equation has the advantage (beyond numerical) of turning a positive defined $f$. In contrast, if one includes quantum terms like in \ref{qvlasov}, this property is lost.

Now, even assuming that the semi-classical limit is a valid approximation, one would want to include exchange correlation effects which can be essential in some systems. For this, there is another way to derive VUU+xc using DFT. Going back to the Section \ref{Kohn-Sham} we see that one inherent assumption of DFT was that $\Gamma_1=\sum_ip_i|\psi_i\rangle\langle\psi_i|$. Performing a Wigner transform on this density matrix and using the Kohn-Sham Eqs. \ref{tdks}, one can arrive in the semi-classical limit at a very similar equation with \ref{qvlasov}, but with a different potential:

\begin{eqnarray}\label{qvuu+xc}
	(\partial_t+\frac{\mathbf{p}}{m}\nabla_\mathbf{r}-\nabla_\mathbf{r}V_{ks}\nabla_\mathbf{p})f(\mathbf{r,p},t)=0
\end{eqnarray}

Now, exchange-correlations are included in a mean field manner in the \textit{VUU+XC} equation simply by the presence of the $v_{xc}$ potential for $v_{KS}$. There are other approaches that try to include the correlations effect by relaxing the initial factorization of the second order density matrix but we do not discuss them here.

Regarding the set of quantum effects recovered in this $VUU$ scheme, we obviously can point out the XC and the Pauli-type collisional term. Beside that, there is another statistical quantum effects which must be included from the initial conditions of the equation: namely the initial value of the $f(\mathbf{r,p},0)$. As we shall see later, this configuration is consistent with the Thomas Fermi approximation and it reflects the idempotency of the density matrix of order 1.

Still, complicated, from the mean field perspective, the Vlasov equation is now feasible for numerical applications using one of the zoo of methods. From there we mention just the test particle method \cite{fennel2004ionization} and the PIC \cite{sonnendrucker1999semi} codes. Moreover, there are even analytical results which can be drawn from this kinetic theory, from which the most stringent is the Landau damping.

The theory of Landau damping is old \cite{landau1946vibrations} and mathematically involved so there is no reason to present it in here, but an essential aspect should be noted: by linearizing the most simple variant of VLasov eq, without xc or collisional effects (only the electrostatic), one can obtain a dispersion relation which shows that any wave in a Vlasov system will damp itself, in principle because of the dispersions of the velocities in the momentum space. Other classical interpretations show wave-particle like interactions, etc.. But it remains crucial (and it will be used as a numerical advantage) that any excited quantum state should have during the dynamics this entropy preserving phenomena which directs it towards the ground state.

Not to be confused with the $H$ theorem that shows that any asymptotic solution of Boltzmann equation will be driven finally towards a Boltzmann distribution. This is a problem since allows numerical propagations up to $ps$ and must be complemented by a high number of pseudo-particle in order to overcome this numerical thermalization \cite{fennel2004ionization}.

Regarding the validity of VUU model, this is driven mainly by the lack of quantum effects. At high temperatures, specific to strong laser fields (above $10^3K$) the nature of the interaction is basically Coulombian with collisions and the phase space picture becomes fully valid, in the classical sense.

\section{Hydrodynamic models}
\label{hydrod}

Even though we have this three possibilities of finding more or less approximative the dynamics of a system of fermions (DFT, HF and VUU) we still can be in trouble. There are very large systems (as we shall see in the numerical result section \ref{Chapter3}) which are not feasible to be studied with single particle methods (HF, DFT). On the other hand, even the VUU can impose seriously numerical difficulties for such system. Or sometimes one simply needs to be faster in computation at the cost of precision in calculations. 

For that reason, we search for further simplification. In classical physics, the simplification of the Vlasov (or any kinetic approach) is called fluid dynamics. Etymologically, the name was assigned because the equations obtained by moments of a kinetic equation have the same structure with the Navier-Stokes equations known from hydrodynamics. 

In quantum systems, the situation is the same. One can derive from VUU, DFT or HF the so called Quantum Hydrodynamic Model (QHM). This has been used extensively in the past years in \cite{gardner1994quantum,crouseilles2008quantum,gardner1998smooth,domps1998time} in applications for metal clusters, fullerenes, semiconductors, etc. We shall describe shortly three ways of derivation it, which, even though give similar results, have some fundamental differences. 

\textbf{Kinetic derivation}

We, as physicists must relate the theoretical work with some practical results and while for a mathematician the study of VUU like equations fullfills his soul, we have to obtain physical, realistic, results from it. For that reason let us see how some local observables can be obtained from distribution function. We list bellow, the density $\rho$, the current $\vec{J}$ and the kinetic pressure tensor $\hat{P}$:

\begin{eqnarray}\label{hydroquant}
	\rho(\mathbf{r})=\int f(\mathbf{r,p},t) d\mathbf{r}\\
	\mathbf{J}(\mathbf{r},t)=\int \mathbf{p}f(\mathbf{r,p},t) d\mathbf{p}\\
	\hat{P}(\mathbf{r},t)=\int f(\mathbf{r,p},t)(\mathbf{p}-\mathbf{u})\otimes (\mathbf{p}-\mathbf{u}) d\mathbf{p}
\end{eqnarray}

Where $\mathbf{u}=\mathbf{J}/\rho$ is the global velocity field. Now, this are moments of the distribution function in the momentum space. The list can  go on forever. Integrating \ref{qvlasov} this way, we obtain an infinite hierarchy of equations which form the hydrodynamic model. We will write down only the first two equations because usually only this two are solved numerically and as the order of the quantity increases the complexity of the corresponding equation is higher:

\begin{eqnarray}
	\label{QHM1}
	\partial_t\rho+\nabla \vec{J}=0\\
	\partial_t\vec{J}+\nabla[\frac{\vec{J}\otimes\vec{J}}{\rho}+\hat{P}]+\rho\nabla V=0
\end{eqnarray} 

Now, as we see, any $n$ order equation, contains the $n+1$ moment of $f$. It is like BBGKY hierarchy all over again. The solution is to find a similar truncation at some point in which the $n+1$ moment is described by the previous $n$ ones. Usually this is done taking the thermodynamic limit $N\to\infty$ and consider an equation of state for the pressure. The most common one is the polytropic (Thomas Fermi) $\hat{P}=P[\rho]\hat{\mathbf{1}}$, where the scalar pressure $P[\rho]\propto \rho^{\gamma}$ but other approximations can be used too. 

This model has some issues: first of all, the Landau damping is no longer present since the features of the momentum space are lost. So linear waves miss the natural kinetic damping. Second, the pressure tensor is unknown. Third, even if the pressure tensor would be exaclty known in terms of density and current, where are the quantum effects? The truth is that deriving it from Vlasov, the QHM contains the quantum effects in the pressure term in some obscure manner by the statistical properties of $f$. A more detailed discussion on the derivation can be found in \cite{gasser1997quantum,degond2003quantum,degond2005quantum}. 
\vspace{1cm}

\textbf{DFT derivation}

Regarding the derivation from DFT, there is not too much to say. Already we have a quantum hydrostatic model, simply from the Euler -Lagrange equation \ref{abstractdft}:

$$g[\rho]+v_{KS}[\rho]=\mu$$

But further, if one looks at the derivation of TDDFT, sees that a set of QHM like equation are obtained \cite{runge1984density} as the continuity equation $\partial_t\rho+\nabla\vec{J}=0$ and a momentum equation $\partial_tJ=\vec{P}=\langle [H,\hat{j}]\rangle$.

Already, the continuity equation is not a surprise anymore, and we are able to obtain it from any angle. The fundamental issue is with the current equation. By default, the \cite{runge1984density} equations state just that the evolution of the current field in a system of $N$ identical particle should be equal with the expectation value of the Hamiltonian-current commutator. Again computing what you can from the commutator, the same formal system of QHM equations is at hand:

\begin{eqnarray}\label{QHM2}
	\partial_t\rho+\nabla \vec{J}=0\\
	\partial_t\vec{J}+\nabla[\frac{\vec{J}\otimes\vec{J}}{\rho}+\hat{P}[\rho]]+\rho\nabla V_{ks}=0
\end{eqnarray} 

Where the tensor $\hat{P}$ is again unknown. As for the Vlasov derived QHM, one can make the assumption that the tensor is isotropic and can be related with the $T_s$ the kinetic energy functional by $\hat{P}=t_s[\rho]\hat{\mathbf{1}}$.

\textbf{Single particle derivation}

This third approach can be found in literature \cite{gasser1995quantum} and can be derived in the frame of DFT or HF since they both work with single particle orbitals. It has the main advantage that reproduces pure quantum effects. Let us consider the set of HF/DFT solutions $\{\psi_i\}$. We use the Madelung transform which is nothing else than a polar form of an wave function $\psi_k=\rho_k^{1/2}exp(iS_k/\hbar)$ in terms of single particle density and phase. Then we introduce it in the HF/DFT equation and separate the real and the imaginary parts resulting, in order to obtain a set of two equation known as Madelung equations and are the first attempt to construct a quantum hydrodynamic theory:

\begin{eqnarray}
	\label{Madelung}
	\partial_t\rho_k+\nabla\vec{j}_k=0\\
	\partial_tS_k+\frac{(\nabla S)^2}{2m^*}+V-\frac{\hbar^2}{2m^*}\frac{\nabla^2\rho^{1/2}}{\rho^{1/2}}=0
\end{eqnarray}

Now, from the definition of the current operator is easy to show that the single particle current $j_k=\rho_k\nabla S_k$. Also we can associate with this the single particle velocity field $\vec{u}_k=\nabla S_k$. With these, one arrives at the density-current hydrodynamics for a single particle: 

\begin{eqnarray}
	\label{Madelung2}
	\partial_tj_k+\nabla (\frac{j_k\otimes j_k}{\rho_k}-\frac{\hbar^2}{2m^*}\rho\nabla\otimes\nabla\ln\rho_k)+\rho_k\nabla V=0
\end{eqnarray}

Now we use the additivity conditions for density $\sum_kp_k\rho_k=\rho$ and for currents $\sum_kp_k\vec{j}_k=\vec{J}$ plus the notation of total velocity $u=J/\rho$ and add the eq. \ref{Madelung2} in the following form: 

\begin{eqnarray}
	\label{QHMs}
	\partial_t\rho+\nabla\vec{J}=0\\
	\partial_tJ+\nabla (\frac{J\otimes J}{\rho})+\nabla\hat{\Pi}+\rho_k\nabla V=0\\
	\label{quantumtensor}\hat{\Pi}=\sum_kp_k(\rho_k(u_k-u)\otimes(u_k-u)-\frac{\hbar^2}{2m^*}\rho_k\nabla\ln\rho_k)
\end{eqnarray}

Again, we have obtained the equations of QHM. Now it is nice to observe that kinetic energy tensor \ref{quantumtensor} contains the Bohm like terms which are pure quantum. These type of effects were not present in the definition of pressure tensor derived under Vlasov equation since the $\hbar\to 0$ limit has been taken. Beside reproducing some quantum effects related with the gradients of density, this form of QHM has the same issues of not knowing explicitly the tensor. Further ways to approximate it will be discussed in the next section.

\subsection{Kinetic energy functional/quantum pressure}

Until now we have stated about DFT that its conceptual power lies in the fact that is exact in principle, written in Euler-Lagrange equations as \ref{abstractdft}. Also, any of the forms of the QHM model derived above are also exact in principle since a QHM is nothing else than the time dependent version of the abstract DFT. Nonetheless, one still doesn't know the exact xc potential, nor the kinetic energy functional/kinetic energy tensor. Trying to approximate these terms (without going on the path of KS equations) we should find ways to express $T_s[\rho]$ or more general the $\hat{\Pi}[\rho]$ functional. To gain more insight about how to make such approximations, let us relate the above mentioned quantities to the KS orbitals, or to the hydrodynamic variables $\{\rho_k,S_k\}$:

\begin{eqnarray}
	T_s=-\frac{\hbar^2}{2m^*}\int\sum_k\psi_k^*\nabla^2\psi_k=\frac{\hbar^2}{2m^*}\int\sum_k|\nabla\psi_k|^2\\
	T_s=\frac{\hbar^2}{2m^*}\int\sum_k(\frac{\rho_k}{\hbar^2}|\nabla S_k|^2+\rho_k^{1/2}\nabla^2\rho_k^{1/2})\\
	\hat{\Pi}(r,t)=\sum_kp_k(\rho_k(u_k-u)\otimes(u_k-u)-\frac{\hbar^2}{2m^*}\rho_k\nabla\times\nabla\ln\rho_k)
\end{eqnarray}

In literature there are a quite various set of approaches to approximate this terms. First of all, historically speaking is the Thomas Fermi approximation [] which is based on the assumption that in the ground state the orbitals behave like free non interacting particles, therefore obey the Fermi-Dirac statistics. From these assumptions, one can easily prove that the local density and the local density of kinetic energy in the ground state can be related by:

\begin{equation}
	t_{TF}[\rho]=\kappa_0\rho^{5/3}
\end{equation}

Going further, Von Wieszacker proposed a correction to TF approximation which takes into account the gradient corrections to the functional:

\begin{equation}
	t_{TF+W}[\rho]=t_{TF}[\rho]+\frac{\nabla^2\rho^{1/2}}{\rho^{1/2}}
\end{equation}

Further there is an extension of all this, with the Bloch density matrix in the frame of Wigner-Kirkwood expansion which gives us the so called gradient expansion functional and it can be written up to the fourth order in $\hbar$ as:

\begin{eqnarray}\nonumber
	\label{semiclass}
	t_{GE}[\rho]&=&\frac{\hbar^2}{2m}\{\kappa_0\rho^{5/3}+\kappa_2\frac{(\nabla\rho)^2}{\rho}+\\&& +\kappa_4[8(\frac{\nabla\rho}{\rho})^4-27(\frac{\nabla\rho}{\rho})^2\frac{\nabla^2\rho}{\rho}+24(\frac{\nabla^2\rho}{\rho})^2]\}
\end{eqnarray}

With $\kappa_0=3/5(3\pi^2)^{2/3}$, $\kappa_2=1/36$, $\kappa_4=(6480(3\pi^2)^{2/3})^{-1}$ for $3D$ systems. 

Nowadays, the most employed orbital free method is the one that uses an exchange-correlation potential with the TF functional corrected by a coefficient dependent Weiszacker term. Pauli terms, linear response based functional, etc. are also useful methods to derive approximations for the kinetic energy of a KS system. 

For the dynamic functional there are some attempts to derive it but the results are not impressive. Essentially, the TDDFT theorems assure us that there is indeed such a tensor only density dependent, fact which allows one to write $\hat{\Pi}[\rho]$. As described in the previous section, many times the only thing that can be done is to consider that some functional for the kinetic energy in the ground state holds also in non-equilibrium situations (which is not true, since by definition an equation of state is designed for equilibrium) so the mathematical approximation is done as $\hat{\Pi}=t_s(\rho)\mathbf{1}$. This is connected also with the kinetic theory in the sense that is implied that the tensor is diagonal and there are some symmetries in the phase space. In particular the geometrical structure of the distribution function is only translated in the momentum space.

Some approximation can be done on the Bohm like part of the \ref{quantumtensor} tensor supposing that all the orbitals have approximately the same spatial amplitude. This approximation has been used in \cite{crouseilles2008quantum}, but it has no solid justification. The only systems in which it could work are the systems with quasi free particles. Imposing it heuristically one can use it in the following sense:

\begin{equation}
	\frac{\hbar^2}{2m^*}\sum_kp_k\rho_k\nabla\otimes\nabla\ln\rho_k\approx\frac{\hbar^2}{2m^*}\rho\nabla\otimes\nabla\ln\rho
\end{equation}

In general, even though the single particle velocities are irotational $\nabla\times\nabla S=0$, the total velocity field can have rotational feature $\nabla\times u\neq 0$. But some anzatz can be done that the rotational feature of the total velocity field is negligible, therefore $\vec{u}\approx\nabla S$. This assumption is not unrealistic since the natural excitation in a cluster has dipolar character and the rotational fields are weak. Moreover, in \cite{domps1998time} there has been performed and analysis of the kinetic solution during the dynamics and it has been shown that in principle, the Fermi sphere from the momentum space is always shifted with an irrotational field. Further, using the approximation of diagonal pressure tensor $\hat{\Pi}_{clas}\approx\rho f(\rho)$, the QHM equation are to be written as:

\begin{eqnarray}
	\label{QHM4}
	\partial_t\rho+\nabla(\rho\nabla S)=0\\
	\partial_tS+\frac{|\nabla S|^2}{2m^*}+f(\rho)+V+\frac{\hbar^2}{2m^*}\frac{\nabla^2\rho^{1/2}}{\rho^{1/2}}=0\\
\end{eqnarray}

As a final step, the pseudo total wave function $\Phi=\rho^{1/2}exp(iS/\hbar)$  can be defined as an inverse Madelung transform and the \ref{QHM4} eqs. can be embedded in a Schroedinger like equation:

\begin{eqnarray}\label{NLS} 
	i\hbar\partial_t\Phi=[-\frac{\hbar^2}{2m^*}\nabla^2+w]\Phi\\
	w=V+f(\rho)
\end{eqnarray}

This approximation has been studied in connection with semiconductors, metal clusters \cite{domps1998time}, fullerenes \cite{palade2014optical}, thin metal foils, with good gross results. For a more detailed discussion about the Orbital-Free DFT see \cite{wang2002orbital}.

\section{Linearised approaches}
\label{linearised}

We, as civilization, have constructed the mathematical theories in pathological way, based on a linear thinking. Unfortunately, the nature proved to be mainly described by non-linear phenomena and many times we have to appeal to the last resort measure: the numerical tools. Still, wanting to capture weak effects or normal modes in non-linear theories, one often does a linearisation of the equation around some known solution to get insight into the problem. The many-body theory makes no exception and every theory from the one discussed until now have been dragged trough the method of linearisation. 

While from an operatorial procedure we could go to things like Dyson series, or to Feynmann diagrams in the many-body theory, in the present section I shall avoid such things, keeping a more practical view on how DFT, HF and Vlasov equations can be linearised into other tools for weak laser cluster interaction. 

\subsection{LRDFT}
\label{LRDFT}

In quantum mechanics, the most basic linearization procedure goes by the name of perturbation theory and is easily understandable in the Dirac representation. For many-body systems one can go to more abstract perturbation series like Dyson series. In principle the linearization of DFT equations can be derived from many angles. In the present work, I shall adopt a derivation based on the fundamentals of the linear response theory. 

In the frame of DFT  one can speak about the single particle response function $\chi_0(\mathbf{r,r'},\omega)$. On the other hand, the true linear response function $\chi(\mathbf{r,r'},\omega)$ can be related ($\delta \rho (\mathbf{r},\omega)$ is the induced density variation in the energetic space) to the external excitation $V_{ext}(\mathbf{r},\omega)$ by:

\begin{eqnarray}
	\label{zerorespfunct}
	\delta\rho (\mathbf{r},\omega)=\int d\mathbf{r'}\chi_0(\mathbf{r,r'},\omega)V_{KS}(\mathbf{r'})\\
	\delta\rho (\mathbf{r},\omega)=\int d\mathbf{r'} \chi(\mathbf{r,r'},\omega) V_{ext}(\mathbf{r'},\omega)
\end{eqnarray}

One can introduce in \ref{zerorespfunct} the expression for $V_{KS}$ and derive a self-consistent equation for the density in the spirit of Dyson series:

\begin{eqnarray}
	\label{self1}
	\delta\rho (\mathbf{r},\omega)=\int d\mathbf{r'}\chi_0(\mathbf{r,r'},\omega)[V_{ext}(\mathbf{r'},\omega)+\int d\mathbf{r''}\frac{\delta V_{KS}(\mathbf{r'})}{\delta\rho(\mathbf{r''})}\delta\rho(\mathbf{r''},\omega)]\\
	\chi_0(\mathbf{r,r'},\omega)=\sum_{i\in occ,\mu\in unocc}(\frac{\psi_i(\mathbf{r})\psi_\mu^*(\mathbf{r})\psi_\mu(\mathbf{r'})\psi_i^*(\mathbf{r'})}{\varepsilon_i-\varepsilon_{mu}-\hbar\omega-i\eta}+\frac{\psi_i^*(\mathbf{r})\psi_\mu(\mathbf{r})\psi_\mu^*(\mathbf{r'})\psi_i(\mathbf{r'})}{\varepsilon_i-\varepsilon_{mu}+\hbar\omega+i\eta})
\end{eqnarray}

This equation must be solved iteratively for the density usually on a real spatial grid. A description and more details on the method can be found in \cite{yabana2006real}. Once you have the density in the frequency domain, one can easily compute quantities as the dipolar response to investigate the optical response of the cluster for example in the linear regime. 

Finally one can recover a formal solution for the response function which will be useful to compare with the one from Vlasov theory:

\begin{eqnarray}
	\chi(\mathbf{r,r'},\omega)=[1-\chi_0(\mathbf{r,r'},\omega)\frac{\delta V_{KS}(\mathbf{r'})}{\delta\rho(\mathbf{r''})}]^{-1}\chi_0(\mathbf{r,r'},\omega)
\end{eqnarray}

\subsection{Random Phase Approximation}
\label{RPA}

The Random Phase Approximation has some controversal history. Originally it has been derived for nuclear physics and condensed matter calculation \cite{bohm1951collective,bohm1953collective,pines1952collective,pines1953collective}. An older version of it can be found as Tamm-Dancoff \cite{hirata1999time} approximation. It can be derived from considerentes of HF theory, but there are some problems related with its interpretation in terms of Feynmann diagrams where it can be proven that RPA is in fact a consequence of the Ring approximation \cite{jansen2010equivalence}. By abuse of language, now both are known as RPA.

In the present subsection we will derive the RPA equation following \cite{reinhard1992rpa} . We start from the HF theory in which the total wave function is thought to be a Slater determinant $|\Psi_0\rangle$ in the ground state. Now, we act with some external perturbation which induces an $\eta \hat{G}$ generator of the dynamics. By means of Thouless theorem \cite{thouless1960stability} one can show that the time dependent state of our system can be written $|\Psi\rangle=exp\{i\eta\hat{G}\}|\Psi_0\rangle$, where $\hat{G}$ is hermitian, and moreover must be a first order particle-hole like operator, $\hat{G}\simeq \hat{a}^\dagger_p\hat{a}_h$. 

In order to find the equations of motion for the system we apply the variation principle of quantum action $\mathcal{A}=\int dt\langle\Psi|i\partial_t-\hat{H}|\Psi(t)\rangle$ minimization $\delta \mathcal{A}=0$. We do this, inserting the expression of $|\Psi\rangle$ in $\mathcal{A}$, expand it to second order in $\eta$ and minimize. Denoting $\{ \hat{a}^\dagger_p\hat{a}_h, \hat{a}^\dagger_h\hat{a}_p\}\equiv \hat{a^\dagger a}$, the equation of motion is obtained:

$$\langle\Psi_0|[\hat{a^\dagger a},i\partial_t\hat{G}]|\Psi_0\rangle=\langle\Psi_0|[\hat{a^\dagger a},[\hat{H},\hat{G}]]|\Psi_0\rangle$$

The approximation $\hat{G}=\sum_n[\hat{C}_n^+exp(-i\omega t)+\hat{C}_n^-exp(i\omega t)]$ is taken in order to find normal modes in a fermionic system. Here $\hat{C}_n^\pm$
is the operator that generates/annihilates the $n-th$ eigenmode. Now, introducing these into the eqs. of motion and separating the linear independent oscillating factors $exp(\pm i\omega t)$, we get two eigenvalue problems for the $n-th $ eigenmode:

\begin{eqnarray}
	\omega_n\langle\Psi_0|[\hat{a^\dagger a},\hat{C}_n^\dagger]|\Psi_0\rangle=\langle\Psi_0|[\hat{a^\dagger a},[\hat{H},\hat{C}_n^\dagger]]|\Psi_0\rangle\\
	-\omega_n\langle\Psi_0|[\hat{a^\dagger a},\hat{C}_n]|\Psi_0\rangle=\langle\Psi_0|[\hat{a^\dagger a},[\hat{H},\hat{C}_n]]|\Psi_0\rangle
\end{eqnarray}

Even more, to arrive at the well known matriceal RPA , we expand $C_n=\sum_{ph}(x_{ph}^{(n)}\hat{a}_p^\dagger\hat{a}_h-y_{ph}^{(n)}\hat{a}_h^\dagger\hat{a}_p)$ and obtain the standard eigenvalue problem

\[\label{RPAeq}\omega_n\left( \begin{array}{c}
X^{(n)} \\
Y^{(n)}\end{array} \right)= \left( \begin{array}{cc}
A & B\\
-B^* & -A^*\end{array} \right) \left( \begin{array}{c}
X^{(n)} \\
Y^{(n)}\end{array} \right)\]

Where $A_{ij,\mu\nu}=\langle i\mu|V|j\nu\rangle$  and $V_{ij,\mu\nu}=\langle \nu\mu|V|ji\rangle$. Solving the RPA eigenvalue problem poses a serious numerical difficulty in the computation of the $A\& B$ coefficients. A discussion on this matter will be given in the Chapter \ref{Chapter3}.  

\subsection{Linearized Vlasov}
\label{linearvlasov}

RPA/LRDFT are methods derived under the single particle picture of the system of $N$ particles with the purpose of determining the linear dynamics in terms of eigenmodes. If one looks at a optical spectra constructed from RPA calculations, it will see a set of lines positioned at the resulting exiting energies. And this is natural, since in the search for eigenmodes we have assumed pure oscillatory behaviour. Sometimes, as using the $\chi_s$ density response function, instead of delta functions, the resonances are fitted with some narrow lorentzians, given a phenomenological damping imposed by a fake imaginary part of the energy. 

Even though essentially a poorer approximation, Vlasov method is able to reproduce some damping effects even in the linear regime in the absence of any collisional integral. The phenomenon is known as Landau damping and its proof can be quite mathematically involved. The derivation will be just sketched further. We start from the Vlasov eq. \ref{VUU}, neglect the collision integral, and assume that in the stationary state the solution is $f_0(r,p,t)$. We take an initial perturbation $f(r,p,t)=f_0(r,p,t)+\eta f_1(r,p,t)$ from where the linearized Vlasov eq becomes:

\begin{equation}
	(\partial_t + v)h -\frac{1}{i\hbar}\int \frac{drdp'}{(2\pi\hbar)^3}e^{-i(p-p')r/\hbar}(V_1(x,t)|^{r+r'/2}_{r-r'/2}f_0(r,p',t)dp'=0
\end{equation}

Where $V_1(r,t)$ is only taken to be only the hartree potential created by the perturbation density $\rho_1(r,t)=\int dp f_1(r,p,t)$, so $\nabla^2V_1=-4\pi\rho_1$. Now we use the Laplace-Fourier transform of this equation and after some algebra the relation between the induced density and the effective field can be written as $\rho_1(k,\omega)=V_1(k,\omega)\Pi(k,\omega)$, where it has been introduced the density response function (similar with the same quantity defined in the frame of DFT).

\begin{equation}
	\label{Lindhard}
	\Pi(k,\omega)=\hbar\int \frac{dp}{(2\pi\hbar)^3}\frac{f_0(p)-f_0(p+\hbar\vec{k})}{\hbar\omega+p^2/2m-(p+\hbar\vec{k})^2/2m+i\delta}
\end{equation}

The relation \ref{Lindhard} is known as Lindhard polarization function and it is equivalent with the RPA polarization function. When one does the classical limit of this eq. (the long wavelength limit $\vec{k}\to 0$) obtains:

\begin{equation}
	\label{Lindhard2}
	\Pi(k,\omega)=-\hbar\int \frac{dv}{(2\pi\hbar)^3}\frac{\vec{k}\frac{\partial f_0(v)}{\partial v}}{\omega-\vec{k}\vec{v}+i\delta}
\end{equation}

The apparent problem with this functions is its continuity in the complex plane. Landau solve this issue and the interesting result is what now is known as \textit{Landau damping}. Just to sketch the idea, one should consider the dielectric function $\Pi$ defined in the complex plane and study its behaviour when the energy (in particular $\delta$) passes through the real axes. Writing :

\begin{equation}
	\tilde{\Pi}(k,\omega,\delta)=\left\{
	\begin{aligned}
		\Pi^R(k,\omega,\delta), \delta<0\\
		\mathcal{P}\int\frac{d\omega'}{2\pi}\frac{\Pi(k,\omega')}{\omega-\omega'}-i\pi\Pi(k,\omega), \delta=0\\
		\Pi^A(k,\omega,\delta)-2\pi i \Pi(k,\omega,\delta), \delta>0
	\end{aligned}
	\right.
\end{equation}	

Where $\Pi^R(k,\omega,\delta)$ and $\Pi^A(k,\omega,\delta)$ are the usual retarded and advanced functions. 

\section{Classical models}
\label{nanoplasmamodels}

When one goes to ultra intense laser fields, the quantum effects start to fade away. This should allow in principle the use of semi-classical or even classical method for the study of dynamics. On the other hand, there is a numerical problem given the fact that pure quantum methods (as DFT is) require the propagation of continuous fields which in turn require refined grids to be represented (also a good representation of the continuum spectrum). The electromagnetic problem itself is a large scale (compared to the electronic scale) problem therefore, the numerical refinement becomes a too hard task for a computer simulation. A last point is the fact that DFT/HF are theories designed for pure states and can be taken into account in the statistical cases using impure states (combination of non-integer occupation numbers). This means that at non zero temperature they describe canonical ensembles which by default can not resolve the microfluctuations arising in the strong field and which might be important for the dynamics.

For all these reasons, the cluster world entered in the domain of very strong laser fields with classical methods. In this sense are two approaches: the molecular dynamics which basically denies any quantum effects and solves a classical point-like problem both for electrons and ions and the nano-plasma model which has been introduce as a schematic model to investigate the macroscopic quantities of a cluster during its interaction with the external field.

\subsection{Molecular dynamics}
\label{moleculardynamics}

In the present, it is almost unanimous accepted that the single solution to solve the problem of cluster dynamics in very strong laser fields and consequently to represent the atomic ionization is the Molecular Dynamics techniques. Various groups have been used this approach and related works can be found in \cite{bauer2004small,belkacem2006coulomb,shao1996multi,ditmire1996interaction,ditmire1998explosion,lezius1998explosion,saalmann2003ionization,saalmann2006mechanisms} etc.

The essence of the theory is to consider that due to large amplitude of the electric field from the laser pulse, the electrons from the core levels of atoms are authomatically driven in the continuum spectrum, therefore, they become equivalent with the valence electrons, i.e. free particles. From this moment on, the MD methods should work taking into account only the Coulomb interaction in the electron-ion plasma formed. The interaction is cut in the near region since it can give unphysical electron-ion recombinations:

\begin{eqnarray}
	U(\mathbf{r,r'})=\frac{q_iq_j}{\sqrt{\mathbf{r^2+r'^2}+a^2}}
\end{eqnarray}

The equations of motion for electrons and nuclei are solved as classical Newton equations. Still, the inner ionizations pose a difficult problem since require large energies and very short time scales. For this reasons, they are not treated explicitly but taken into account using probabilistic rate equations for the rate ionizations.

For clusters which contain a few hundreds of electrons and nuclei, the MD problem is moderate from numerical point of view. When you pass to large clusters with thousands of atoms the problem is not directly tractable any more in the sense that the interaction becomes very expensive if it is evaluated with a direct method (scales with $n^2$). In turn one has to implement tree-based methods [] or to compute continuously the electromagnetic field using Particle-In-Cell techniques.

\subsection{Nano plasma model}
\label{nanoplasma}

The last level in the chain of approximation is the nano-plasma model developed in \cite{ditmire1996interaction}. Its main assumption is that the rapid ionization creates a quasihomogeneous plasma for large clusters at the level at $nm$. This spatial scale requirement is given by the fact that the Debye length $\sqrt{\epsilon_0k_BT/(e^2\rho)}\simeq 5 \mathring{A}$ should be less smaller than the spatial extension of the cluster.  In turn, this motivates the treatment of global quantities and not microscopic ones. 

The main quantities to be investigated are:  $N_j$ the number of ions with the charge $j$,  $N_e$ the number of inner-ionized electrons,  $E_{int}$ the internal energy of the cluster and  $R$ its radius. These must be complemented by the $W_j$, the ionization rate for ions in charge state $j$, $Q$ the total charge of the cluster,  $P_C=3Q^2e^2/(8\pi R^4)$ the coulomb pressure and  $P_H=n_ek_bT_e$ the hot electron gas pressure. With all these, the rate equations for nano-plasma model can be resumed at:

\begin{eqnarray}
	\frac{dN_j}{dt}=W_{j}N_{j-1}-W_{j+1}N_j\\
	\frac{dN_e}{dt}=\sum_jj\frac{dN_j}{dt}-\frac{dQ}{dt}\\
	\frac{d R}{dt}=\frac{P_C+P_H}{n_im_i}\frac{5}{R}
\end{eqnarray}

The total ionization rate can be decomposed in tunnelling, laser and thermal rate:

\begin{eqnarray}
	W_j=W_j^{tun}+W_j^{las}+W_j^{th}\\
	W_j^{tun}=C_{n,l,m}/\hbar I_p^{(j)}(\zeta)^{3n-m-3/2}e^{-\zeta}\\
	\zeta=\sqrt{\frac{32m_eI_p^{(j)3}}{9e^2\mathcal{E}_{int}^2\hbar^2}}\\
	W_j^{las}=n_e\langle\sigma_jv\rangle_{las}\\
	W_j^{th}=n_e\langle\sigma_jv\rangle_{th}
\end{eqnarray}

Of course, $n_e=3N_e/(4\pi R^3)$ is the concentration of electrons, $I_p{(j)}$ is the ionization potential for the charge state $j$ and $C_{n,l,m}$ is a factor which depends on the quantum numbers describing the electrons involved in ionization. The term involving the variation of total charge can be further decomposed in:

\begin{eqnarray}
	\frac{dQ}{dt}=\int_{v_esc}^\infty dv v^2\rho(v)\int_{\Sigma}d\vec{S}\vec{v}\\
	v_{esc}=\sqrt{\frac{2K_{esc}}{2m_e}}\\
	K_{esc}=\frac{(Q+1)e^2}{2R}\\
	\rho(v)=n_e\frac{1}{4\pi}(\frac{2k_BT}{\pi m_e})^{3/2}exp(-\frac{m_ev^2}{2k_BT_e})
\end{eqnarray} 

For the internal energy and its energetic coupling with the external laser field, we can write:

\begin{eqnarray}
	\frac{dE_{int}}{dt}=P_{abs}-\frac{2E_{int}}{R}\frac{dR}{dt}-\sum_jI_p^{(j)}\frac{dN_j}{dt}-P_{loss}\\
	P_{abs}=-\frac{V\epsilon_0}{2}\mathcal{E}_{int}^2Im[\varepsilon(\omega_{las})]
\end{eqnarray}

Where $\epsilon(\omega)=1-\omega_p^2/(\omega^2+i\nu\omega)$, the Mie frequency $\omega_p^2=4\pi n_ee^2/m_e$ and the Landau damping term $\nu=\nu_{ei}+Av/R$. 

Even though heavily simplified, the nano-plasma model captures almost all the important effects that take place during the dynamics. Therefore, it should not be used as a state of art/precision model, but as a starting point for the gross properties and trends to be expected.

\section{Final words about electron-nuclei interaction}

\subsection{Pseudopotentials}
\label{pseudopotential}

All the methods descried above that should be used for electrons assumed that the electron-nuclei interaction it is contained in the external potential. Also, in Section \ref{md}, the nuclei-nuclei interaction has been described generically with an effective potential. Now is the time to give more details about those and to present further simplifications that can be used. 

Naturally, the first assumption (the correct assumption) is that since the nuclei are considered classical, their interaction can be modelled with electromagnetic forces, i.e. the Lorentz force:$
\vec{F}_{en}=q\vec{E}[\rho,\vec{j}]+\vec{v}\times\vec{B}[\rho,\vec{j}]$. The electric and the magnetic field are subject to Maxwell's equations and depend on the total charge density and total current density in the system. So, during the classical dynamics of ions, the Maxwell's equation must be solved to compute the electromagnetic field created by the electrons. Fortunately, in clusters, due to their form and structure, the total angular momentum and the magnetic self consistent field are very small, and in this approximation, the only relevant aspect of the interaction is the electric, Coulombian one. This can be obtained as solution of the Poisson equation. 

The nuclei-nuclei interaction, must be treated as classical two-body electrostatic interaction and the force acting on the $I-th$ nuclei due to the other $N-1$ nuclei is :

$$F=\sum_{J\neq I}\frac{R_I-R_J}{|R_I-R_J|^2}$$

Only in the hydrodynamic models that treat the nuclei as a continuum distribution [] this force is also subject to Poisson eq. Conversely, the external potential created by the nuclei on the electrons is calculated as :

$$v_{Ext}^{nuclei}=\sum_I\frac{1}{|r-R_I|}$$

But it happens that in the atoms from a cluster the core electrons from the filled shells suffer almost no dynamics if the external laser field has moderate intensities. To gain some insight in why is so, let us just look at the energetic levels in the $Na$ and $C$ atoms which are around $40eV$ and $20eV$ for the $2p$ core electrons. Therefore, it would be a major simplification to get read of this core electrons considering them fixed with the nuclei. This gives us access to a treatment of ions instead of nuclei. 

But now, there is a question of how to treat the remaining electron-ion interaction. At first site, if the core electrons are considered bound in the point like region of nuclei, it should be straight forward to simply modify the $q$ charge of the an ion. But keeping in mind that even bound on small spatial regions, the core electrons still have a tail of density at larger distances due to their wave function, it becomes clear that the screening problem should be carefully treated. Anyhow, the resulting mean field potential created by nuclei-core electrons complex is called pseudopotential. 

The first simplified models that treat this problems are solid core model \cite{ashcroft1966electron} that uses the anzatz of solid sphere of charge for the core elctrons. A more refined model uses a density of core e in terms of superposition of two gaussians charge distributions. This gives us a pseudopotential of the form:

\begin{eqnarray}
	V_{PsP}(\mathbf{r})=\sum_Iv_{PsP}(|\mathbf{r-R_I}|)\\
	v_{PsP}(\mathbf{r})=\sum_jc_j\frac{Erf(|\mathbf{r}|/\sigma_j)}{|\mathbf{r}|}
\end{eqnarray}

There are other extensions to generalize this simplified approach. Generally speaking, a local pseudopotential as the one above, is good only for alkali metals when you need to treat only the valence which have zero angular momentum. For other types of elements, a pseudopotential must reflect the angular character of the wave function. The practical way of describing the pseudopotential in a non-local form:

\begin{eqnarray}
	\hat{V}_{PsP}(r)=V_{loc}(r)+\sum_LV_L(r)\hat{\Pi}_L\\
	\hat{\Pi}_L=\sum_M\int d^2\Omega'Y_{LM}(\Omega)Y_{LM}(\Omega')
\end{eqnarray}

Further "ultra-soft" pseudopotential can be constructed by the use of some parametrized solutions of KS equations for the valence electrons included in the general formula of :

\begin{eqnarray}
	\hat{V}_{PsP}=\sum_c|\psi_c\rangle(\varepsilon_v-\varepsilon_c)\langle\psi_c|
\end{eqnarray}

\subsection{Jellium model}
\label{jellium model}

Even further simplification can be worked out for the pseudopotential. In general, a PsP breaks any symmetry and imposes a full 3D treatment which can be quite expensive. On the other hand, there is a lot of experimental knowledge \cite{de1993physics} about the almost symmetric (spherically, or axial) clusters, therefore, an approximation should be used in this sense to impose the symmetry in terms of the PSP. The method is called "symmetry averaged ". 

Almost straightforward we define the spherically averaged pseudo-potential-SAPS (for spherical symmetric clusters) and the cylindrically averaged pseudo-potentials-CAPS (for azimuthal symmetric clusters):

\begin{eqnarray}
	U^{SAPS}_{back}(r)=\int d^2\Omega\sum_IV_{PsP}(r\vec{e}_{\Omega}-\vec{R}_I)\\
	U^{CAPS}_{back}(r,z)=\int d\phi\sum_IV_{PsP}(\{rcos(\phi),rsin(\phi),z\}-\vec{R}_I)
\end{eqnarray}

Now, as long as the effective pseudopotential is continuous and second-order derivable, it can be linked to a pseudo-density of charge by the trivial $\nabla^2 U_{back}\propto \rho_{ps}$. This idea is not just a mathematical thing, but has a good physical sense due to the fact that the PsP is an artefact of the complicated electric interaction and screening of the nuclei-core electrons complex. This charge can be also averaged in the same sense as the effective resulting potential to give access to so called jellium model. The model was used for the first time in the condensed matter physics (for metals) where it was considered that the electronic density and the ion density cancel eachother in an exact way. 

For clusters, metal clusters, it is customary to use the jellium model as some distribution of positive charge $\rho_{jel}$ from which the external potential for electrons can be computed as $\nabla^2V_{back}(r)=-4\pi\rho_{jel}$. In a more general sense, it can be putted in relation with a two particle PsP interaction in the form:

$$V_{back}=\int dr' V_{PsP}(r-r')\rho_{jel}(r')$$

For spherical, metal clusters a lot of success has been obtained with the step jellium model [] in which :

\begin{eqnarray}
	\rho_{jel}=\rho_0\Theta(|R_{jel(r)-r}|)\\
	R_{jel}=R_0(1+\sum_{lm}\alpha_{lm}Y_{lm}(\Omega))
\end{eqnarray}

Where, obvously, $Y_{lm}$ are spherical harmonics and $\alpha_{lm}=\alpha^*_{l,-m}$. In order to improve the results including diffusion[] effects at the surface of the cluster, soft jellium models[] have been proposed:

\begin{eqnarray}
	\rho_{jel}=\rho_0[1+exp(-\frac{R_{jel}(r)-r}{\sigma_{jel}})]^{-1}
\end{eqnarray}

The normalization condition must be satisfied by the jellium density: $\int dr \rho_{jel}(r)=Nw-q$. Where $q$ is the ionization of the cluster.

\subsection{Atomic forces}
\label{atomicforces}

The pseudopotential method is a good method to take into account the screening effects when neglecting core electrons and usually gives good results regarding the structure of a cluster. Still, one might be less interested in the fine details of electronic orbitals or simply wants to work in intense regimes where the details are not important. For such cases, empirical interactions can be introduced to model ion-ion repulsion/attractions.

Generically, this is the case of rare gase clusters. If such an inter-atomic force is considered, then it can be use to obtain both the numerical optimized geometry and in molecular dynamics simulations also. To mention just two such examples, we write down the Lennard-Jones and Gupta potentials:

\begin{eqnarray}
	V_{LJ}(r)=4\varepsilon[(\frac{\sigma}{r})^{12}-(\frac{\sigma}{r})^6]\\
	U_{G}(r_i)=A\sum_{j\neq i}v_{i,j}-\zeta\sum_i\sqrt{\rho_i}\\
	v_{i,j}=exp[-p(\frac{r_{ij}}{r_0}-1)]\\
	\rho_i=\sum_{j\neq i}exp[-2q(\frac{r_{ij}}{r_0}-1)]
\end{eqnarray}

% Chapter 1

\chapter{Numerical results} % Write in your own chapter title
\label{Chapter3}

Now that the main phenomena specific to laser cluster interaction have been presented in the second chapter and the main theoretical tools for study in the third, it is time to present the numerical methods that have to be used in connection with these theories to reproduce experimental data. 

It is my intention in this chapter to present personal results from as many different theories and regimes as possible. While almost each one of them, has been done long ago and can be found in different studies (to be cited), this chapter is a reflection of the personal work done in this direction.

As structure, I have chosen to present the results divided in regimes and quantities of interest and then, for every regime many methods have been applied. Where possible, the simulations have been performed on each of the four cases presented in the introduction: $Na$, $C$ , $Ar$ and $Xe$ clusters. 

All the simulations have been performed on different calculus machines with computing powers comparable with a standard personal computer. Depending on the type of simulation to be done and on the numerical methods to be applied, the FORTRAN programming language or the WOLFRAM MATHEMATICA system of computation were employed. As a general thumb rule, for methods that implied test particle methods, Fortran codes have been developed while for partial differential equations, Mathematica. All figures have been created with the astonishing features provided in the Wolfram system.

While times of computation could be specified individually, in general, there has been a concious sloppiness in the implementation in the sense that there have been used the minimum numerical refinement (in spatial grids, time steps, etc.) in order to gain computational speed. As a wide picture, semi-classical calculations as Thomas Fermi for ground states or calculation of static quantities are usually subject of a few seconds of CPU time, while at the other pole, TD-LDA or Vlasov methods had even days of computing CPU time.

\section{Numerical generalities}

Following chapter \ref{Chapter2}, almost every theory discussed in there ends with one or more partial (integro)differential equations (PDEs). Sometimes there are even systems of such equations coupled through some macroscopic quantity as the mean field potential. The truth is that this type of PDE's are the practical tool to investigate the reality1 in any sub field of physics, therefore, there is a long history of numerical, approaches more or less general with different levels of accuracy. 

Trying to form a clear picture on the logic and the relationships between different numerical methods that must be used in the laser-cluster physics, it would be nice to understand the problem from a more general, abstract point of view. Following basic single particle quantum mechanics, one can easily understand a PDE from the algebraic perspective. More precisely, any function $f(r,t)$ that we encounter in our eqs. is in fact an abstract vector $|f\rangle$ in a Hilbert space. Any operation that is done on this function in our eq. is nothing else than the action of an operator on the function associated with that Hilbert space. 

Being abstract objects, we don't have any understanding on the vectors, unless we represent them in a natural space for us. The most common space is the coordinate space, described by the infinite continuum basis $\{|r\rangle\}$. The same is done with the operators and eqs like $\hat{D}|f\rangle=|u\rangle$ become in the coordinate representation $\hat{D}(r)f(r)=u(r)$ where $f(r)$ is to be understood as the scalar product between our state $|f\rangle$ and some element $|r\rangle$ from the basis. Operators like momentum, are also local in $r$ space but they become differential operators (gradient, laplacian, etc.). The choice of coordinate representation is done usually because is a natural link with what we \emph{see} in real life, but otherwise, is not necessary. Other representations as the momentum representation can be useful since it can diagonalize gradient or laplacian operators. 

In general, having the set called orthonormal basis $|e_i\rangle$ in which we make our representation of the reality, one can speak about a "box of interest" $\mathcal{B}$. This is the subset $|e_j^*\rangle$ from the entire basis on which we expect our quantities to have significant values (also it is motivated by the fact that we cannot work in practice with infinite number of elements). In terms of $|r\rangle$ basis, the box can be visualized like an actual box, cube for example, in 3D. Also, we can speak  about the boundary of the box $\partial\mathcal{B}$ which is a subset of $\mathcal{B}$ with some specific property. For example: if $|e_i^*\rangle$ are vectors associated with angular variables (spherical harmonics) the boundary of the box is represented by the vector with the higher and the lower angular momentum. Another intuitive example is the actual box in the $\mathbf{R}^3$ space which can be the surface of a sphere, or the surface of a cube, etc. depending on the actual coordinate system that has been employed.

\subsection{Numerical representations}
\label{numericalrepr}

Now, the most natural numerical method, namely the Finite Difference Method (FDM) can be understood as a representation of a PDE in the coordinate space taking just "some" position vectors from a box. In terms of functions is to be understood as an expansion in delta functions. When the chosen subset of elements has certain regularity, one can speak about "structured grids". For example, a standard 3D grid for finite difference consists from a box $\mathcal{B}=[-L,L]\times[-L,L]\times[-L,L]$ where the chopped basis of dimension $n^3$ is the set of $\{(x_i,y_j,z_k)\}, i,j,k=1,n$ with $x_i=-L+2(i-1)L/n$ and the state $|f\rangle$ becomes fully described by the set of coefficients $f_{i,j,k}$. In the frame of this representation, operators like gradient or laplacian can be approximated with stencil configurations:

$$\partial_x f(r) \approx \frac{f_{i+1,j,k}-f_{i-1,j,k}}{hx}$$
$$\partial_{xx} f(r) \approx \frac{f_{i+1,j,k}+f_{i-1,j,k}-2f_{i,j,k}}{hx^2}$$

A more general representation is used in HF or in DFT methods. Even though there are many approaches that use real space with Eulerian grids more sophisticated than the one described above, there is another way to solve equations like Kohn-Sham. Instead of choosing the $\{|r\rangle\}$ basis, one uses another finite basis $|e_j^*\rangle$. Usually this basis has nice properties with analytical behaviour regarding how different operators act on it and more important is chosen in such a way to meet the criteria of symmetry and asymptotic values for the specific system. I will enter later in more details about how this method is applied on KS eqs., but essentially, the final result is numerically the same with FDM.

Just as a remainder, the Fourier/Laplace/Fourier-Laplace/Hermite/etc. transforms are all spectral methods that can and will be used since can make a lot more easier the life of anyone. 

All this type of spectral methods can be regarded as Eulerian methods because, even though may not be explicit, they have in the background a "fixed" grid representation. But there is another type of spectral-like method which is more related with the Lagrangian methods or particle based methods. Essentially, it uses a basis with some unfixed parameter which is free to change under time evolution. To be more intuitive, let us imagine that we have $f(r)$ and we choose a basis which can be also $r$ represented as a function $W(r-r_j;h)$ where $h$ is a parameter to control how narrow is function $W$. Based on the identity $f(r)=\int dr' f(r')\delta(r-r')$ and assuming that $\int W(r;h)dr=1$ and $\lim\limits_{h\to 0}W(r;h)=\delta(r)$, one can approximate:

$$f(r)\approx \sum_i m_i W(r-r_i;h)$$

This approximation is heavily used in what is called Smoothed Particle Hydrodynamics (SPH) \cite{monaghan1992smoothed} where usually the kernel $W(r;h)\propto exp(-r^2/h^2)$. The same exponential is used in test particle methods (TPD) \cite{rostoker1964test} employed in the Vlasov equation. The main advantage of this method is that the due to the structure of the PDE's involved (and their linearity), one can show that the entire time evolution of the quantity $f(r,t)$ can be reduced to the classical motion of the particles represented by the $W_i$. The single aspect is that this particles move in a mean field acceleration field created by themselves accordingly with the type of interaction present in the system.

\subsection{Types of equations}
\label{typesofeq}

\textbf{\emph{Poisson equation}} is the most pregnant appearance in all the theoretical methods discussed since it models one of the basic type of interaction, the Coulombian interaction which usually supersedes any other one (like xc). In essence, the equation can be written $\nabla^2\phi = 4\pi \rho$. Working under FDM, one can write $\phi$ and $\rho$ as vectors of dimension $n^d$ (where $d$ is the dimensionality of the problem: 1D, 2D or 3D) and the laplacian as a $n^d\times n^d$ matrix. Further, you get a matrix equation which should be solved: a) by inversion of the laplacian matrix, which even though is sparse is the worst approach ever; b) by iterative methods as Gauss-Seidel, Jacobi, Succesive-Over-Relaxation (SOR), acceptable methods (for reviews and extensive discussions see \cite{saad2003iterative,kelley1995iterative}). Nowadays, one can find out there an entire zoo of numerical libraries that solve the Poisson eq. so you don't have to do it by yourself. Working in full 3D, it becomes very expensive to use FDM. There are other approaches like finite element \cite{zienkiewicz1977finite}, etc[], but the one that I have used during the simulations is the Fourier Transform. Taking the Poisson equation and performing a FT, one obtains a natural diagonalized solution and actually the solution can be written formally as $\phi(r)=\mathcal{F}^{-1}[\mathcal{F}[\rho(r)]/k^2]$. The boundary conditions (BC) are essential in the sense that if you do not impose the right BC, you may end up with incredibly wrong results. In general, the boundary of the box is far enough to consider that the entire charge is contained in the box, therefore, a multipole expansion is used to compute the BCs.

\textbf{\emph{Eigenvalue Problems}} appear natural in HF and DFT as stationary Schrodinger equations: $\lambda_i|f_i\rangle=\hat{H}|f_i\rangle$. Beyond any particular representation, let us consider that we take the spectral box as $\mathcal{B}=\{|e_j\rangle\}$ which is large, finite and \emph{not necessary} orthonormal. We can expand more like an approximation and not as a consequence of the nuclear spectral theorem the eigenvectors as $|f_i\rangle=\sum_jc_i^j|e_j\rangle$, where $c_i^j=\langle e_j|f_i\rangle$. Now let us introduce the following notations:$H_{i,j}=\langle e_i|\hat{H}|e_j\rangle$, $S_{i,j}=\langle e_i|e_j\rangle$, $C_{i}=\{c_i^j,j=1,M\}$ the latter being the vector of coefficients for the $i-th$ eigenvector. Now the problem can be rewritten as a matrix eigenvalue problem $\lambda_iSC_i=HC_i$. In literature, this type of equations can be found as Roothaan-Hall equations \cite{roothaan1951new,hall1951molecular}. Again, for this problem, many standard numerical methods can be used \cite{saad1992numerical}.

\textbf{\emph{Time-dependent equations}} arise naturally when you go from ground state towards dynamic regimes and you have to solve equations like time dependent KS. Their general form can be expressed in abstract as $i\partial_t|f_j\rangle=\hat{H}|f_j\rangle$. Being basically a Cauchy problem on time, this eqs. have the formal solution:

$$|f_j(t)\rangle=exp[-i/\hbar\int d\tau \hat{H}(\tau)]|f_j(0)\rangle$$

For numerical purposes, one needs to take a discrete sequence of time moments $t_i$ separated by a step $\delta t$ (which is not necessary constant) and than the integral from the exponential  can approximated as $\int_{t_k}^{t_{k+1}} d\tau \hat{H}(\tau)]\approx \delta t\hat{H}(t_{k+1/2})$. Than, keeping in mind that we can expand the hamiltonian operator in kinetic and potential terms $\hat{H}=\hat{T}+\hat{V}$, one can use what is called the time-split method :

$$|f_j(t+\delta t)\rangle=exp[-\frac{i\delta t}{2\hbar}\hat{V}]exp[+\frac{i\delta t}{\hbar}\hat{T}]exp[-\frac{i\delta t}{2\hbar}\hat{V}]|f_j(0)\rangle$$

First step is to construct the potential $V(t)$ and apply the $exp[-i\delta t\hat{V}/2\hbar]$ operator on $|f(t)\rangle$ which is easy to the in the real space if the potential is local. Then take advantage of the fact that in the momentum space, $\hat{T}$ becomes local $\hat{T}(k)=-k^2$ and take the FT of the new result, multiply it with $exp[-i\delta tk^2\hbar]$, perform inverse FT and multiply again with $exp[-i\delta t\hat{V}/2\hbar]$, this time, the potential being computed with the intermediar solution obtained after the inverse FT. This method is semi-implicit and second order accurate, the main source of error being the fact that $[T,V]\neq 0$. 

\textbf{\emph{Boundary conditions}}. Solving a PDE implies the knowledge of the boundary conditions on the coordinate space. Usually we work with isolated clusters so one could expect a Dirichlet null boundary condition for almost all quantities. This is not the case. Even solving the Poisson equation in large boxes and neutral clusters, there are B.C.s that must be employed due to the fact that there can be non-zero multipolar moments of charge inside the box. Therefore, when one computes the electrostatic potential must keep in mind that at the frontier of the box Dirichlet conditions for the potential must be imposed using the multipolar expansion of Coulomb potential outside a charge distribution

$$\phi(r)|_{r\in\partial\mathcal{B}}=\frac{1}{4\pi\varepsilon_0}\sum_{l=0}^\infty\frac{4\pi}{2l+1}\frac{1}{r^{l+1}}\sum_{m=-l}^l(-1)^mY_l^{-m}(\hat{r})Q_{lm}$$
$$Q_{lm}=\int_{\mathcal{B}}drd\Omega r^{l+2}\rho(r)Y_{lm}(\Omega)$$

In practice, just the first few moments are taken into account and in the present thesis just the dipole and quadrupole moments have been used. In practice, introducing higher moments is not an expensive procedure, but is redundant since higher multipolar moments are neglijable, the most important being the dipolar one in laser fields.

Regarding the boundary conditions for lagrangian methods as test particle method (for Vlasov) or SPH (for QHM) one can force spurios particles which manage to reach $\partial\mathcal{B}$ to be reflected inside $\mathcal{B}$. This should be done in order to keep the normalization of the total charge and should be used just when the laser regime is weak enough to be sure that there is no ionization in the system. When the laser field is strong, electrons and even ions are free to leave the cluster therefore, particles which pass outside $\partial\mathcal{B}$ should be discarded from calculations. The same is true for quantum methods as TDDFT in which Crank Nicholson method forces the normalization and the reflection of electron by  construction. On the dark side, in order to achieve ionization of the cluster, a mask function [ C.-A. Ullrich, Time-dependent density-functional approach to atoms in strong laser pulses, Ph.D. Thesis, J.-M.
UniversitaKt WuKrzburg, 1995] is applied at the frontier of the box. More precisely, one has to propagate in time the wavefunctions by the above described method and than, in a shell region on $\partial\mathcal{B}$ the following procedure is applied:

$$\psi\to cos^{1/4}(\pi\frac{|x_{max}-x|}{\Delta})cos^{1/4}(\pi\frac{|y_{max}-y|}{\Delta})cos^{1/4}(\pi\frac{|z_{max}-z|}{\Delta})\psi$$

\textbf{\emph{Optimization problem}}. It is clear that in the time dependent regime, one needs to solve simultaneously with the equations for electron system, the ones for nuclei (ions) \ref{clasion}. About how to propagate correctly a particle in time I shall speak bellow, but there is a problem about how to find its stationary positions.

Even though there are various methods to tackle large sparse eigenvalue problems, it happens that they can be quite slow in 3D situations. Therefore, it would be desirable to find another method for the stationary solutions of KS eqs. One solution which, even though not particularly fast in convergence, is easy to implement and requires to store small amounts of data, is the Imaginary Time method \cite{davies1980application}. The first step is to take some guessed KS orbitals, and than iterate until convergence the following operation:

$$\psi_i^{n+1}=\mathcal{O}[\psi_i^{n}-\delta (H^{n}-\langle H^{n}\rangle_i)\psi_i^{n}]$$

In here, $\delta$ is a small, damping parameter which must be $\delta<1/|E_{max}|$ in the sense that must be smaller that the inverse of largest eigenvalue and $\langle H^{n}\rangle_i$ is in fact the energy of the $i-th$ orbital computed with the $n-th$ hamiltonian. The operator $\mathcal{O}$ is a orthonormalization operator.

During those iterations, the optimization geometry for the ions can be handled following a similar scheme. Together with the initial guessed orbitals, one guesses some positions for ions $\{R_I^0\}$.  After the first iteration has been performed one starts some kind of Monte-Carlo sampling applying the technique of simulated annealing \cite{van1987simulated}. More exactly, All the positions of ions are changed by a small variation $\delta R_I^0$ for each such configuration the total energy of the \emph{electrons+ions} system is computed. After a reasonable number of such random samplings, one chooses the configuration $\{R_I^1=R_I^0+\delta R_I^0\}$ which gave the lowest total energy. Than the second iteration is done for electrons, and loop is repeated until convergence.  

This process of simulated annealing can be applied also for rare gas clusters in which the stationary configuration is found using the Lennard-Josen interatomic potential discussed in Sect. \ref{atomicforces}.

\textbf{\emph{Classical single particle equations of motion}} appear in the frame of SPH for QHMs or TestParticle methods for Vlasov - like equations and can be generically written as $\dot{r}_i=v_i$ and $\dot{v}_i=a_i[\{r_i\}]$. The acceleration is usually computed from a grid based approach of the Coulombian interaction (solving Poisson equation and interpolating the resulting force). The issue that has to be here discussed is how to solve appropriately the generic eqns. of motion. Being a system of apparently independent ODEs, one could at first sight employ a Runge-Kutta (RK) method of arbitrary order (depending on the necessary accuracy). Taking a discrete time sequence $\{t_i\}$, I have used the more employed cousin of RK in molecular dynamics, namely the Verlet algorithm (leap-frog):

\begin{eqnarray}\label{Verlet}
	r(t+\delta t)=r(t)+v(t)\delta t+a(t)/2\delta t^2\\
	v(t+\delta t)=v(t)+\delta t\frac{a(t)+a(t+\delta t)}{2}
\end{eqnarray}

Basically, one has to propagate the positions accordingly with the first eq. from \ref{Verlet} than compute the macroscopic quantities from this positions, from that the new accelaration $a(t+\delta t)$ and than propagate the velocity. Global errors in this scheme are of $\mathcal{O}(\delta t^2)$ which usually is enough for most applications of ionic dynamics or test particle dynamics. 

As we shall see in more detail later, one of the technique of simulating binary collisions in complicated systems at finite temperatures (this being a signature of intense lasers regimes) is to induce a random component in the velocity. The motivation is that treating large systems requires high numbers of test particles. In turn, if one wants to use a VUU like scheme, the collision operator becomes quite time consuming, and the improvements might not be balanced by the high costs. On the other hand, at finite temperatures, the Pauli principle for electrons (which is derived under the Fermi-Dirac statistics) is relaxed. A last reason for the use of such forces is that when you want to compute the ground state in the Vlasov theory  some initial guess on the ground state is done and than the system is relaxed towards a numerical stable state. This procedure can be highly accelerated with a small random collision which dissipates any possible long term collective waves that might appear. Therefore, it is plausible to use instead of complicated collision operators, simple random forces drawn from some probability distributions that act on our virtual particles. Doing this numerically is easy: just add to the acceleration a random component. But there are two questions regarding the magnitude and the probability distribution functions from which they should be drown this matters will be discussed later.

\textbf{\emph{Iterations}} As mentioned many times through this thesis, self-consistent problems as KS eqns. are to be treated either through an imaginary time method or through an iterative scheme in which any solution gives in turn an effective KS potential which is used to compute the next generation of solutions. This map has nothing special to be spoken about, just that it usually doesn't work. The reason is that a stupid initialization of the density can put the iterative method on a divergent track, therefore, some physical state-of-art guesses are needed for its success. On the other hand, the iterative method can be generalized and protected of any divergence. This is done using the $n-th$ solution $\rho^n$ and the solution generated with it $\rho^*$ to construct $\rho^{n+1}$
using a control parameter : $\rho^{n+1}=\alpha\rho^{n}+(1-\alpha)\rho^{*}$. Where $\alpha\in (0,1)$.

\section{Ground state}

Even though the main purpose of the present thesis is to deal with laser interactions, it is mandatory to have knowledge about the ground state of the cluster, usually as $t=0$ state. There are indeed methods regarding the ultra-intense laser interaction where the initial state is not taken necessary to be GS, or it does not even matter too much the specific details, but in general, for smaller systems it can be crucial. For these reasons we present in this section results regarding the ground state of some clusters computed with different methods.

The most simplified method which works with metal and carbon clusters is the jellium model defined in Sect. \ref{jellium model} in which the ions (nuclei) are fixed and we only focus on the electron configuration. Since I will use this model further in different results including linear optical response, more details should be given. First of all, there are 2 decades old studies that show how well the jellium model works at least for metal clusters \cite{brack1993physics}. In their case, the simplest jellium is the one with spherical symmetry, where the following parametrization has been used for the jellium density $\rho_{jel}(r)=\rho_0 (1+Exp[(r-r_0)/\sigma ])^{-1}$, where $r_0=r_sZ^{1/3}$ is the approximate radius of the cluster and due to the normalization condition, $\rho_0\approx 3Z/4\pi r_0^3$, for very low $\sigma$. Usually, in sodium clusters $\sigma\approx 0.2r_0/Z^{1/3}$. For fullerenes, due to their spherical symmetry a spherical shell has been used \cite{madjet2008photoionization} to model the density of jellium. Roughly speaking $\rho_{jel}^{fullerene}\propto \Theta[(r-r_1)(r_2-r)]$.

For fullerenes the spherical symmetry is an approximation of the $I_h$ symmetry of the geometrical structure. In turn, for sodium clusters this is associated only with closed shell cases where the number of atoms follows one of the magic numbers $2,8,10,18,...$. In general cases, the better symmetry to employ is the azimuthal one where an anisotropic parameter $q$ is introduced as a measure of the axial deformation of the cluster and the jellium density can be written as $\rho_{jel}(r,z)=\rho_0 (1+Exp[\sqrt{(r-r_0)+(z-z_0)}/\sigma ])^{-1}$. This is the so called deformed-jellium model which again has been used extensively in cluster physics \cite{lyalin2000hartree,brack1993physics,hirschmann1994spheroidally,yannouleas1995electronic}

Even though atomic units are the natural choice when doing atomic and molecular physics, I have chosen to scale all the equations in terms of the characteristic radius of the cluster $r_0$ spatially, with a computational time $t_0$ temporally and the energetic quantities with an $e^2/4\pi\varepsilon_0r_0$. 

Another method to \emph{cheat} is to take, lets say for $C_{60}$ fullerene cluster the ionic positions from other studies, given the fact that they have been intensively studied and have specific geometries. Having the positions of the GS specified, one keeps the ions fixed and again, only the electrons are studied.

The full, fair treatment involves, as described in Sect. \ref{typesofeq}, an optimization procedure done in parallel for ions and for electrons. Using pseudopotentials to simulate the electron-ion interaction makes the optimization of the ionic geometry a quite time-consuming part of the simulation. The best numerical approach that has been found by the author was to split the problem in two parts and the pseudopotential in a local and a non-local part $V_{PsP}(\mathbf{r})=V_{loc}(\mathbf{r})+\hat{V}_{nloc}(r)$. The splitting between ionic and eelctronic problem has been already discussed. The splitting of the PsP puts two different problems.

If you have the electron density fixed, than the local part of the PsP potential can be computed very elegant with the Fast Fourier Transform, using the Convolution Theorem, since :

$$V_{loc}(\mathbf{r})=\int\rho(\mathbf{r'})v_{psp}(\mathbf{r-r'})d\mathbf{r}'$$ 

Which is nothing else than the convolution of the local psp with the density. Therefore I have used $V_{loc}(\mathbf{r})=\mathcal{F}^{-1}[\mathcal{F}[\rho(\mathbf{r})]\mathcal{F}[v_{psp}(\mathbf{r})]]$. This sequence is particularly easy to implement when working in the real space on a structured grid. 

\subsection{Ionic structure}

\begin{figure}[h]
	\centering
	\includegraphics[width=1.\linewidth]{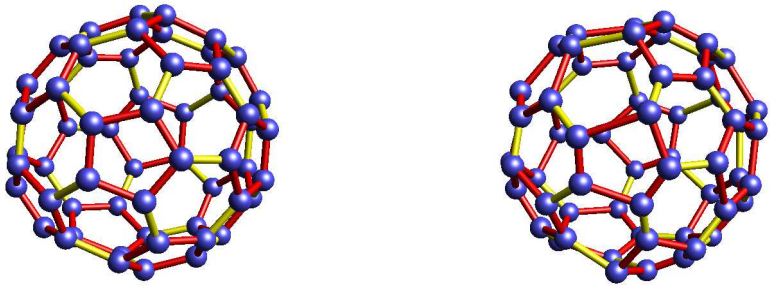}
	\caption{Comparison between the true ionic configuration of $C_{60}$ (left) and the one obtained after geometry optimization and solving KS equations. }
	\label{fig:fullerene}
\end{figure}

Let us now go to the specific case of fullerene. As already said, the positions of the $60$ carbon ions obey the $I_h$ symmetry and they are represented in Fig. \ref{fig:fullerene} on the left side. On the right is represented the ionic configuration obtained with LDA method (Gunnarson-Lundqvuist xc potential) and taking into account all-electron. It can be seen that the present results have same small deviation from the true configuration but the results are under $5\%$ (when comparing the norm of the position vectors). The difference comes from the insufficient
refinement of the spatial grid as well from the errors induced by the parametrizations used for xc potential. Similar calculations can be found in \cite{ohno1996stability,pavlyukh2010kohn}.

\begin{figure}[!h]
	\centering
	\includegraphics[width=0.7\linewidth]{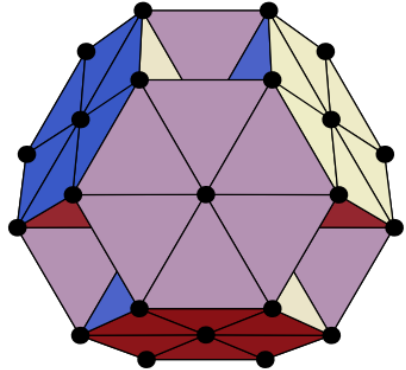}
	\caption{Ionic structure of $Xe_{38}$ obtained with simulated annealing method and Lennard-Jonnes atomic potential}
	\label{fig:Xe}
\end{figure}

Some results regarding the ionic structure of $Xe$ clusters can be found in \cite{doye2002entropic,wales1997global}. In here we should discuss only some case of Lennard-Jones cluster which was obtained with the simulated -annealing methods coupled with Monte Carlo sampling of the configuration space. The bright side of these types of clusters is that the electronic problem is discarded since the full treatment with DFT with be very difficult to handle. On the not so bright side, the LJ potentials are known to have a number of local minima increasing exponentially with the number of constituents, therefore, the computational time increases dramatically with the size of the cluster. For those reasons, I shall present in Fig. \ref{fig:Xe} the ionic structure obtained after the geometry optimization was performed on a random $Xe_{38}$ cluster.

\subsection{Electron densities}

Going further from the ground state ionic configuration, the interest moves to the electronic densities in those stable states. While many calculations have been performed with ionic structure, I choose in this section to plot merely profiles of electron density obtained in the frame of jellium model due to its possible radial symmetry. 

\begin{figure}[!h]
	\centering
	\includegraphics[width=0.8\linewidth]{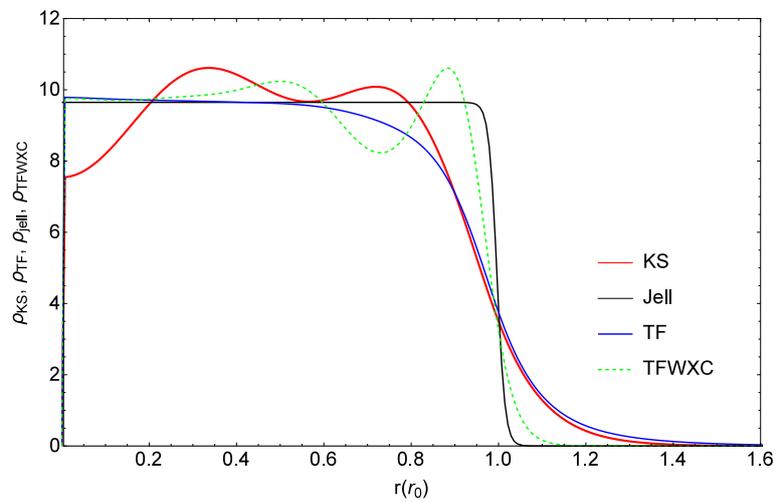}
	\caption{Ground state of $Na_{40}$ cluster, obtained with TF (black solid line) and DFT (dashed line); in dot-dashed is plotted the radial profile of the jellium density}
	\label{fig:sodium40grs}
\end{figure}

As methods, I have used: Orbital-Free DFT (Thomas Fermi +extensions) and DFT-LDA. For other methods as Vlasov, the ground is consistent with TF solution. Some things worth mentioned about the solution of TF equation. In its simple form can be obtained in an iterative manner solving \ref{abstractdft} in which the $v(\mathbf{r})$ potential contains the jellium electrostatic potential, the xc potential and the self consistent electrostatic potential from the e-e interaction. Still, if one wants to use extended TF methods as the one described in \ref{semiclass}, the best numerical approach is to embed the Euler-Lagrange eq. \ref{abstractdft} in a Schrodinger like equation, similarly with eq. \ref{NLS}. This time the equation is solved also by iteration, but it can be subject to the imaginary-time method which is faster than the standard approach which is known to impose some problems regarding the normalization and the positiveness of density. 

As a numerical aspect, it has been observed that the xc potential is crucial in the calculation. Neglecting xc gives too high energies for orbitals and incorect density profiles. The detail of which type of approximation is used is less important, but the presence is mandatory.

\begin{figure}
	\centering
	\includegraphics[width=0.9\linewidth]{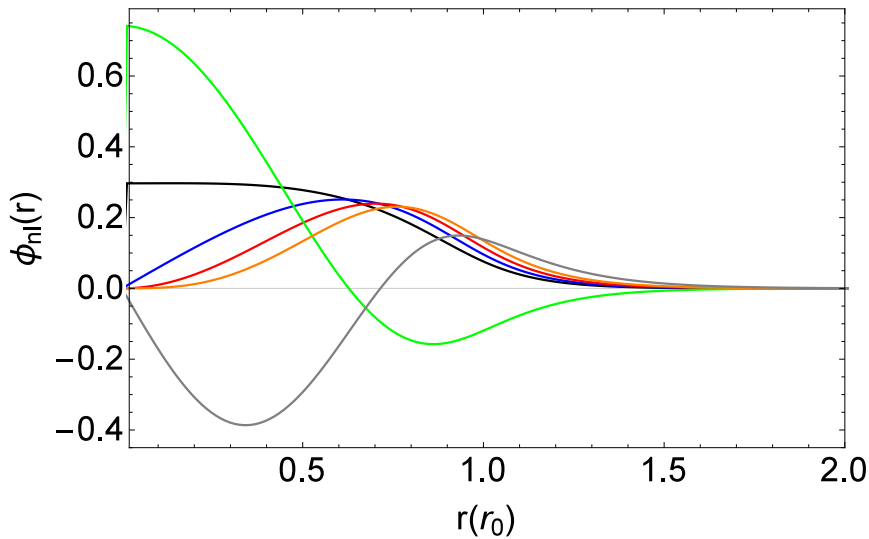}
	\caption{Radial profiles of the occupied KS orbitals in the ground state of $Na_{40}$ cluster}
	\label{fig:orbitals}
\end{figure}

\begin{figure}[h!]
	\centering
	\includegraphics[width=0.8\linewidth]{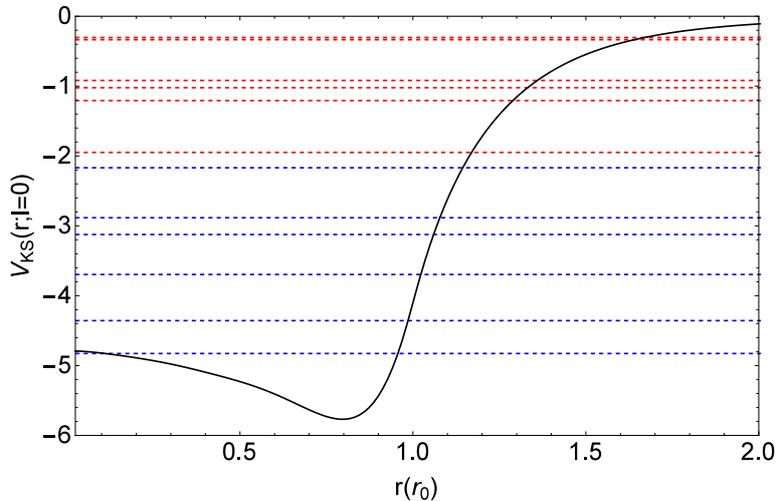}
	\caption{In black is plotted the $V_{KS}$ potential for $Na_{40}$ corresponding to $l=0$ orbital expressed in $eV$. Horizontal blue lines are the energies of the occupied orbitals and with red the bounded unoccupied orbitals}
	\label{fig:KSpot}
\end{figure}

In Fig. \ref{fig:sodium40grs} is plotted the electronic density vs the jellium density obtained with TF (blue line), DFT-LDA (red line) and TF+WXC (green line). This particular case was chosen due to its close to spherical symmetry. As one can see... In literature there are many works that have solved DFT-LDA problem in the jellium frame for sodium clusters. As I cannot cite them all, some references should be given \cite{rubio1990static,chou1984total,beck1984self,ekardt1984dynamical,calaminici1999static,calvayrac1997spectral}. Consequently, one can find an also impressive number of Thomas-Fermi (or related) calculations on metal clusters \cite{blaise1997extended,rusek2000cluster,fennel2004ionization,kresin1988electronic,serra1989static,palade2014general}.

To have an image about the single particle components of the electronic structure, in Fig. \ref{fig:orbitals} there are plotted the radial profiles of the single particle KS orbitals in the ground state, only for the occupied levels. Further, the energetic levels are plotted explicitly in Fig. \ref{fig:KSpot} in respect with the KS effective potential. In blue, the occupied levels and in red the bounded unoccupied ones.

\begin{figure}[h]
	\centering
	\includegraphics[width=0.9\linewidth]{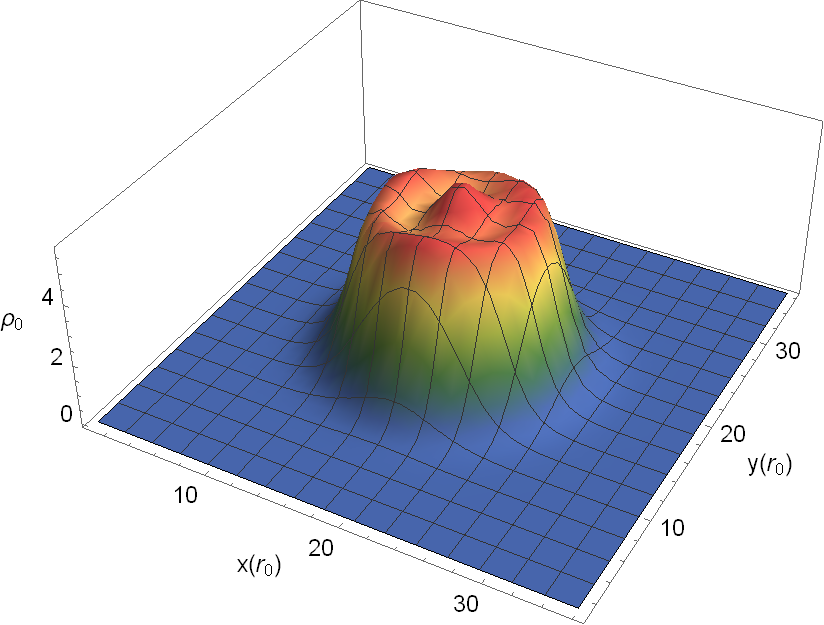}
	\caption{Electronic density in section at $z=0$ obtained with DFT-LDA and local pseudopotentials in $Na_{20}$}
	\label{fig:3ddft-Copy}
\end{figure}

Finally, about the $Na_{20}$ clusters, in Fig. \ref{fig:3ddft-Copy} I have plotted a section at $z=0$ from the electronic density computed on a 3D grid in the presence of jellium parametrization with a small deformation. 

\begin{figure}[!h]
	\centering
	\includegraphics[width=0.8\linewidth]{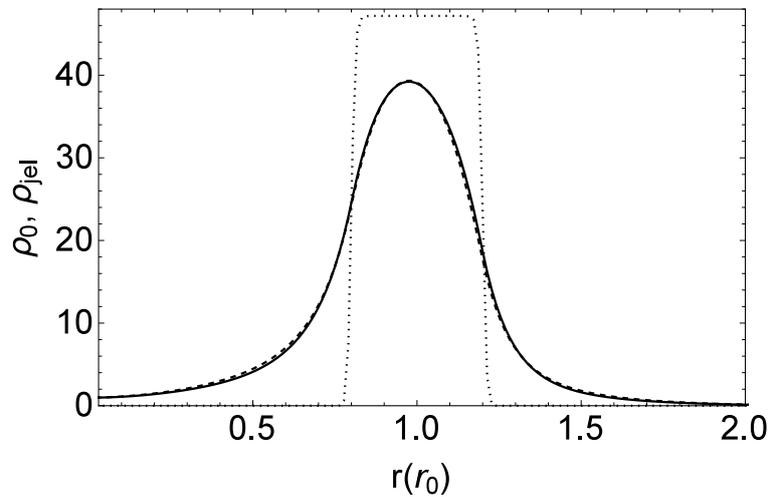}
	\caption{Ground state of $C_{60}$ cluster in the frame of jellium model}
	\label{fig:C60grst}
\end{figure}

\begin{figure}[!h]
	\centering
	\includegraphics[width=0.8\linewidth]{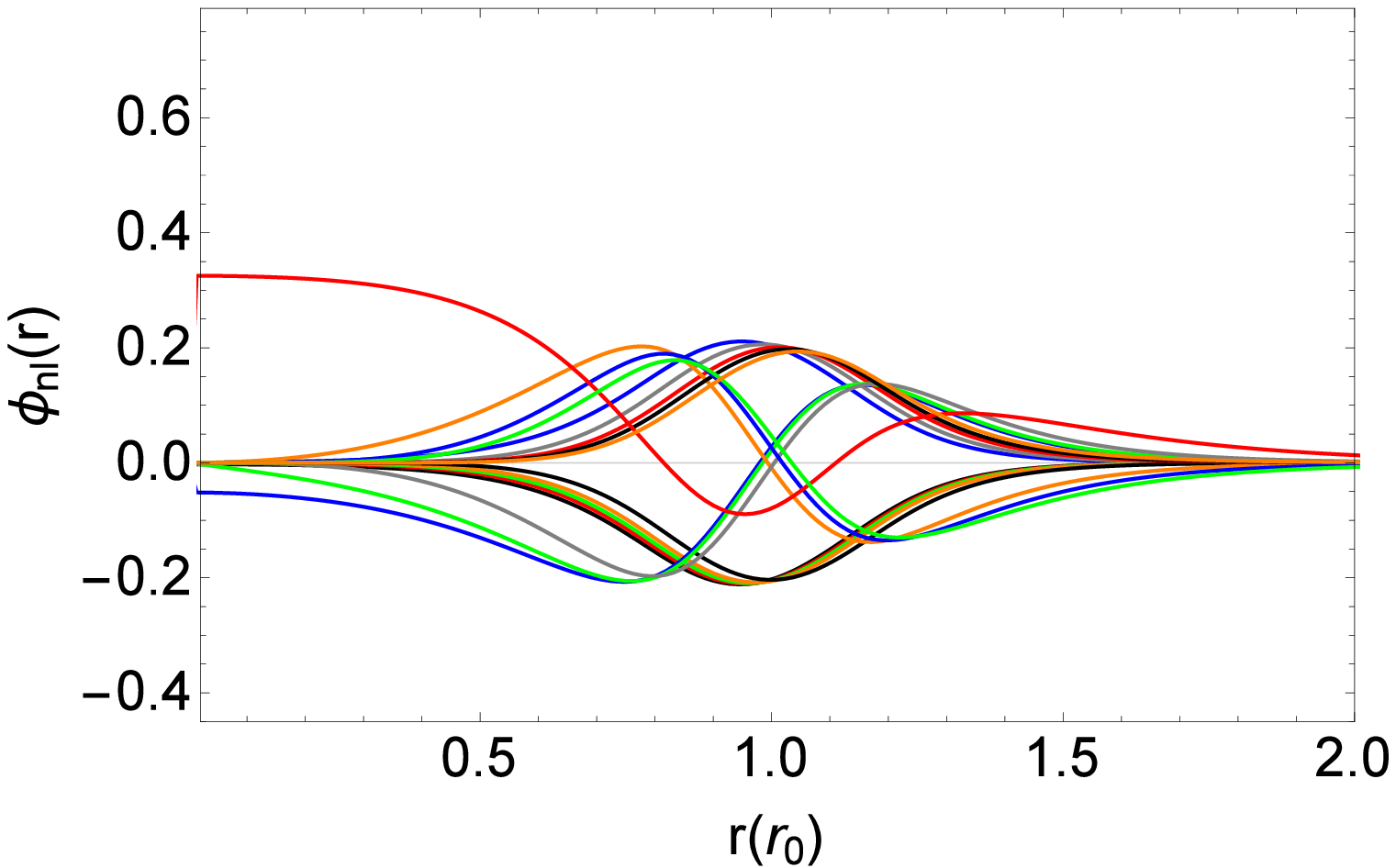}
	\caption{Radial profiles of the KS orbitals of the occupied states in the $C60$ fullerene. Results obtained in the frame of jellium model}
	\label{fig:C60orbital}
\end{figure}

\begin{figure}[h]
	\centering
	\includegraphics[width=0.8\linewidth]{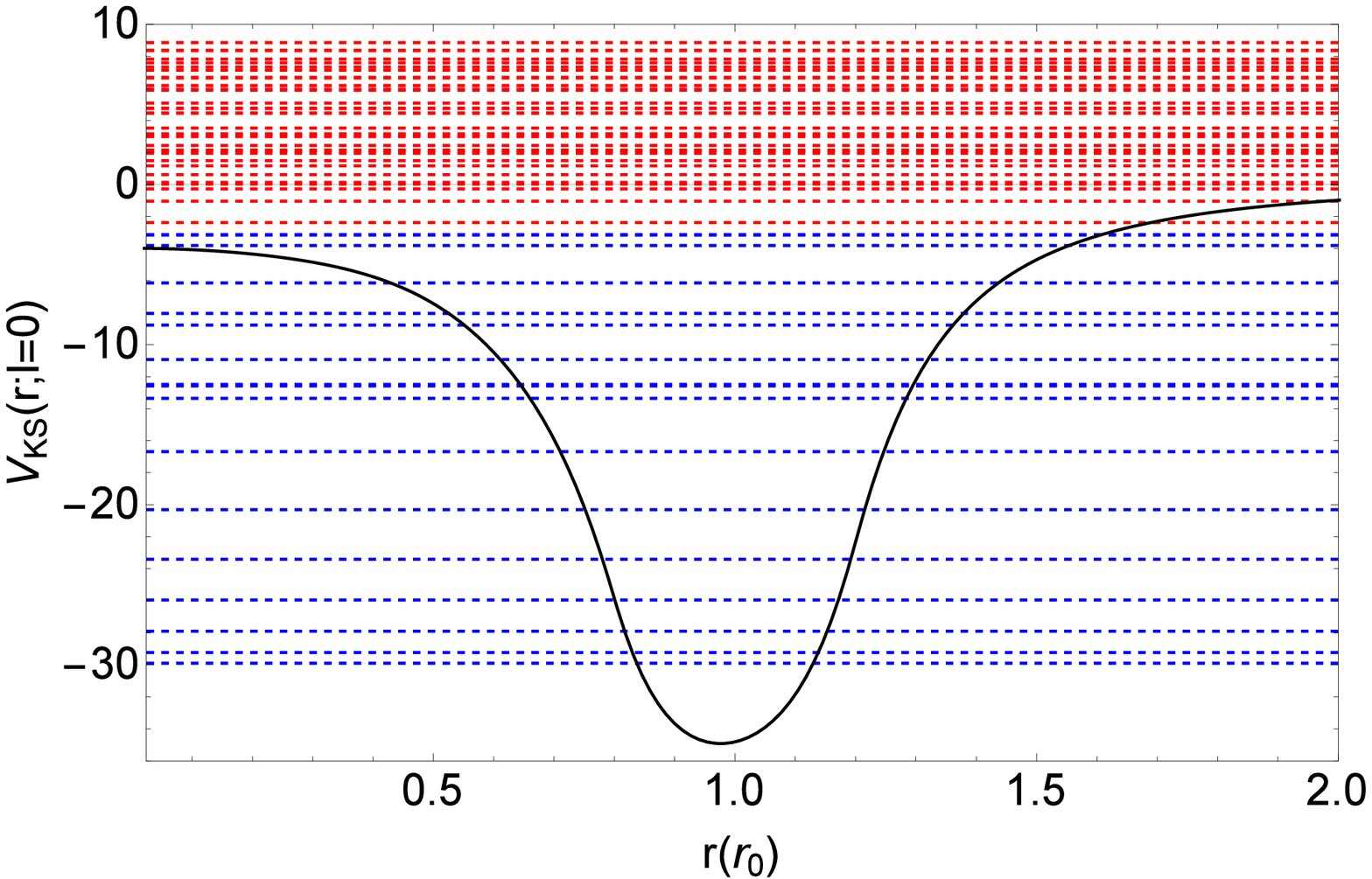}
	\caption{In black is plotted the $V_{KS}$ potential for $C_{60}$ corresponding to $l=0$ orbital expressed in $eV$. Horizontal blue lines are the energies of the occupied orbitals and with red the bounded unoccupied orbitals}
	\label{fig:KSpotc60}
\end{figure}

Similar results can be drawn for the $C_{60}$ fullerene using the jellium model. More details about the problems and solutions for the jellium-fullerene coupling can be found in \cite{yannouleas1994stabilized,verkhovtsev2012hybridization,ivanov2001photoionization}. The same figures as the ones for sodium are plotted in Figs. \ref{fig:C60grst}, \ref{fig:C60orbital} and \ref{fig:KSpotc60}.

\subsection{Static polarizabilities}

A first observable to be computed from the numerical simulation is the static polarizability as a response to an external electric field. We have taken into account only dipolar electric fields and in here there are two methods to compute this quantity: either you take only a slow external electric field and in this case, linearisation techniques are at hand, or the strength of the external field goes beyond linear and than full computations must be done. 

An example of linearized method for polarizability have been developed in \cite{palade2014general,serra1989static} in the frame of TF theory. This method can be used in connection with sodium clusters and fullerenes. Given the fact that its results are not impressive and are merely connected with systems with spherical symmetry in the GS, I shall not discuss it here. A single particle approach which is also linear takes into account the single particle effects is to use the linear response DFT described in Sect. \ref{LRDFT} or Sect. \ref{RPA} taking the linear response function in the limit $\omega\to 0$. 

Even though a little more involved numerically, the best approach is to take the KS equations and solve them in an external field. It has some advantages over the linearised methods: you can include any type of external field, not only a dipolar one and it works also for stronger fields which do not enter in the category of linearisations. On the downside, it can provide excitation-dependent results, which in the linear regime are not realistic.

\begin{figure}
	\centering
	\includegraphics[width=0.9\linewidth]{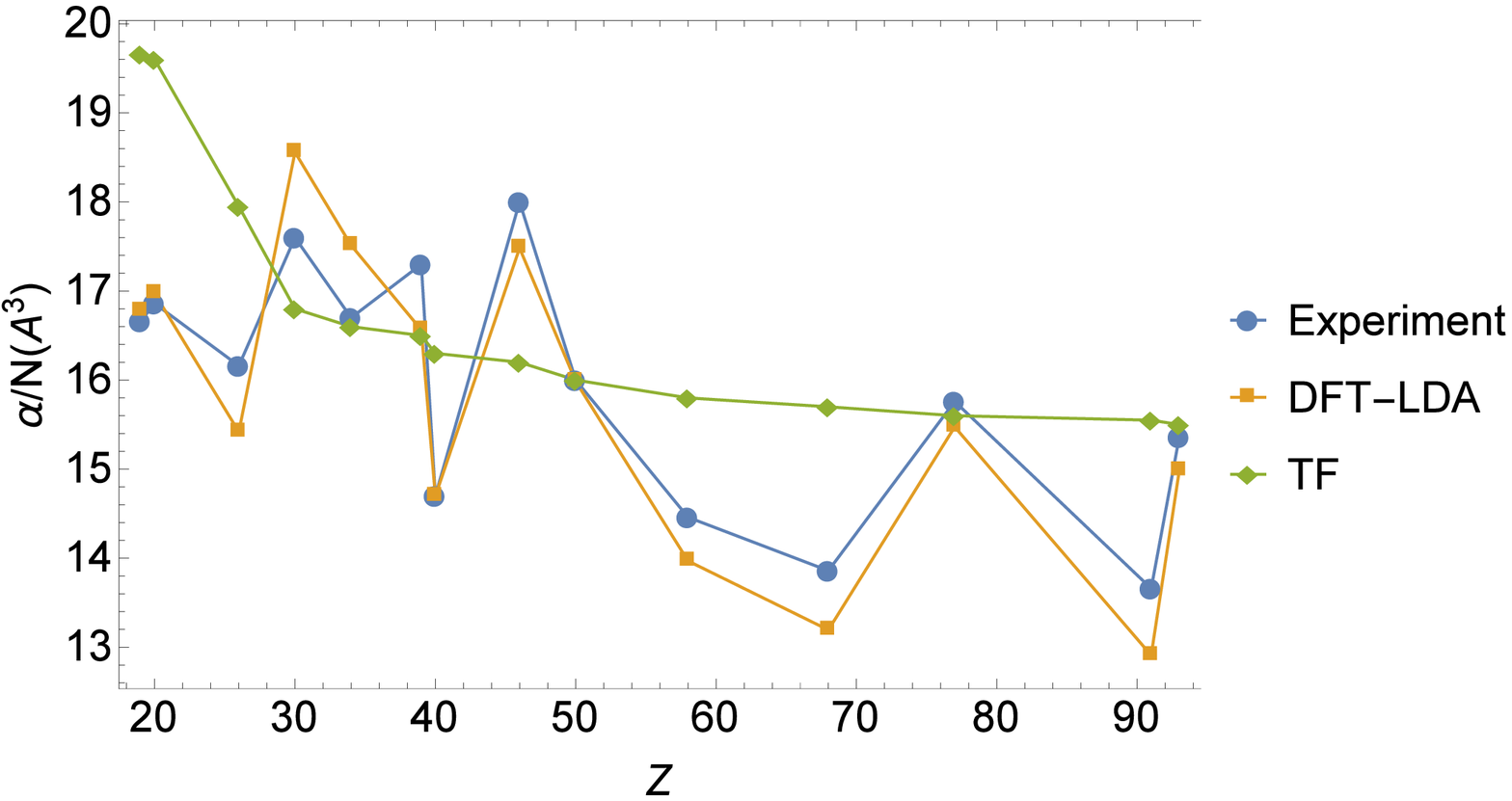}
	\caption{Comparison between the theoretical obtained polarizabilites for sodium clusters and the experiment \cite{tikhonov2001measurement}. The Orange points represent the results obtained solving the DFT-LDA with ionic structure. The green points are obtained with the linearized Thomas-Fermi in the frame of jellium model}
	\label{fig:polarizabilities}
\end{figure}

Again, the two pyllons of electronic structure computation have been employed: the TF and the DFT methods.

From the numerical experiments performed, an important aspect has been retained. While one could introduce from the start an external dipolar electric field and solve the stationary TF or KS equations, this approach induces some spurious numerical instabilities and the converges is harder to achieve. Perhaps is a consequence of the fact that the external potential has finite values on the boundary of the spatial box. Instead, I have solved first the ground state KS/TF equations and then the electric field has been introduced in a new series of iterations. This time the convergence is easily achieved since the "initial guess" is the true ground state and already close to the dipole-induced states.

To summarize the capabilities of TF and DFT-LDA methods, in Fig. \ref{fig:polarizabilities} I have plotted the results obtained in the frame of these two theories vs the experimental data, imported from \cite{tikhonov2001measurement}.

\section{Weak and moderates regimes}

As many times stated, the informations obtained about the ground state of a cluster are important from many points of view, especcially regarding the stationary properties of the system. Still, from the laser perspective, their main purpose is to serve as initial condition in the Cauchy type dynamical problem posed by the time dependency of the phenomena involved. 

For this reasons I shall go further in parallel with the classification from the first chapter and 
investigate the dynamics for weak and beyond strong regimes. Several quantities connected with specific types of experiments will be in the centre of this section: the optical response, the density of states obtained from photoelectron spectroscopy and the angular cross section obtained from photo angular spectroscopy.

As theoretical methods, DFT, with its time dependent extension, remains the main tool to investigate any of the above mentioned quantity since it gives direct access to real time-real space single particle behaviour. It is complemented for large systems by the Vlasov equation and Quantum Hydrodynamic Models.  When the laser fields are weak we can resort to the linearised approaches as RPA. If the laser fields are not weak, one should employ the ionic dynamics also in order to capture the deformations of the ionic backgrounds and their effect on different quantities.

\subsection{Vlasov interlude}

When going up on the size or laser intensity scale, Vlasov equation becomes the favourite theoretical approach. While it is a genuine differential equation in a $6+1D$ space, natural methods that are used for Schrodinger equation for example, do not work any more due to the high dimensionality. Therefore, one must resort to lagrangian methods described briefly in Sect. \ref{typesofeq}. I consider that some supplementary words should be spoken about the ideas behind this method. Just as a reminder, let us rewrite the VUU eq.:

\begin{equation}
	\label{vlasovnum}
	(\partial_t+\frac{\vec{p}}{m^*}\nabla_{\vec{r}}-\nabla_{r}U(r)\nabla_p)f(r,p,t)=I(f)
\end{equation}

Where $U$ is a mean field potential constitued from the external electric potential (from the laser) the LDA exchange correlation potential and the Hartree self consistent potential of electrostatic interaction between electrons. Further, we have mentioned in Sect. \ref{numericalrepr} that a numerical approach used for this equation is the lagrangian test particle method. To expand the idea, we choose a chopped super-complete basis of $\mathcal{N}$ gaussian functions in the phase-space $exp(-(\vec{r}-\vec{r}_i(t))^2/2\chi)exp(-(\vec{p}-\vec{p}_i(t))^2/2\phi)$. With these, the distribution function can be expanded (to obey the normalization condition):

\begin{eqnarray}
	f(r,p,t)=\frac{N}{\mathcal{N}(4\pi^2\chi\phi)^{3/2}}\frac{1}{(2\pi\hbar)^2}\sum_i^\mathcal{N}exp(-\frac{(\vec{r}-\vec{r}_i(t))^2}{2\chi})exp(-\frac{(\vec{p}-\vec{p}_i(t))^2}{2\phi})
\end{eqnarray} 

Without entering in details, if one introduces this expression in the eq. \ref{vlasovnum} the equations of motion are obtained $\dot{\vec{r}}_i=\vec{p}_i/m^*$ and $\dot{\vec{p}}_i=-\nabla_{r_i}U(\vec{r}_i)$. Now the problem in solving this system is how to evaluate the potential $U$ at the position of every particle. For this, we need to express the density (and suplementarry the current) in terms of test particles::

\begin{eqnarray}
	\rho(r)=\frac{N}{\mathcal{N}(4\pi^2\chi)^{3/2}}\frac{1}{(2\pi\hbar)^2}\sum_i^\mathcal{N}exp(-\frac{(\vec{r}-\vec{r}_i(t))^2}{2\chi})
\end{eqnarray}

Having the relation between density and particles positions it is easy (at least conceptually) to represent the density on a grid from the lagrangian particles, to compute the potential $U$ which is density dependent and then to interpolate the resulting force, in order to construct the force acting on each test particles.

In all the simulations performed for this thesis which involved test particles, the density was constructing using an auxiliary very refined grid. This grid was constructed as being regular with a cell length of $\le 5\chi$ and a numerical procedure of counting the particles inside each cell was applied. Due to the large number of test particles $\mathcal{N}\sim 10^5$, the density can be constructed with very small errors (around $1\%$) with an interpolation between the number of particles present in each cell. Finally, the interpolation function is applied on a less accurate grid on which the Poisson equation is solved and the exchange correlation potential. From this, in a finite difference description the forces are computed on the same second grid and  then interpolated to be computed on each pseudo-particle. This technique must be applied on each time step. 

\subsection{Optical response: plasmons}

The most stringent observable in connection with laser interaction in clusters is the optical response. Moreover, in the linear regime there are some specific signatures of different types of clusters, from which the most important are the giant resonances called plasmons.

On the other hand, choosing between full pseudopotentials or jellium model, the latter works good only for metal clusters and carbon. Moreover, even in here, if the cluster is small, generically with a number of atoms bellow $10^2$, there is a small shift of the plasmonic resonance towards red.

Plasmons can be seen in metal clusters especially, but also in fullerene, etc.  There is a long history in the plasmonic physics, these collective phenomena being firstly studied in macroscopic systems as metal interfaces where an incoming electromagnetic wave is known to induce a surface oscillation of the electronic charge. Further more, this oscillation can couple with other photon yielding a polariton.

\begin{figure}[h]
	\centering
	\includegraphics[width=0.8\linewidth]{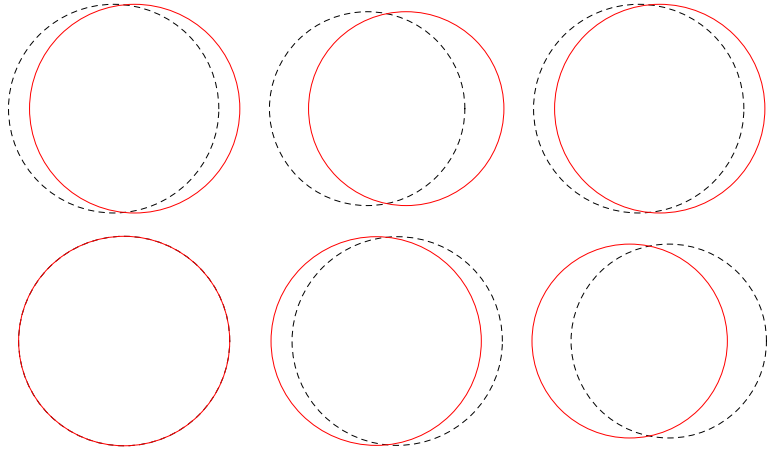}
	\caption{The ionic core with black dashed line and the electronic sphere with red line. Stages of the oscillation against each other are represented for intuitive image}
	\label{fig:mieplasmon}
\end{figure}

In clusters, the idea of plasmon has been introduced by analogy especially due to the Coulomb interaction. To give a classical picture of what happens during the plasmonic oscillation, imagine two charged sphere, one positive for ions crossed over by a negative one for electrons. Applying a short uniform electric field on this system, the ions will be driven in the sense of the electric field, while the electrons in opposite direction. The external interaction stops and then, an internal coulomb attraction appears between the ionic and electronic spheres and consequently, a dipolar oscillation of charge. This picture is represented in Fig. \ref{fig:mieplasmon}. In a quantum view, there are many components that contribute to the plasmon. In essence it can be obtained in the frame of RPA theories as a genuine collective excitation given by the coherent superposition of many single particle-hole excitations. It is the analogous of the GDR \cite{myers1977droplet} in nuclear physics. 

\begin{figure}[!h]
	\centering
	\includegraphics[width=0.8\linewidth]{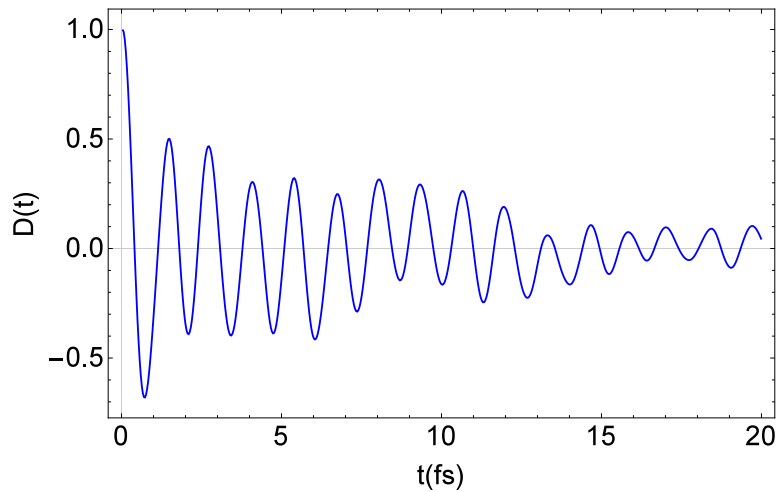}
	\caption{Dipole evolution in the first $20 fs$ in the $Na_{20}$ cluster after a dipolar initial shift. The results are computed with the Vlasov method}
	\label{fig:dipole}
\end{figure}

Due to the fact that the metal clusters have usually a spherical like geometry, this oscillations can be indeed viewed as variations of charge density on the surface, therefore, are called "surface plasmons". They are visible in photo-absorption or photo-ionization experiments. On the other hand, there are collective dipolar oscillations that induce density variations inside the volume of the clusters and these are called volume plasmons. They differ from the surface ones due to the fact that they have other specific resonance energies and can be excited only by non-uniform electric fields as the impact with charged projectiles in Electron Energy Loss experiments \cite{verkhovtsev2013plasmon}.

\begin{figure}[h]
	\centering
	\includegraphics[width=0.8\linewidth]{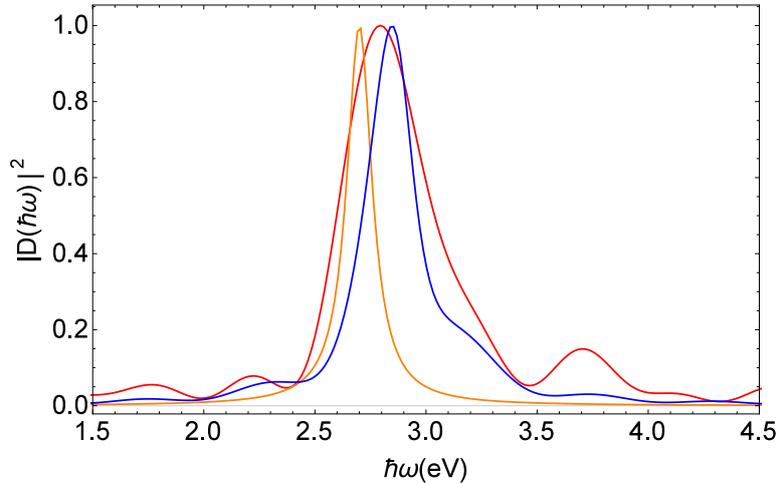}
	\caption{Optical response in the semi-linear regime (no ionization) of the $Na_{20}$ cluster obtained with three different methods: TD-DFT (red), Vlasov equation (blue) and TD-TF (orange). In all of them the jellium model have been used.}
	\label{fig:opticalna}
\end{figure}

The presence of the plasmon can be seen in the photo-ionization spectra which can be computed using the linearised approaches as RPA or LRDFT described in Sect. \ref{linearised} or using the link between the cross section and dipole moments $D(t)$ :

$$\sigma(\omega)\propto \omega \Im[\tilde{D}(\omega)]$$

As an example of typical dipole moment evolution obtained with full propagation of TDTF method, one can see Fig. \ref{fig:dipole}. This is a good example on which we can examine how good is each theoretical approach. First of all, they give roughly the same result on the position of the peak in the power spectra. This is not a surprise since all are capable to capture the main types of interaction: Coulomb and exchange while the interaction with the back-ground is the same (fixed in time) for each. It is a surprise that the Vlasov results are so close to the DFT one since first does not capture part of the quantum effects which are missed due to the $\hbar\to 0$ limit. On the other hand, TDTF is the method which gives the smallest damping (taken as the width of the peak). Again, this is not a surprise given the fact that the equation of state used in TF cancels the phase space features from the dynamics, therefore neglects the basic Landau damping. Not to mention the collisions which can be introduced only in an empirical manner. Vlasov and TDDFT present some smaller bumps further from the $2.8 eV$ position of the surface plasmon. This are not real collective phenomena, but are spurios numerical results due to the fact that the system was not investigated on a long enough period of time. Typically, I have used a $t_{max}=10 fs$, total time. 

\begin{figure}[h]
	\centering
	\includegraphics[width=0.8\linewidth]{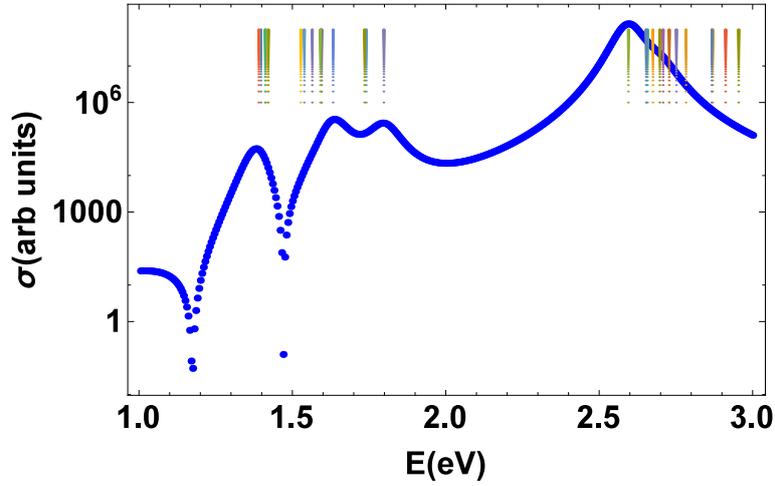}
	\caption{Absorption spectrum for $Na_8$ obtained in the frame of RPA calculations. In colour line there are plotted the obtained eigenvalues}
	\label{fig:RPAa}
\end{figure}

A set of RPA results can be seen in Fig. \ref{fig:RPAa} for the $Na_8$ cluster. In blue there is plotted the cross-section while in rainbow colours, the eigenvalues of the RPA matrix. In practice, the eigenvalue problem \ref{RPAeq} has been solved using the DFT ground state solutions. Any coefficient $A_{i\mu,j\nu}$ was computed as a two-body expectation value of the kernel of the KS potential:

$$\frac{\delta V_{KS}(\mathbf{r})}{\delta\rho(\mathbf{r'})}=\frac{e^2}{4\pi\varepsilon_0}\frac{1}{|\mathbf{r-r'}|}+\frac{\delta V_{xc}(\mathbf{r})}{\delta\rho(\mathbf{r'})}\delta(\mathbf{r-r'})$$

As it can be seen, the xc potential is taken in the frame of LDA, therefore, the kernel is local in the coordinate space. The particle-hole elementary excitations are taken to be combinations of occupied - unoccupied pairs of KS orbitals \cite{yabana2006real}. The major computational difficulty is the calculation of the kernel associated with the Coulomb interaction which is non-local and requires $6D$ integrals. In order to reduce the calculation costs, in practice one computes the Coulomb potential from a particle-hole pair and then the integral is computed in the $3D$ space:

\begin{eqnarray}
	\mathcal{C}_{i\mu}(\mathbf{r})=\int d\mathbf{r'}\frac{\psi_\mu(\mathbf{r'})\psi_i^*(\mathbf{r'})}{|\mathbf{r-r'}|}\\
	A_{i\mu,j\nu}=\int d\mathbf{r}\psi_j(\mathbf{r})\mathcal{C}_{i\mu}(\mathbf{r})\psi_\nu^*(\mathbf{r})
\end{eqnarray}

The case of $Na_8$ is a nice selected case since is one of the \emph{magic} clusters and it has almost spherical symmetry (reflected in the jellium profile). In general, deformed clusters, or clusters in which the symmetry is broken, require a fine grid for the computation of the orbitals. Moreover, the degeneracy of the electronic levels is removed and the space of particle-hole excitations is considerably enlarge. With these comes a multiplication of the computational effort which may not worth if one is interested in simple quantities as the position of the plasmon centroid. Some simplified approaches on RPA (as separable schemes) can be used \cite{kleinig1998plasmon,yannouleas1993evolution,nesterenko1997multipole,muta2002solving}.

\begin{figure}
	\centering
	\includegraphics[width=0.8\linewidth]{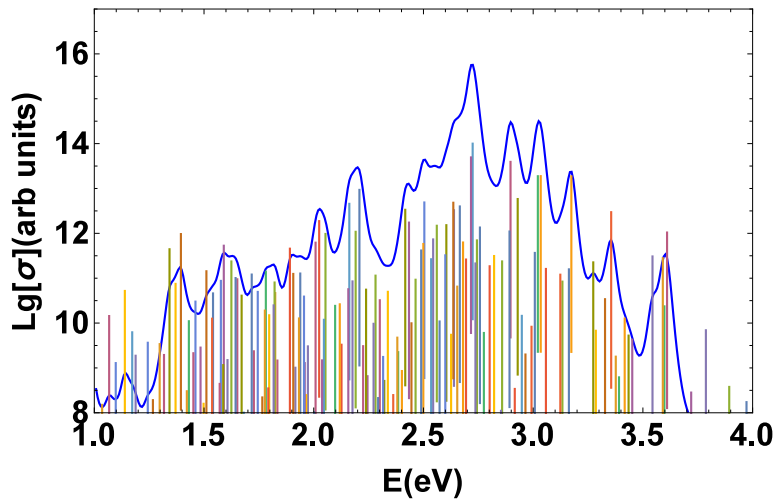}
	\caption{Absorption spectrum of the deformed $Na_{21}^+$ obtained in the frame of RPA calculations}
	\label{fig:rpa20}
\end{figure}

For the ground state calculations the shell effects make the net charge density somehow small relative to the total density and so the Coulomb potential has the same magnitude with the XC potential. In the dynamic regime, this is not the case anymore. If one would neglect the Coulombian self-consistent field in the time dependent regime, the resonances obtain would be around $1eV$ for $Na$ clusters, almost three times lower than the experimental value. On the other hand, neglecting the xc potential, the plasmon peak is recovered within a $\sim 0.1eV$ error. This gives us a good sense about the importance of the electrostatic interaction and the fact that it couples strongly with the dipolar motion of the electron cloud. 

\subsection{PES \& PAD}

As discussed in the first chapter, the single photoelectron processes are connected naturally with the electronic structure of the cluster. In order to analyze the density of states in the electron system from a dynamic perspective, we should resort to some specific laser-interaction and a quantity to be measured. In respect to the first, in PES experiments there are used laser pulses in trains with frequency \cite{gao2015analysis} in the UV or IR regimes. 

\begin{figure}[!h]
	\centering
	\includegraphics[width=0.8\linewidth]{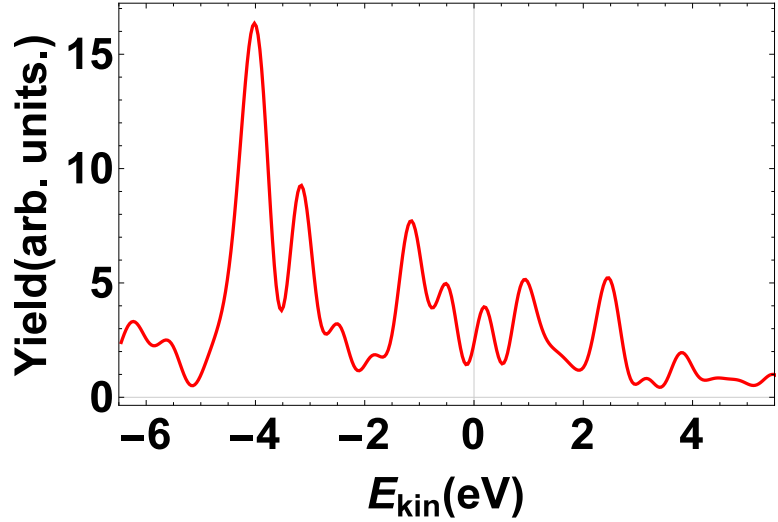}
	\caption{PES spectra obtained with TD DFT-LDA for the $Na_{41}^+$ cluster}
	\label{fig:PES}
\end{figure}

The quantity to be studied for PES spectra is \cite{gao2015analysis} the yield:

\begin{eqnarray}
	\mathcal{Y}_{\Omega_{r_\mathcal{M}}}(E_{kin})\propto\sum_{i=1}^N|\tilde{\psi}_j(r_\mathcal{M},E_{kin})|^2
\end{eqnarray}

Given the fact that the point $r_\mathcal{M}$ is supposed to be \emph{at infinity} and in practice sufficiently far from the cluster, we can consider that the mean field KS potential is zero, therefore the orbitals are represented through free particles. This implies a dispersion relation $\omega=k^2/2$ which allows us to represent the yield in terms of the kinetic energy but to compute it from the Fourier transform on time.   

The PES of a cluster can be computed also from kinetic approaches as Vlasov equation, following a similar recipe and investigating the Fourier Transformed of the density at the far-point $r_{\mathcal{M}}$: $\mathcal{Y}_{\Omega_{r_\mathcal{M}}}(E_{kin})\propto|\tilde{\rho}(r_\mathcal{M},E_{kin})|^2$. Some detailed discussions can be found in \cite{de2012ab}. 

\begin{figure}[h]
	\centering
	\includegraphics[width=0.8\linewidth]{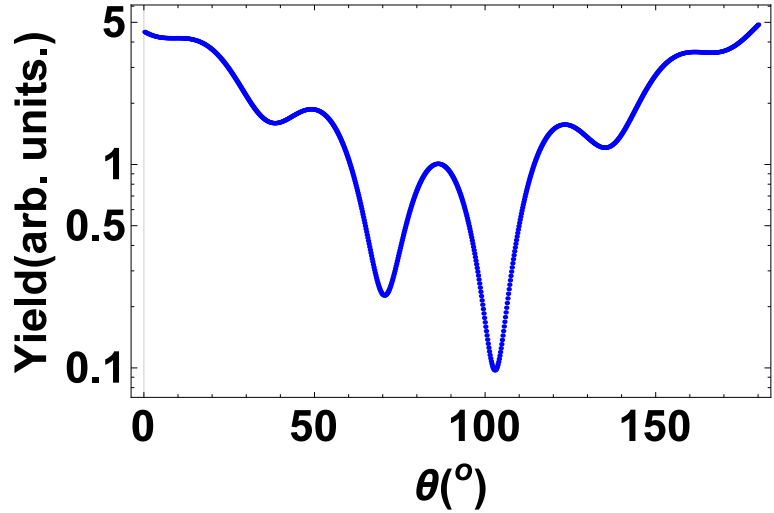}
	\caption{PES spectra obtained with TD DFT-LDA for the $Na_{9}^+$ cluster}
	\label{fig:PAD}
\end{figure}

In Fig. \ref{fig:PAD} is presented a generic PAD result obtained in the frame of DFT for the $Na_9^+$ cluster. More results about the PAD spectra can be found in \cite{bartels2009probing,seideman2002time}.

\section{Strong regime}

Already for moderate intensities of the laser field, any tentative to apply linearized methods breaks totally and one has to use full time propagation schemes. As discussed in \ref{nanoplasma} section, even these methods break due to the problem of multiple scale that arise when you go in the strong regime. Therefore, the last resort measure is to implement expensive (for large clusters) molecular dynamics techniques or to extract gross properties of the system using the nano-plasma model.

\begin{figure}[!h]
	\centering
	\includegraphics[width=0.8\linewidth]{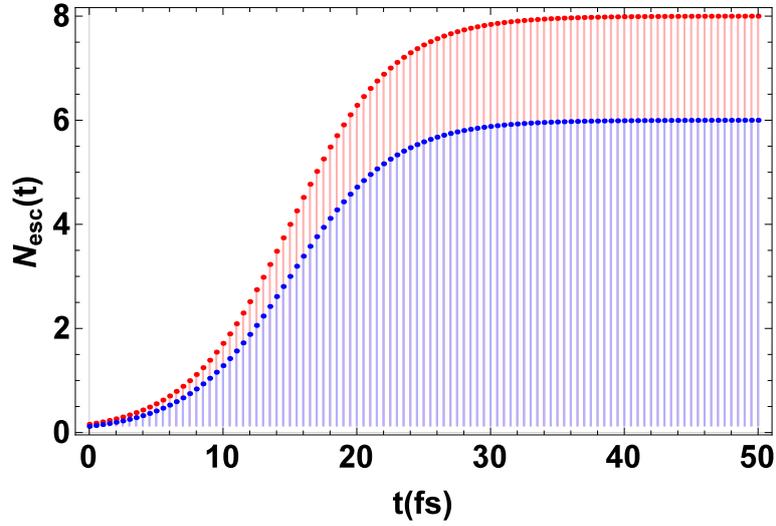}
	\caption{Results for the electron ionization of the $Na_{139}$ cluster obtained with VUU (red) and DFT(blue) in a laser field of $I=10^{16}$}
	\label{fig:excape}
\end{figure}

For an intermediate situation, in Fig. \ref{fig:excape} there are presented the results obtained with TDLDA (blue) and VUU method (red) regarding the number of escaped electrons in a metal cluster $Na_{139}$. The system has been excited with a gaussian laser pulse of peak intensity $I=10^{16}W/cm^2$ and a time width of $1fs$.  Interesting enough, the DFT method predicts a lower ionization of the cluster. This might be explained from many differences that are present between these two theories. First of all, the initial distribution function $f(\mathbf{r,p},0)$ has been chosen identical with the solution of the Thomas Fermi theory which means that the kinetic distribution of the electronic velocities might be larger than the one prescribed by DFT. 

\begin{figure}[!h]
	\centering
	\includegraphics[width=0.8\linewidth]{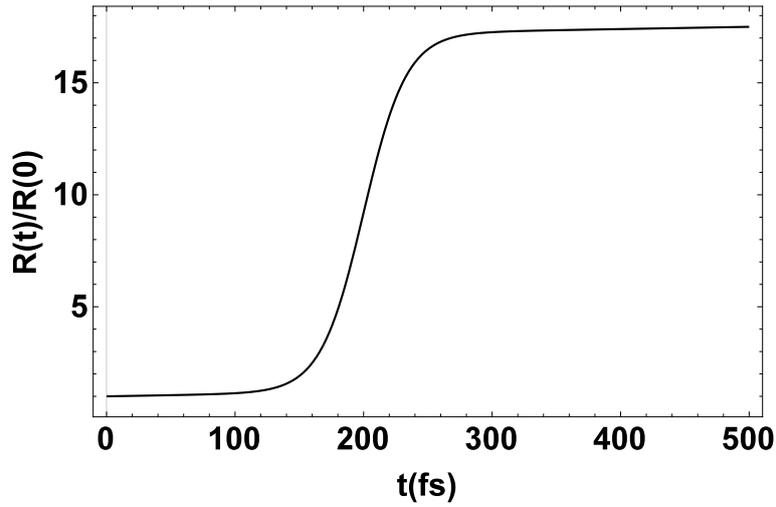}
	\caption{Time evolution of the total radius for the $Xe_{1400}$ in a laser field of $I=10^{18}W/cm^2$ peak intensity; the results are computed with the nano-plasma model \ref{nanoplasma}}
	\label{fig:radius}
\end{figure}

\begin{figure}[!h]
	\centering
	\includegraphics[width=0.8\linewidth]{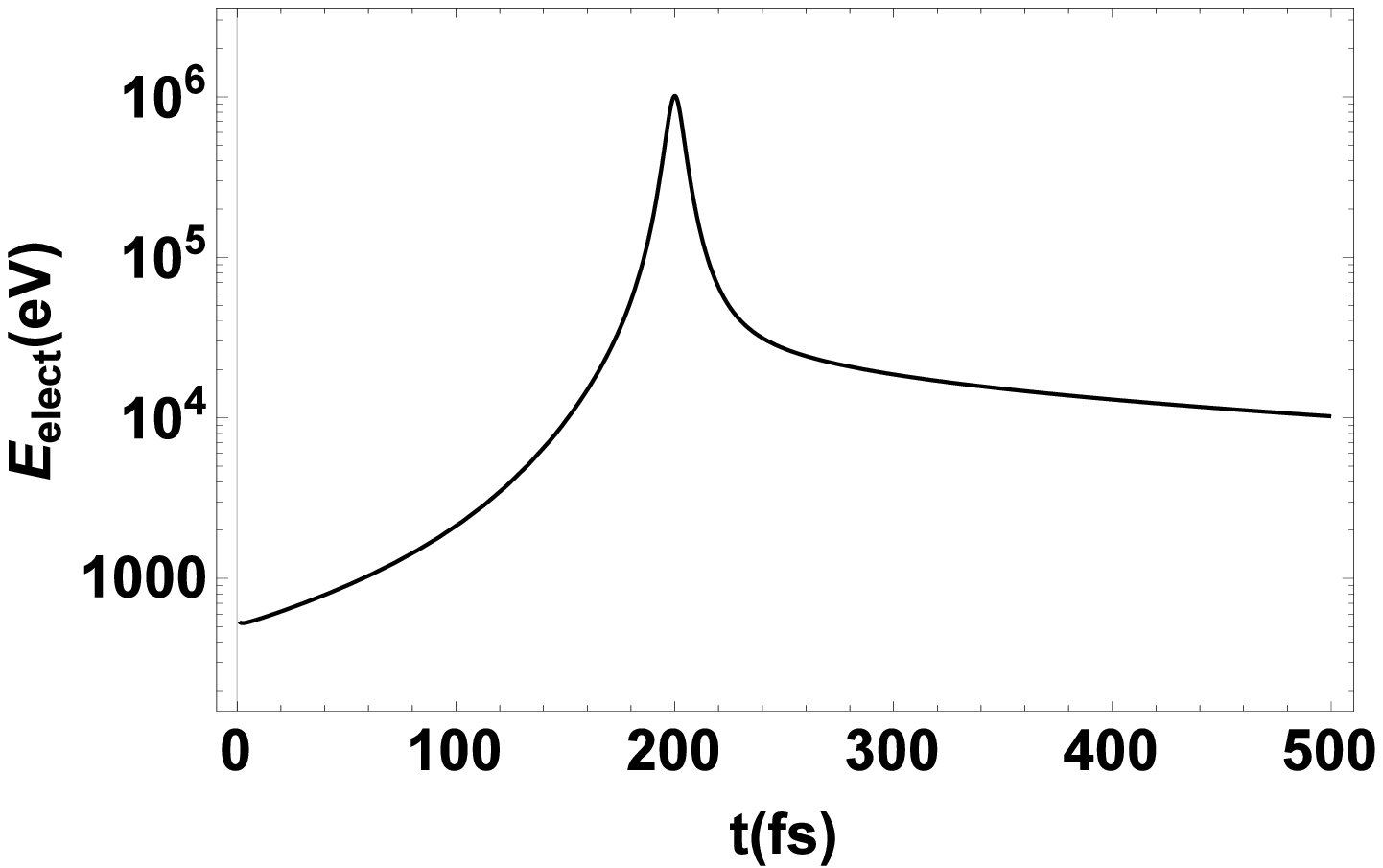}
	\caption{Electron energy distribution for $Xe_{1400}$ in a laser field of $I=10^{18}W/cm^2$ peak intensity; the results are computed with the nano-plasma model}
	\label{fig:nanoplasma}
\end{figure}

Second, the UU collision integral can be a source of ionization at the surface of the electron cloud giving rise to some kind of electron evaporation. Finally, there might be numerical artefacts inherent in the method which could explain this differences. In principle, the ionization in DFT is obtained using a mask on the boundary of the domain which is very sensitive to various parameters of input. In VUU, this is not the case, the particles being able to fly out the system. 

\begin{figure}
	\centering
	\includegraphics[width=0.8\linewidth]{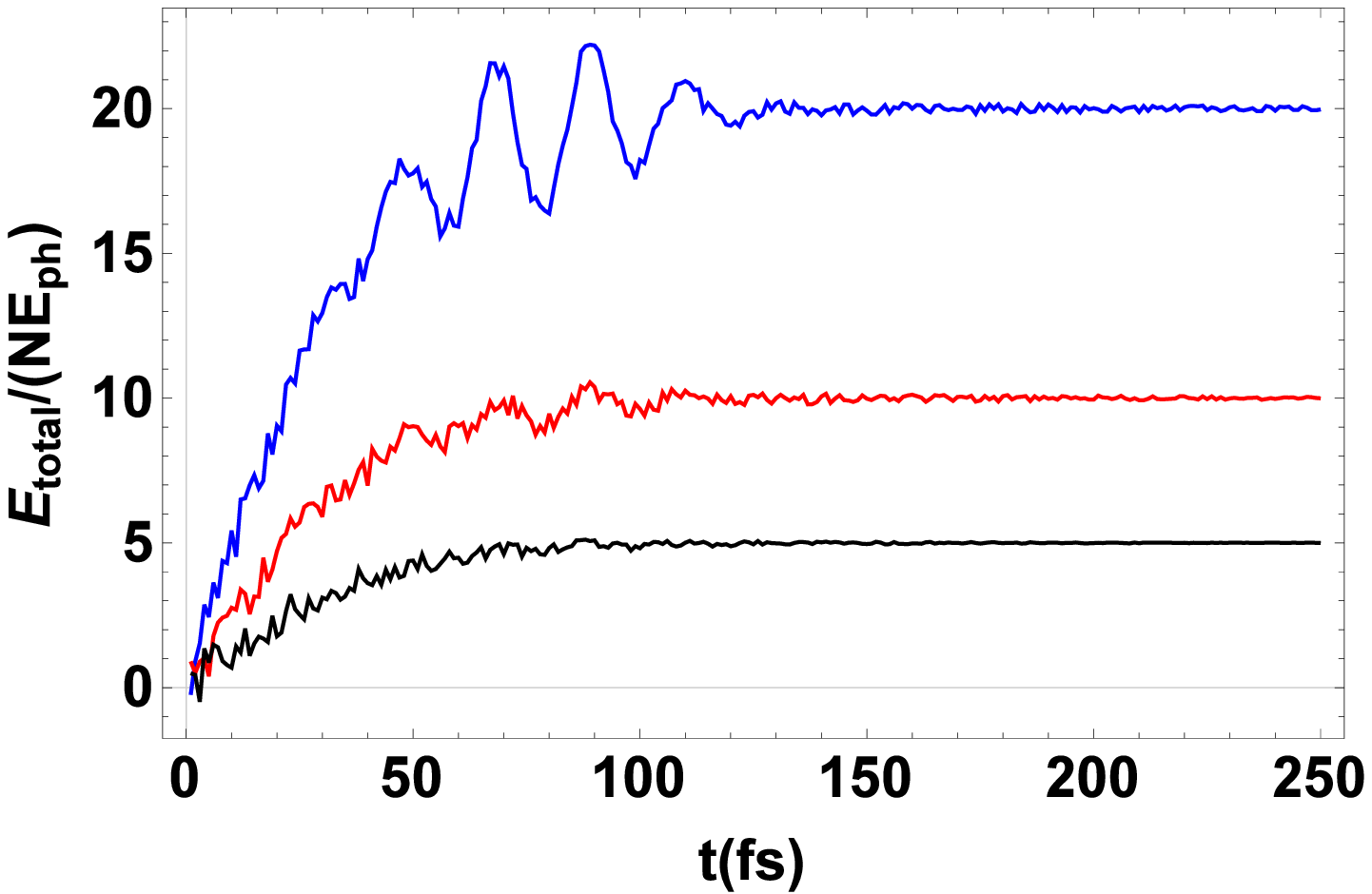}
	\caption{Time evolution of the absorbed energy in $Xe_{147}$ (blue), $Xe_{88}$ (red) and $Xe_{55}$ clusters calculated with the lagrangian TDTF method}
	\label{fig:rusek1}
\end{figure}

Going to the nano-plasma model the system of equations presented in \ref{nanoplasmamodels}, for the $Xe_{400}$ in a gaussian laser pulse of intensity $I=10^{18}W/cm^2$ we have obtained the time dependency of the radius of the nano-plasma in good qualitative agreement with other theoretical and experimental studies.

Finally, the spectrum of emitted electrons from this case has been computed and is presented in fig. \ref{fig:nanoplasma}. 

\begin{figure}
	\centering
	\includegraphics[width=0.8\linewidth]{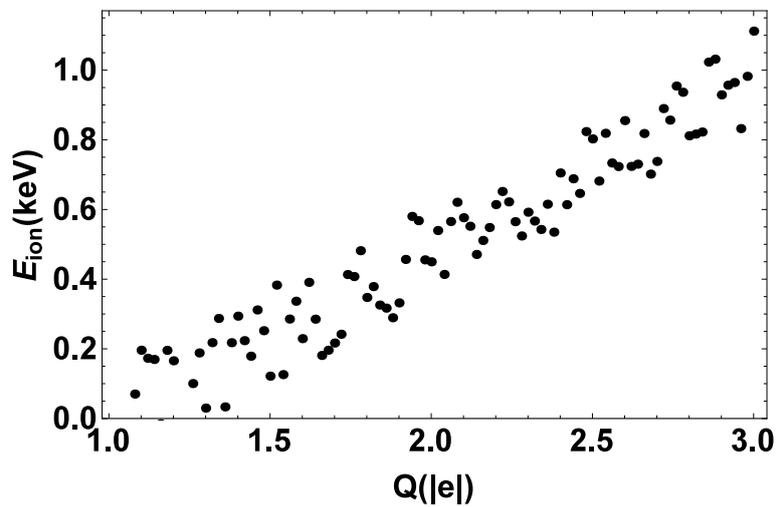}
	\caption{Ionic energy distribution from $Xe_{147}$ cluster vs. various initial ionizations.}
	\label{fig:rusek12eps}
\end{figure}

Further numerical studies on the Molecular Dynamics method should be done to complete this numerical section, but to the date the codes written in this direction are not numerically stable to give reliable results. 

In turn, a hybrid method (presented in \cite{rusek2000cluster,rusek2005different}) based on Smoothed Particle Hydrodynamics for the Time Dependent Thomas Fermi Theory has been reproduced numerically for the study of rare-gas clusters, in particular $Xe_{n}$. Such results can be seen in \ref{fig:rusek1} for the total energy absorption in an UV laser field and in \ref{fig:rusek12eps} for the associated distribution of ionic energies resulting under the Coulomb explosion. 

% Chapter 1

\chapter*{Conclusions and perspectives} % Write in your own chapter title
\label{Conclusions}

The present thesis tackles the problem of laser-cluster phenomena from an overview perspective emphasising the theoretical methods to be used. The first chapter is designed to be a short introduction in the concept of cluster, laser and describes the expected characteristic scales. Further, following the concept of ionization and the intensity of the laser field, three main regimes of interactions are discussed: weak (linear) dominated by single-photon processes, moderate dominated by multi-photon processes and strong in which the nano-plasma state appears and the dynamics of the cluster is followed by a Coulomb explosion. 

The second chapter is the core of this thesis since tries a schematic hierarchy of the theoretical tools that are capable of modelling the cluster structure and dynamics. It starts with an axiomatic view of the density matrix formalism and the Quantum Liouville equation and then, passes through various approximations to obtain Hartree-Fock, Density Functional Theory, Vlasov equation, the Quantum Hydrodynamic Models and the classical models in terms of Molecular Dynamics and nano-plasma model.

Finally, the third chapter, is focused on the how the equations developed in various theories can be tackled numerically. After a short discussion about numerical representations generic types of equations (Poisson, TD Schrodinger, Vlasov, etc.) are detailed since they require various tricks that should be implemented in the numerical scheme. Starting with the ground state properties of metal clusters, fullerene and $Xe$ clusters and ending in the strong regime, up to my present capabilities, some numerical simulations have been performed and results are presented to emphasize the capabilities of the theoretical methods described in chapter 2.

Obviously, there is much more to learn for the author in the subject and a lot more to construct in term of numerical codes and results. The main perspective is to investigate the intense and ultra-intense regime and to study how the problem posed by the multi-scale phenomena could be tackled.

\end{document}